\documentclass[12pt]{article}

\usepackage{verbatim,color,amssymb}
\usepackage{amsmath}					
\usepackage{amsthm}					
\usepackage{natbib}
\usepackage{multirow}
\usepackage{setspace}
\usepackage[mathscr]{euscript}
\usepackage{fancyhdr}
\usepackage{enumitem}
\usepackage{graphicx}
\usepackage{geometry}
\usepackage[bookmarks=false]{hyperref}
\usepackage{dsfont}
\usepackage[table]{xcolor}
\usepackage{tikz}
\usetikzlibrary{arrows}
\usetikzlibrary{patterns}
\usepackage{algorithm, algorithmicx}

\usepackage{multibib}
\newcites{latex}{References}

\usepackage{tabularx, booktabs}
\newcolumntype{Y}{>{\centering\arraybackslash}X}
\usepackage{lscape}

\usepackage{subfig}
\usepackage{float}
\usepackage{lineno}

\makeatletter
\def\thickhline{%
  \noalign{\ifnum0=`}\fi\hrule \@height \thickarrayrulewidth \futurelet
   \reserved@a\@xthickhline}
\def\@xthickhline{\ifx\reserved@a\thickhline
               \vskip\doublerulesep
               \vskip-\thickarrayrulewidth
             \fi
      \ifnum0=`{\fi}}
\makeatother
\newlength{\thickarrayrulewidth}
\newtheorem{Thm}{\underline{\bf Theorem}}

\newtheorem*{Proof*}{Proof}

\def\eE{\mathbb{E}}

\def\F{{\cal F}}

\def\H{{\cal H}}

\def\X{{\cal X}}
\def\Z{{\cal Z}}

\def\data{\mathrm{data}}

\def\log{\hbox{log}}

\def\Dir{\hbox{Dir}}

\def\Ga{\hbox{Ga}}

\def\HC{\hbox{C}^{+}}

\def\IG{\hbox{Inv-Ga}}

\def\MVN{\hbox{MVN}}

\def\Normal{\hbox{Normal}}

\def\Unif{\hbox{Unif}}
\def\Mult{\hbox{Mult}}

\def\P_25_ICML{{\it Proceedings of the 25th international conference on Machine learning}}

\def\bse{\begin{eqnarray*}}
\def\ese{\end{eqnarray*}}
\def\be{\begin{eqnarray}}
\def\ee{\end{eqnarray}}
\def\bq{\begin{equation}}
\def\eq{\end{equation}}

\def\trans{^{\rm T}}

\def\th{^{th}}

\def\ba{{\mathbf a}}
\def\bA{{\mathbf A}}
\def\bb{{\mathbf b}}
\def\bB{{\mathbf B}}

\def\bD{{\mathbf D}}
\def\b1e{{\mathbf e}}
\def\b1f{{\mathbf f}}

\def\bI{{\mathbf I}}

\def\bP{{\mathbf P}}
\def\br{{\mathbf r}}

\def\bx{{\mathbf x}}

\def\by{{\mathbf y}}

\def\bz{{\mathbf z}}

\def\bzero{{\mathbf 0}}

\newcommand{\bmu}{\mbox{\boldmath $\mu$}}

\newcommand{\bpi}{\mbox{\boldmath $\pi$}}

\newcommand{\btheta}{\mbox{\boldmath $\theta$}}

\newcommand{\bbeta}{\mbox{\boldmath $\beta$}}

\newcommand{\bzeta}{\mbox{\boldmath $\zeta$}}

\newcommand{\bSigma}{\mbox{\boldmath $\Sigma$}}

\newcommand{\abs}[1]{\left\vert#1\right\vert}
\newcommand{\norm}[1]{\left\Vert#1\right\Vert}

\renewcommand\footnoterule{\kern-3pt \hrule \textwidth 2in \kern 2.6pt}

\newcommand\Algphase[1]{%
\vspace*{-.5\baselineskip}\Statex\hspace*{\dimexpr-\algorithmicindent-2pt\relax}\rule{\textwidth}{0.4pt}%
\Statex\hspace*{-\algorithmicindent}\textbf{#1}%
\vspace*{-.5\baselineskip}\Statex\hspace*{\dimexpr-\algorithmicindent-2pt\relax}\rule{\textwidth}{0.4pt}%
}

\def\boxit#1{\vbox{\hrule\hbox{\vrule\kern6pt \vbox{\kern6pt \textcolor{blue}{#1}\kern6pt}\kern6pt\vrule}\hrule}}

\def\authorfootnote#1{{\let\thefootnote\relax\footnotetext{#1}}}

\allowdisplaybreaks

\begin{document}
\thispagestyle{empty}
\baselineskip=28pt

\begin{center}
{\Large{\bf Bayesian Semiparametric\\ 
\vskip-10pt Hidden Markov Tensor Partition Models\\ 
\vskip-10pt for Longitudinal Data with\\
\vskip-3pt Local Variable Selection}}
\end{center}
\baselineskip=12pt
\vskip 20pt

\begin{center}
Giorgio Paulon$^{a}$ (giorgio.paulon@utexas.edu)\\
Peter M\"uller$^{a,b}$ (pmueller@math.utexas.edu)\\
Abhra Sarkar$^{a}$ (abhra.sarkar@utexas.edu)\\

\vskip 7mm
$^{a}$Department of Statistics and Data Sciences, \\
The University of Texas at Austin,\\ 2317 Speedway D9800, Austin, TX 78712-1823, USA\\
\vskip 8pt 
$^{b}$Department of Mathematics, \\
The University of Texas at Austin,\\ 2515 Speedway C1200, Austin, TX 78712-1202, USA\\
\end{center}

\vskip 20pt 
\begin{center}
{\Large{\bf Abstract}} 
\end{center}
\baselineskip=12pt

We present a flexible Bayesian semiparametric mixed model for longitudinal data analysis in the presence of potentially high-dimensional categorical covariates. 
{{Building on a novel hidden Markov tensor decomposition technique, }
our proposed method allows the fixed effects components to vary between dependent random partitions of the covariate space at different time points. 
The mechanism not only allows different sets of covariates to be included in the model at different time points 
but also allows the selected predictors' influences to vary flexibly over time. 
Smooth time-varying additive random effects are used to capture subject specific heterogeneity. 
{We establish posterior convergence guarantees for both function estimation and variable selection.}
We design a Markov chain Monte Carlo algorithm for posterior computation. 
We evaluate the method's empirical performances through synthetic experiments and demonstrate its practical utility through real world applications.

\vskip 20pt 
\baselineskip=12pt
\noindent\underline{\bf Key Words}:  
Factorial hidden Markov models (fHMM), 
Higher order singular value decomposition (HOSVD), 
Local variable selection, 
Longitudinal data,
Partition models

\par\medskip\noindent
\underline{\bf Short/Running Title}: Longitudinal Functional Mixed Models

\par\medskip\noindent
\underline{\bf Corresponding Author}: Abhra Sarkar (abhra.sarkar@utexas.edu)

\pagenumbering{arabic}
\setcounter{page}{0}
\newlength{\gnat}
\setlength{\gnat}{16pt}
\baselineskip=\gnat

\newpage

\section{Introduction}

We propose a novel statistical framework for modeling longitudinally varying continuous response trajectories in the presence of categorical covariates. 
{Building on novel hidden Markov tensor decompositions, 
our approach is especially suited to high-dimensional settings, efficiently eliminating the redundant covariates thereby allowing time-varying variable selection
while also parsimoniously representing higher order interactions between the selected predictors. 
}

The settings analyzed here may be viewed as longitudinal adaptations of static analysis of variance (ANOVA) designs 
and hence are very generic and almost ubiquitously encountered in modern scientific research in many diverse fields, 
examples from recent statistics literature including pharmacodynamics \citep{de2004anova}, mass spectroscopy \citep{morris2006wavelet}, early pregnancy loss studies \citep{maclehose2009nonparametric}, etc.
In such scenarios, assessing the local variations in the response profiles, 
including especially 
how the associated predictors might influence the response differently in different stages of the longitudinal process, 
can provide valuable insights into the underlying data generating mechanisms.
{Figure \ref{fig: intro_example} shows a synthetic illustrative example where the mean profiles of a continuous response $y$ vary smoothly over time. 
Two associated covariates, $x_{1}$ and $x_{3}$, from a set of ten total available $\{x_{1},\dots,x_{10}\}$ are important 
and they jointly influence the response means differently in different longitudinal stages.
The goal of this article is to understand such complex dynamics from data on response and covariate values.} 
\begin{figure}[!ht]
	\centering
	\includegraphics[width=0.55\linewidth]{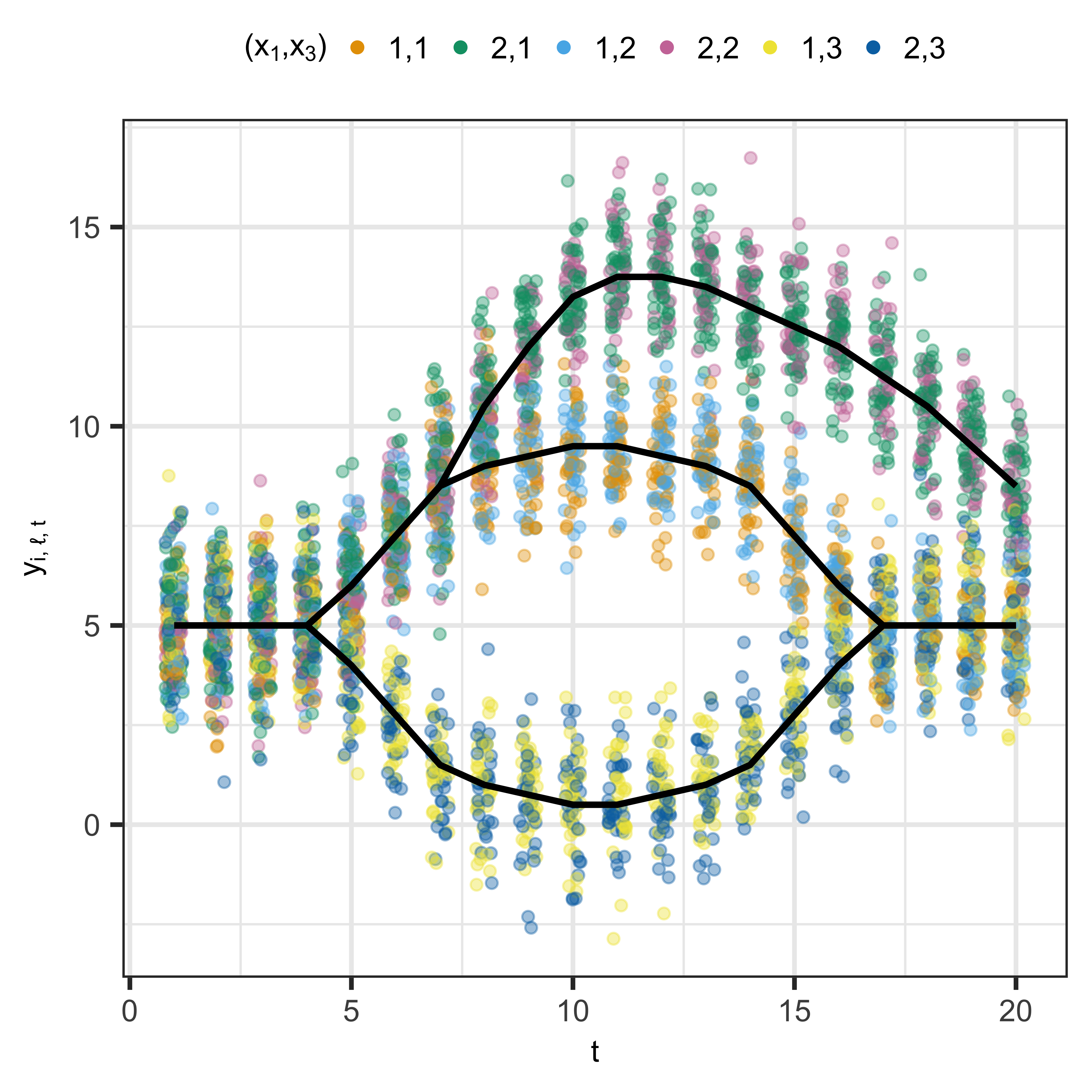}
	\caption{\baselineskip=10pt 
	A synthetic scenario with ten covariates $(x_{1},\dots,x_{10})$. 
	No covariate is globally important 
	but $(x_{1}, x_{3})$ are locally important: 
	They have no influence on $y$ for $t \in [1,4]$ but have a complex pattern of joint influence on $y$ for $t \in (4,20]$.
	The response values $y_{i,\ell,t}$ are represented here as slightly jittered points for all combination of the levels of the significant predictors $(x_{1}, x_{3})$. 
	The true underlying mean functions are superimposed (black lines).
	}
	\label{fig: intro_example}
\end{figure}

{\bf Existing Methods:} 
The literature on longitudinal data analysis is really vast 
\citep[see, for example, books by][and the references therein]{diggle2002analysis, singer2003applied, fitzmaurice2008longitudinal}. 
Bayesian methods for longitudinal data have also been extensively developed \citep[][etc.]{daniels2002bayesian, chib2002semiparametric, li2010bayesian, muller2013bayesian, quintana2016bayesian}.
However, the problem of characterizing dynamically varying variable importance in such settings has not received much attention. 
This article presents a novel Bayesian semiparametric method that addresses such needs.

Our work in this direction was inspired by 
the existing sparse literature on local clustering in functional data \citep{duan2007generalized, petrone2009hybrid, nguyen2010inference, nguyen2011dirichlet}. 
These Bayesian nonparametric approaches assume that the mean functions of interest can be represented by a smaller set of canonical curves that are in turn modeled, 
for instance, as independently and identically distributed (i.i.d.) realizations from a stationary Gaussian process.
\citet{gelfand2005bayesian} specify an infinite mixture of these global functional atoms in which each observation is a noisy realization around a draw from the set of canonical curves. 
Such an approach allows for curves that are either completely different or completely identical across the entire functional domain, capturing only their global difference patterns. 
Many applications, however, involve data exhibiting local heterogeneity. 
Local clustering in such cases could 
greatly improve estimation and prediction, borrowing information across locally homogenous regions, 
as well as interpretability and inference, providing potentially interesting insights into the underlying causes of local heterogeneity.  
Toward this goal, \citet{duan2007generalized} proposed a solution by defining a stick-breaking construction at each location, which allows for local selection of curves.
\citet{petrone2009hybrid} assumed that the individual curves can be obtained as hybrid species defined as recombinations of different portions of the canonical curves. 
Both these approaches define the local allocation rules by means of a single hidden labeling process that indicates which canonical curve is chosen at each time stamp. 
Additional challenges are represented by the choice of functional dependence in the labeling process, whose theoretical properties have been studied by \citet{nguyen2011dirichlet}.
\cite{suarez2016bayesian} proposed an alternative approach, 
using independent priors at different time points to cluster wavelet basis coefficients first, 
but then using these local features to find global functional clusters as the final inference goal.

The approaches mentioned above have limitations that deserve attention.
First, defining the mean functions as recombinations of canonical curves implies that 
these curves are discontinuous, which can be an impractical assumption in most applications. 
Continuous curves may be desired, for instance, in dose-response relationships \citep{de2004anova}.
Second, the inclusion of covariates in these models has only been accomplished via an additive term in the mean function.
Even when flexible random effects are used, the linearity assumption of the covariate effects can be quite restrictive in practice.
Furthermore, the problem of dynamically characterizing variable importance in these settings has not been addressed.

Alternative approaches to model time-varying predictor effects and interactions in longitudinal data include tree based methods. 
Bayesian additive regression trees (BART) \citep{chipman2010bart} perform well when the regression function consists of low order nonlinear interactions.
With time as an additional covariate, these models can be adapted to capture longitudinally varying influences of the predictors \citep{sparapani2016nonparametric}. 
Separate ideas involving a single tree have also been proposed \citep{taddy2011dynamic,gramacy2013variable} where the tree structure evolves when new data streams become available.
\citet{linero2018bayesiansmooth} and \citet{starling2020bart} proposed smoothing the covariate effects 
which yields more appropriate results when the outcome is expected to vary smoothly over time.
{These models, albeit flexible, do not directly assess the local influence of each individual predictor 
but measure variable importance by calculating their contributions to reducing the in-sample mean squared error. 
With such heavy emphasis on prediction, they often include many weakly informative or even spurious predictors 
in the ensembles and lack parsimony and interpretability as a result \citep{breiman2001statistical,efron2020prediction}.}

Yet another related strategy comprises varying coefficients (VC) regression models 
where the regression coefficients are allowed to smoothly vary over a set of chosen modifiers \citep{hastie1993varying}. 
VC models have been adapted to longitudinal data by considering time as the only modifier \citep{hoover1998nonparametric}.
More recently BART priors \citep{deshpande2020vc} and variable selection techniques \citep{koslovsky2020bayesian} have also been adapted to VC settings.
While VC models allow for an easy assessment of the predictors' importance, 
they are restricted in their ability to accommodate interactions between predictors. 
For example, for $p$ categorical predictors $x_{j} \in \{1,\dots,x_{j,\max}\}, j=1,\dots,p$, it is necessary to include $\sum_{j=1}^{p} (x_{j,\max} - 1)$ dummy variables for the main effects, $\sum_{j_{1} \neq j_{2}} (x_{j_{1},\max} - 1) (x_{j_{2},\max} - 1)$ for the first order interactions, and so on.

{\bf Our Proposed Approach:} 
We propose a longitudinal functional mixed effects model that combines predictive power and interpretability 
by addressing the limitations of the local clustering approaches cited above. 
Most existing methods imply a tension between the main goals of statistical analysis \citep{breiman2001statistical}, 
namely estimation, attribution and prediction \citep{efron2020prediction}.
Our proposed approach tries to strike a balance - 
it is highly flexible, being able to accommodate higher order interactions between the predictors, 
but also favors parsimony, modeling these complex effects implicitly and compactly, 
while also allowing some ease of interpretation, including explicitly encoding each predictor's varying overall significance at different time points. 
Our method also comes with theoretical guarantees for both function estimation and variable selection.

The construction of our proposed model proceeds by characterizing the longitudinal evolution of both the predictor dependent fixed effects and the subject specific random effects 
as flexible functions of time \citep{ramsay2007applied, morris2015functional, wang2016functional}
modeled by mixtures of locally supported spline bases \citep{de1978practical, eilers1996flexible}.
The fixed effects model spline coefficients are allowed to vary with the associated predictors' level combinations, thereby accommodating all order interactions between them. 
Structuring these coefficients as multi-way tensors and applying a novel higher order singular value (HOSVD) type decomposition 
\citep{tucker:1966, de_lathauwer_etal:2000,kolda2009tensor}, 
we reduce the high-dimensional problem of modeling the complex joint influence of many different predictors to that of estimating much smaller-dimensional core coefficients. 
In effect, this induces a local partitioning of the joint covariate space 
such that the different predictor level combinations belonging to the same partition set will have a similar effect on the response variable. 
The local partitions constructed this way can in fact be indexed by combinations of separate latent allocation indicators, 
one for each level of the associated categorical predictors, 
facilitating separate assessment of the influences of each individual covariate \citep{sarkar2016bayesian}.  
To induce dependence between the adjacent local partitions, we allow the latent allocation indicators evolve according to a factorial hidden Markov model (fHMM) \citep{ghahramani1996factorial}. 
In constructing the model this way, we break free from the assumption of separate canonical curves of the previously existing Bayesian nonparametric literature cited above 
but allow the dependencies across adjacent temporal locations 
be further informed by the associated local partition configurations through 
a novel conditionally Markov prior on the core spline coefficients, conditional on the partition structure, improving model interpretability and estimation efficiency.  
The proposed functional approach also has the important advantage of avoiding to have to impute missing data when they are missing under simple mechanisms \citep{little2019statistical}. 
{We establish theoretical results on posterior consistency of the proposed method for both function estimation and variable selection.} 
We evaluate its numerical performance in simulation experiments where it significantly outperformed its competitors 
not just on average but also uniformly in all simulation instances. 
Finally, we illustrate the method's practical performance in real data applications from diverse domains.

{The methodology presented here is highly generic and broadly adaptable to diverse problems. 
For instance, \cite{paulon2020driftdiff} developed a similar local clustering method in the presence of a single categorical predictor $x$ with a small number of levels 
for a specific application with a complex drift-diffusion likelihood function. 
The focus of this article, however, is on developing a general methodology 
with an emphasis on the multivariate case $(x_{1}, \dots, x_{p})$ which presents significant and unique additional modeling and computational challenges. 
For instance, redefining the $\prod_{j=1}^{p}x_{j,\max}$ level combinations of $(x_{1}, \dots, x_{p})$ as the levels of a new single predictor $x$, 
while conceptually straightforward, does not provide a practically effective solution as it 
does not allow separate characterization of the local importances of the different predictors and, 
with $\prod_{j=1}^{p}x_{j,\max}$ increasing exponentially fast with $p$, 
quickly becomes computationally inefficient even in small to moderate dimensional problems. 
The strategy is practically useless, for instance, in a real data applications we discuss in Section \ref{sec: applications}, where $\prod_{j=1}^{p}x_{j,\max} = 580,608$. 
Our proposed dynamic HOSVD based approach, in contrast, not only provides a flexible and highly efficient tool for dimension reduction and simultaneous variable selection 
but also does this locally at each time point while borrowing information across a number of levels.}

At a basic level, the proposed methodology operates on a very simple idea. 
By way of our construction, a predictor $x_{j}$ taking values in $\{1,\dots, x_{j,\max}\}$ is selected to be important in influencing the response, 
if its levels are clustered in at least two different sets, 
the levels belonging to any particular cluster influencing the response similarly 
but the levels belonging to different clusters influencing the response differently.   
If, on the other hand, there is no clustering, or, put differently, all its $x_{j,\max}$ levels are clustered together, 
that would imply that there is no influence of the predictor on the response, 
and $x_{j}$ will then not be selected as an important predictor of the response. 
Modeling the joint influences of the important predictors varying flexibly over time is still a daunting challenge 
and is achieved via our novel use of dynamic tensor factorization, fHMM, smoothing splines, etc.

Our proposed approach does not partition the response values directly, 
which has been considered by many in the static setting \citep{hartigan1990partition,denison2002bayesian,quintana2003bayesian} 
and some in the dynamic setting \citep{barry1992product,page2020dependent}. 
Instead, we partition the covariate space according to their influences on the response. 
Separately, the literature on HMMs and fHMMs is also vast \citep{Rabiner:1989,Scott:2002,fruhwirth2006finite,zucchini2017hidden}.  
To our knowledge, however, they have never been adapted in the novel ways proposed in this article 
to dynamic variable selection problems. 
There is also a growing body of literature on regression methods for tensor valued predictors 
with tensor factorization techniques used as a dimension reduction tool. 
These methods, however, apply tensor factorizations with all continuous components, 
where the general Tucker decomposition runs into identifiability and interpretability problems. 
To avoid these issues, the literature has focused on parallel factor (PARAFAC) type decomposition \citep[see, e.g.,][etc.]{guhaniyogi2017bayesian, papadogeorgou2019soft}, 
a much simpler but restrictive special case of the Tucker. 
Aside from the development of sophisticated dependence models for the tensor components in a longitudinal setting, 
our proposal is also novel in that we employ a compact HOSVD, a flexible but interpretable version of the Tucker decomposition, 
where the core tensors take continuous values 
but the mode matrices comprise specially structured binary elements, 
resulting in interpretable partition structures that allow dynamic variable selection.

{\bf Outline of the Article:} The rest of this article is organized as follows. 
Section \ref{sec: lfmm} develops the generic longitudinal mixed model framework. 
Section \ref{sec: post inference} develops Markov chain Monte Carlo (MCMC) algorithms for posterior computation. 
Section \ref{sec: post con} establishes posterior convergence guarantees for the proposed model, for both function estimation and variable selection.
Section \ref{sec: sim studies} presents the results of simulation experiments.
Section \ref{sec: applications} presents real data applications. 
Section \ref{sec: discussion} contains concluding remarks.
Substantive additional details 
are presented in the supplementary materials.

\section{Longitudinal Functional Mixed Model} 
\label{sec: lfmm}
In this section, we develop a novel generic statistical framework for longitudinal functional mixed model (LFMM), 
where a response $y$ is generated under the influence of $p$ categorical predictors $x_{j} \in \{1,\dots,x_{j,\max}\} = \X_{j}, j=1,\dots,p$ longitudinally over time. 
To be precise, data $y_{i,\ell,t_{i}}$, available for individuals $i \in \{1,\dots,n\}$ and trials $\ell \in \{1,\dots,L_{i,t_{i}}\}$ at time points $t_{i} \in \{t_{i,1},\dots,t_{i,T}\}$, 
are generated under the influence of the predictors $x_{j}, j=1,\dots,p$. 
Importantly, we are not only interested in assessing the overall global influences of the predictors but also how they affect the responses locally 
at various times of the longitudinal studies.

We consider the following generic class of LFMMs 
\vspace{-5ex}\\
\be
& \{y_{i,\ell,t} \mid x_{j,i,\ell,t} = x_{j}, j=1,\dots,p \}= f_{x_{1},\dots,x_{p}}(t) + u_{i}(t) + \varepsilon_{i,\ell,t}, ~~~~~\varepsilon_{i,\ell,t} \sim f_{\varepsilon}, \label{eq: function 1}
\ee
\vspace{-5ex}\\
where $f_{x_{1},\dots,x_{p}}(t)$ denotes time-varying fixed effects due to associated predictors $\bx = (x_{1},\dots,x_{p}) \in \X_{1} \times \cdots \times \X_{p} = \X$, 
$u_{i}(t)$ are time-varying subject specific random effects, 
and $\varepsilon_{i,\ell,t}$ are random errors, i.i.d. from $f_{\varepsilon}$, satisfying $\eE_{f_{\varepsilon}}(\varepsilon_{i,\ell,t}) = 0$. 
We assume that $f_{x_{1},\dots,x_{p}}(t)$ and $u_{i}(t)$ evolve continuously with time. 
In this article, we focus on normally distributed errors with an inverse-Gamma prior on the error variance as
\vspace{-5ex}\\
\bse
f_{\varepsilon} = \Normal(0,\sigma_{\varepsilon}^{2}),~~~\sigma_{\varepsilon}^{2} \sim \IG(a_{\sigma}, b_{\sigma}). 
\ese
\vspace{-5ex}

For ease of exposition, we assume in (\ref{eq: function 1}) and henceforth that the data points are measured at a common set of equidistant time points $\{t_{1},\dots,t_{T}\}$, 
denoted simply as $\{1,\dots,T\}$. 
With some abuse of notation, 
generic values taken by the response $y$, the predictors $x_{j}$ are also denoted by $y$, $x_{j}$, etc.  
Without loss of generality, we also assume henceforth the same number of replicates $L_{i, t} = L$ for all $i, t$.
To further simplify notation, generic data recording time stamps in $\{1,\dots, T\}$ as well as other generic time points in $[1,T]$ will both be denoted by $t$.  

For longitudinal data observed on a regular time grid, 
as in the setting considered in this article, 
continuous functional parameter trajectories may still be more appealing and interpretable to a practitioner. 
A functional approach to modeling longitudinal data also does not require 
to impute missing data when they are missing at random \citep{little2019statistical}. 

The focus of this article is on continuous responses with categorical predictors. 
In many applications, the covariates are exogenous, that is, for each $i$, the $x_{j,i,\ell,t}$'s equal some fixed level $x_{j}$ for all $\ell,t$. 
When they are time-varying, we assume that all levels of $x_{j}$ are present in the sample at each $t$ for each $j$.  
{An easy, highly robust and practically useful approach to include continuous and ordinal predictors in model (\ref{eq: function 1}) would be to 
categorize them by binning their values into intervals 
(for example, using their quantiles) 
or by ignoring their order. 
Non-continuous responses of various types can likewise be conveniently analyzed via latent continuous variable augmentations \citep{albert1993bayesian, dunson2000bayesian, polson2013bayesian}.}

\subsection{Fixed Effects Model} \label{sec: fixed effects}

We propose a novel approach to model the latent functions $f_{x_{1},\dots,x_{p}}(t)$ using basis decomposition methods that allow them 
to flexibly vary with time $t$ 
while also locally depend on the predictor combinations $(x_{1},\dots,x_{p})$. 
Specifically, we let 
\vspace{-5ex}\\
\be
\textstyle f_{x_{1},\dots,x_{p}}(t) = \sum_{k=1}^{K} \beta_{k,x_{1},\dots,x_{p}} b_{k}(t), \label{eq: fixed effects function}
\ee 
\vspace{-5ex}\\
where $\bb(t) = \{b_{1}(t),\dots, b_{K}(t)\}\trans$ are a set of known locally supported basis functions 
and $\bbeta_{x_{1},\dots,x_{p}} = \{\beta_{1,x_{1},\dots,x_{p}},\dots,\beta_{K,x_{1},\dots,x_{p}}: (x_{1},\dots,x_{p}) \in \X \}$ are unknown coefficients to be estimated from the data. 
We use B-spline bases \citep{de1978practical} which are nonnegative, continuous and have desirable local support properties (Figure \ref{fig: b-splines}). 
Allowing the $\beta_{k,x_{1},\dots,x_{p}}$'s to vary with all predictor combinations $(x_{1},\dots,x_{p})$, 
the model also accommodates all order interactions among the predictors. 

While other higher order B-splines can also be used, in this work we use linear B-splines 
whose local support properties result in locally linear approximations of the fixed effects function (Figure \ref{fig: b-splines}). 
In the following, 
we use knots at the observed locations, hence $K = T$.
This allows local clustering at the set of all observable time points. 
In the case of of irregularly spaced data, a suitable fine grid can be chosen where such inference is desired. 
As shown in \cite{ruppert2002selecting}, when smoothing is controlled by data adaptive penalty parameters, 
the number of knots $K$ is not a crucial parameter as long as it is larger than a minimum threshold.

\begin{figure}[!ht]
	\centering
	\includegraphics[width=.49\linewidth]{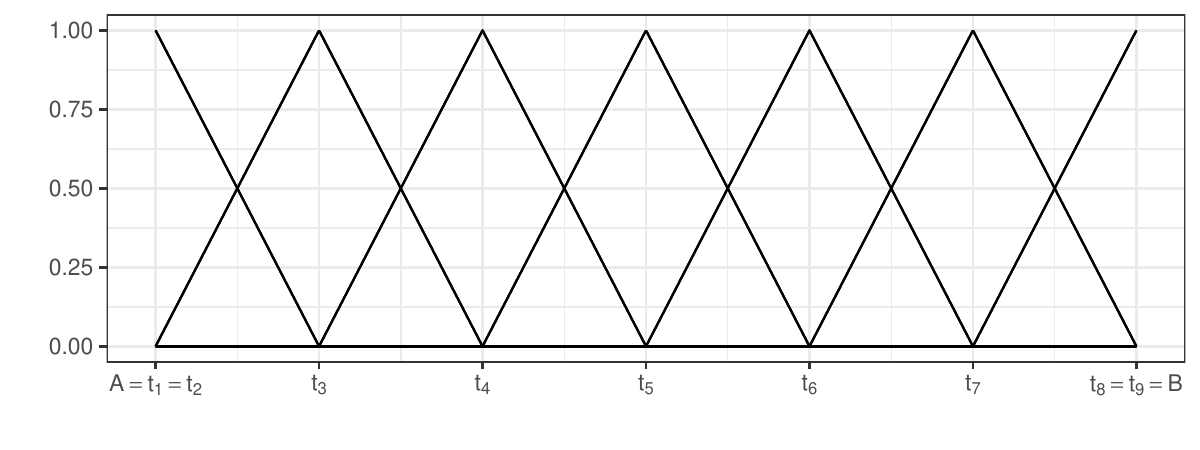}
	\includegraphics[width=.49\linewidth]{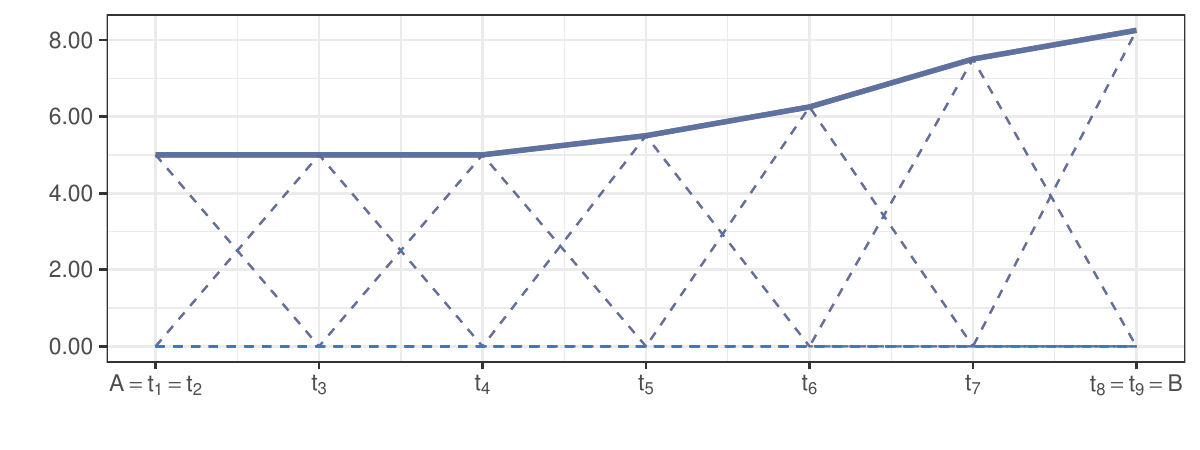}
	\includegraphics[width=.49\linewidth]{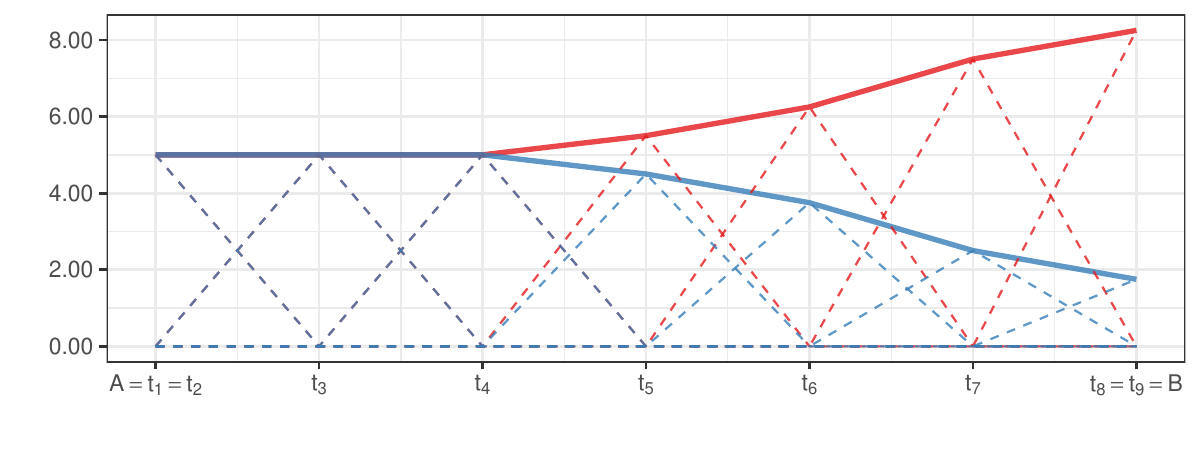}
	\includegraphics[width=.49\linewidth]{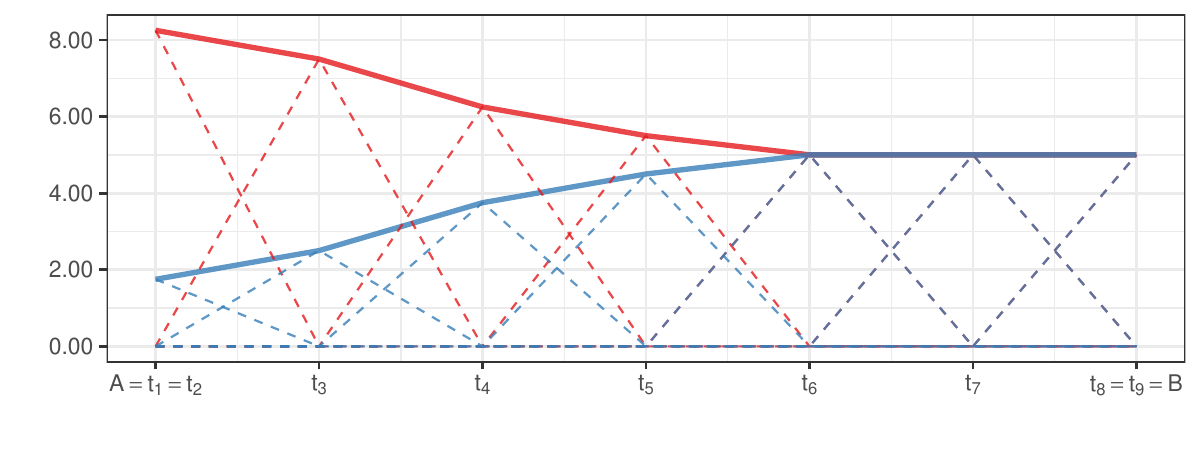}
	\caption{\baselineskip=10pt 
	Top left panel: Plot of $7$ linear B-splines on an interval $[A,B]$ defined by $9$ equidistant knot points that divide $[A,B]$ into $K=6$ equal subintervals.
	{Top right panel: Example of global clustering of two curves with shared spline coefficients $\bbeta_{x_{1}, \dots, x_{p}} = \bbeta_{x_{1}^{\prime}, \dots, x_{p}^{\prime}} = (5, 5, 5, 5.5, 6.25, 7.5, 8.25)\trans$; the solid lines represent the estimated functions, the dashed lines represent the weighted B-spline bases.
	Bottom left panel: Example of splitting of two curves with partially shared spline coefficients $\bbeta_{x_{1}, \dots, x_{p}} = (5, 5, 5, 5.5, 6.25, 7.5, 8.25)\trans$, $\bbeta_{x_{1}^{\prime}, \dots, x_{p}^{\prime}} = (5, 5, 5, 4.5, 3.75, 2.5, 1.75)\trans$.
	Bottom right panel: Example of merging of two curves with partially shared spline coefficients $\bbeta_{x_{1}, \dots, x_{p}} = (8.25, 7.5, 6.25, 5.5, 5, 5, 5)\trans$, $\bbeta_{x_{1}^{\prime}, \dots, x_{p}^{\prime}} = (1.75, 2.5, 3.75, 4.5, 5, 5, 5)\trans$.}
	}
	\label{fig: b-splines}
\end{figure}

For most practical applications, the size $K\prod_{j=1}^{p}x_{j,\max}$ of the unstructured model (\ref{eq: fixed effects function}) may be too big to allow efficient estimation of the parameters. 
It is also difficult to assess local influences of the predictors using such unstructured models. 
A potentially efficient solution that can greatly reduce dimensions while also facilitating the assessment of predictors' importance is 
to cluster the parameters by allowing them to have common shared values across different predictor combinations.  
If, for example, $\bbeta_{x_{1},\dots,x_{j,1},\dots,x_{p}} = \bbeta_{x_{1},\dots,x_{j,2},\dots,x_{p}}$ for all combinations of $(x_{1},\dots,x_{j-1},x_{j+1},\dots,x_{p})$, 
then not only have we reduced the number of parameters to be modeled but have also established that 
the two levels $x_{j,1}$ and $x_{j,2}$ of $x_{j}$ have no differential effect on the data generating mechanism. 

Such global clustering of all elements of $\bbeta_{x_{1},\dots,x_{p}}$ together will still be highly restrictive in most practical applications. 
More realistically, the elements of $\bbeta_{x_{1},\dots,x_{p}}$ should be allowed to cluster locally. 
In the following, we exploit local support properties of B-splines in a novel way to achieve this desirable property. 
In principle, other basis decomposition methods whose bases have compact (local) support can also be used in a similar way.

{\bf Dimension Reduction and Local Clustering via HOSVD:}
To achieve simultaneous dimensionality reduction and local clustering, 
we structure the parameters for different predictor combinations at each location $k$
as a $x_{1,\max} \times \dots \times x_{p,\max}$ dimensional tensor $\bbeta_{k} = \{\beta_{k, x_{1},\dots,x_{p}}: (x_{1},\dots,x_{p}) \in \X \}$ 
and then apply an HOSVD-type \citep{tucker:1966, de_lathauwer_etal:2000} factorization, arriving at
\vspace{-5ex}\\
\be
	& \textstyle \{\beta_{k, x_{1},\dots,x_{p}} \mid z_{j,k}^{(x_{j})} = z_{j,k}, j=1,\dots,p \} = \beta_{k,z_{1,k},\dots,z_{p,k}}^{\star}, 	
	\label{eq: TFM2A}
\ee
\vspace{-5ex}\\
where $z_{j,k}^{(x_{j})}$'s are cluster indicator variables associated with each covariate $j$ for its specific value $x_{j}$ at the knot-location $k$, 
and the $\beta_{k, z_{1,k},\dots,z_{p,k}}^{\star}$'s are the associated unique cluster specific spline coefficients. 
Our construction using locally supported linear B-splines then implies
\vspace{-5ex}\\
\bse
	\textstyle \{f_{k, x_{1},\dots,x_{p}} \mid z_{j,k}^{(x_{j})}=z_{j,k}, j=1,\dots,p \} = \beta_{k,z_{1,k},\dots,z_{p,k}}^{\star},
\ese
\vspace{-5ex}\\
allowing simple interpretations for the allocation variables $z_{j,k}^{(x_{j})}$'s and also easier theoretical treatment and posterior computation.

Let the $z_{j,k}^{(x_{j})}$'s take values in $\Z_{j,k} = \{1,\dots,\ell_{j,k}\}$ for different possible values $x_{j} \in \X_{j} = \{1,\dots,x_{j,\max}\}$. 
\footnote{For notational simplicity, here we assumed 
that the values taken by the cluster allocation variables $z_{j,k}^{(x_{j})}$'s 
for different values of $x_{j}\in \X_{j} = \{1,\dots,x_{j,\max}\}$
are sequentially ordered without gaps, i.e., $\Z_{j,k}=\{1,\dots,\ell_{j,k}\}$. 
In what follows, we allow other general configurations of $z_{j,k}^{(x_{j})}$'s that induce the same equivalent partition of $\X_{j}$.
\\ 
\hspace*{15pt} 
For example, consider some $x_{j} \in \X_{j} = \{1,2,3\}$ with $x_{j,\max}=3$ 
partitioned into $\{\{1,3\},\{2\}\}$ at location $k$.  
Here our notation allows the configurations $(1,2,1)$ or $(2,1,2)$ 
of the corresponding cluster allocation variables $(z_{j,k}^{(1)},z_{j,k}^{(2)},z_{j,k}^{(3)})$ with $\Z_{j,k} = \{1,\ell_{j,k}\} = \{1,2\}$. 
Going forward, we also allow other general configurations 
$(1,3,1)$ or $(3,1,3)$ with $\Z_{j,k} = \{1,3\}$, 
or $(2,3,2)$ or $(3,2,3)$ with $\Z_{j,k} = \{2,3\}$ 
which induce the same partition $\{\{1,3\},\{2\}\}$ of $\X_{j}$ 
with $\ell_{j,k}=\abs{\Z_{j,k}}=2$. 
In this example, the effects of $x_{j}=1$ and $x_{j}=3$ on the response curve are the same and hence these levels are clustered together, 
but these effects are different from the effect of $x_{j}=2$ which therefore forms its own cluster. 
The predictor $x_{j}$ therefore is important at location $k$.
\\
\hspace*{15pt} 
Consider also the example when the levels $\X_{j} = \{1,\dots,3\}$ are partitioned into a single cluster at location $k$. 
Here our notation only allows the configuration $(1,1,1)$ of $(z_{j,k}^{(1)},z_{j,k}^{(2)},z_{j,k}^{(3)})$ with $\Z_{j,k} = \{\ell_{j,k}\} = \{1\}$. 
Going forward, we also allow the configurations 
$(2,2,2)$ or $(3,3,3)$ with $\Z_{j,k} = \{2\}$ and $\Z_{j,k} = \{3\}$ respectively  
which induce the same partition $\{\{1,2,3\}\}$ of $\X_{j}$ 
with $\ell_{j,k}=\abs{\Z_{j,k}}=1$. 
In this example, the effects of all three levels are the same on the response curve. 
The predictor $x_{j}$ therefore 
is unimportant at location $k$.
} 
Separately, $z_{j,k}^{(x_{j})} \in \Z_{j,k}$ therefore forms $\ell_{j,k} \leq x_{j,\max}$ marginal clusters of the predictor levels $x_{j} \in \X_{j}$, 
while jointly, $(z_{1,k}^{(x_{1})},\dots,z_{p,k}^{(x_{p})}) \in \Z_{k} = \Z_{1,k} \times \dots \times \Z_{p,k}$ forms 
$\prod_{j=1}^{p} \ell_{j,k}  \leq \prod_{j=1}^{p}x_{j,\max}$ joint clusters of the predictor level combinations $(x_{1},\dots,x_{p}) \in \X =\X_{1} \times \dots \times \X_{p}$.

To see the HOSVD formulation behind this, note that (\ref{eq: TFM2A}) can then be rewritten as
\vspace{-5ex}\\
\be	
& \textstyle \{\beta_{k, x_{1},\dots,x_{p}} \mid z_{j,k}^{(x_{j})}, j=1,\dots,p \} = \sum_{z_{1,k}} \cdots\sum_{z_{p,k}} \beta_{k,z_{1,k},\dots,z_{p,k}}^{\star} \prod_{j=1}^{p} 1\{z_{j,k}^{(x_{j})}=z_{j,k}\}, 
	\label{eq: TFM2B}
\ee
\vspace{-5ex}\\
where $\bbeta_{k}^{\star} = \{\beta_{k, z_{1,k},\dots,z_{p,k}}^{\star}: (z_{1,k},\dots,z_{p,k}) \in \Z_{k}\}$ is a $\ell_{1,k} \times \dots \times \ell_{p,k}$ dimensional core tensor, $\bz_{j,k}=\{{{\scriptstyle 1\{z_{j,k}^{(x_{j})}=z_{j,k}\} }}: x_{j} \in \X_{j}, z_{j,k} \in \Z_{j,k}\}$ are $x_{j,\max} \times \ell_{j,k}$ dimensional mode matrices (Figure \ref{fig: dynamic HOSVD}).

\begin{figure}[ht!]
	\centering
	\hspace*{-0.25cm}\includegraphics[width=\linewidth, trim=1cm 1cm 1cm 1cm]{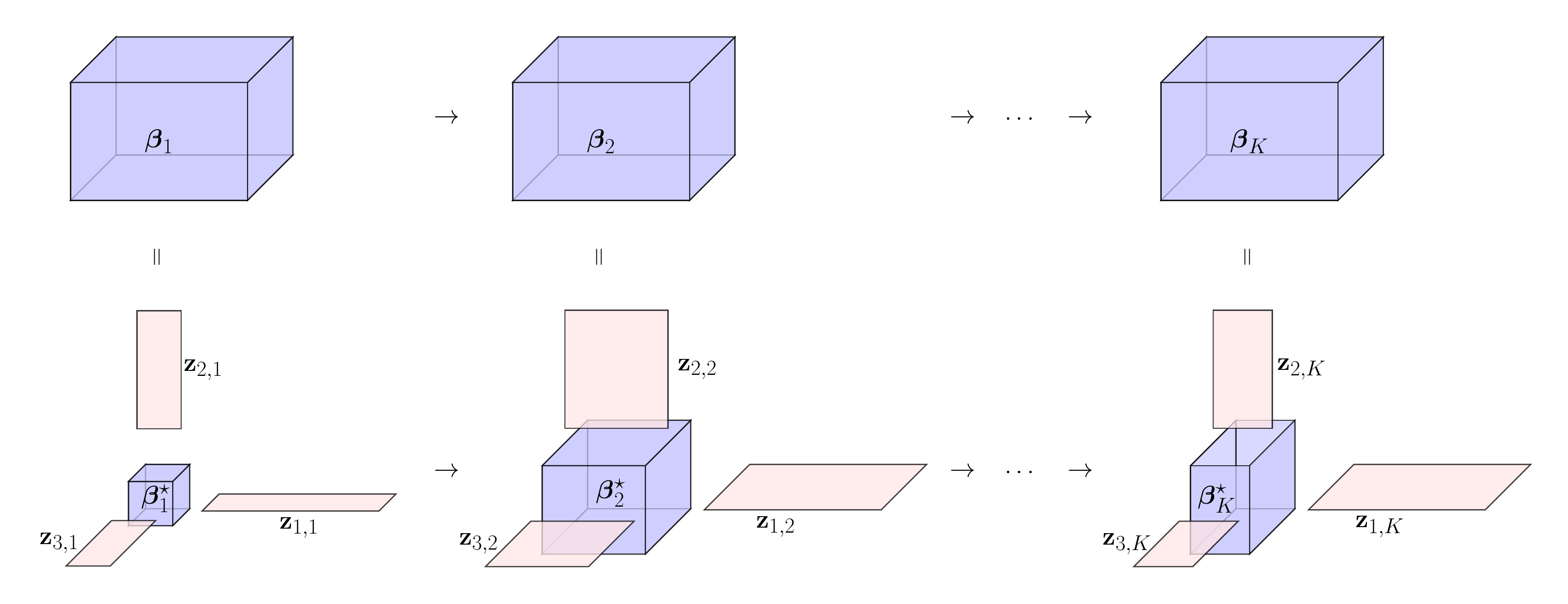}
	\caption{\baselineskip=10pt 
	Model \eqref{eq: TFM2B} with three covariates viewed as a dynamic HOSVD. 
	}
	\label{fig: dynamic HOSVD}
\end{figure}

The allocation variables are assigned probability models supported on $\X_{j}$, so that the number of distinct values taken on by the $z_{j,k}^{(x_{j})}$'s, 
namely $\ell_{j,k}$, 
lies between $1$ and $\abs{\X_{j}} = x_{j,\max}$.
If $\ell_{j,k} = x_{j,\max}$, the $z_{j,k}^{(x_{j})}$'s take on different values for different levels of $x_{j}$, implying that the spline coefficients are all different for different levels of $x_{j}$ at location $k$. 
In this case, all levels of $x_{j}$ differently influence the response generating mechanism at location $k$. 
If $\ell_{j,k} \leq x_{j,\max}$, local clustering of the predictor's effects is performed and the problem of modeling the original parameter tensors $\bbeta_{k}$ is effectively reduced to that of modeling the smaller-dimensional core tensors $\bbeta_{k}^{\star}$.
For instance, when $z_{j,k}^{(x_{j,1})} = z_{j,k}^{(x_{j,2})} = z_{j,k}$ for two different levels $x_{j,1}$ and $x_{j,2}$ of the $j\th$ predictor $x_{j}$, 
the spline coefficients at location $k$ do not differ between $x_{j,1}$ and $x_{j,2}$, 
i.e., $\beta_{k,x_{1}, \dots, x_{j,1}, \dots, x_{p}} = \beta_{k,x_{1}, \dots, x_{j,2}, \dots, x_{p}} = \beta_{k,z_{1,k}, \dots, z_{j,k}, \dots, z_{p,k}}^{\star}$. 
There is thus no significant difference between how the two levels $x_{j,1}$ and $x_{j,2}$ influence the response $y$ at location $k$. 
Importantly, when $\ell_{j,k} = 1$, the $z_{j,k}^{(x_{j})}$'s all take on the same value for all different levels of $x_{j}$, 
characterizing the scenario when $x_{j}$ has no influence on $y$ at location $k$ and local variable selection is achieved.  
The set of important predictors at location $k$ is thus $\{j: \ell_{j,k}>1\}$. 
Significant reduction in model size is achieved at the location $k$  
when 
$\prod_{j=1}^{p}\ell_{j,k} \ll \prod_{j=1}^{p}x_{j,\max}$, 
i.e., when the size of the core tensors is much smaller than the original coefficient tensor (Figure \ref{fig: dynamic HOSVD}).
The varying side lengths $\ell_{j,k}$ 
of the core tensors $\bbeta_{k}^{\star}$ 
at different locations $k$ (Figure \ref{fig: dynamic HOSVD}) 
also crucially allow the model to identify different sets of important predictors at different locations $k$.

In effect, our HOSVD formulation of the continuous coefficient tensors $\bbeta_{k}$ in (\ref{eq: TFM2B}) 
into a continuous core but binary mode matrices thus induces local random partitions 
of the joint covariate space $\X$ 
into $\ell_{k} = \prod_{j=1}^{p}\ell_{j,k}$ sets at each knot location $k$. 
This is different from traditional PARAFAC decompositions of continuous tensors 
into all continuous components as in \cite{guhaniyogi2017bayesian, papadogeorgou2019soft}, etc.
Section \ref{sec: sm tensor factorization} in the supplementary materials provides some additional discussions 
on the novelty and advantages our formulation over these other existing approaches.

{\bf Second-Layer Clustering:}
We note, however, that the partitions of the joint covariate space $\X=\X_{1} \times \cdots \times \X_{p}$ 
induced by the HOSVD in (\ref{eq: TFM2B}) may still lead to some overparametrization as 
they are constructed as the product of $p$ marginal partitions of $\X_{j}$ into $\ell_{j,k}$ sets 
(Figure \ref{fig: two_layers_latent}, left panel).
To eliminate this limitation and obtain an unrestricted partition, say $\rho_{k}= \{ S_{k,1}, \dots, S_{k,m_{k}}\}$, 
we further cluster the elements of the core tensors 
using a second layer of latent variables $z_{k}^{(z_{1,k}, \dots, z_{p,k})} \in \{1,\dots,\ell_{k}\}$ such that
\vspace{-5ex}\\
\bse
\textstyle \beta_{k,z_{1,k},\dots,z_{p,k}}^{\star} = \sum_{z_{k}=1}^{\ell_{k}} \beta_{k,z_{k}}^{\star\star} 1\{z_{k}^{(z_{1,k}, \dots, z_{p,k})} = z_{k}\}.
\ese 
\vspace{-5ex}\\
Such clustering further refines the model (Figure \ref{fig: two_layers_latent}, right panel), making the final partition structure of the covariate space fully flexible\footnote{
Consider, for example, two drugs A and B, each with two dosage levels $\X_{A} = \X_{B} = \{1,2\}$. 
Assume further that for both drugs their two levels have different marginal effects, i.e., for both of them the effect of dosage level 1 is significantly different from that of dosage level 2. 
The basic idea of our approach to cluster the levels of the predictors (here drugs) according to their effects 
therefore would produce a clustering of $\{\{1\},\{2\}\}$ for both drugs. 
The tensor product of these sets then gets us to the joint clustering 
\bse
\{\{1\},\{2\}\} \times \{\{1\},\{2\}\} = \{\{(1,1)\},\{(1,2)\},\{(2,1)\},\{(2,2)\}\}~~~\text{(four clusters)}.
\ese 
It may be possible, however, that there are significant interactions between the levels of A and B, so that the effects of the dose combinations $(1,2)$ and $(2,1)$ are in fact the same. 
The correct final joint cluster configuration therefore should be 
\bse
\{\{(1,1)\},\{(1,2),(2,1)\},\{(2,2)\}\}~~~\text{(three clusters)}.
\ese 
The second layer, which further clusters $(1,2)$ and $(2,1)$ together, allows us to perform such inference.}.

\begin{figure}[ht!]
	\centering
	\includegraphics[height=4.5cm, trim=2cm 1.25cm 1cm 1.25cm]{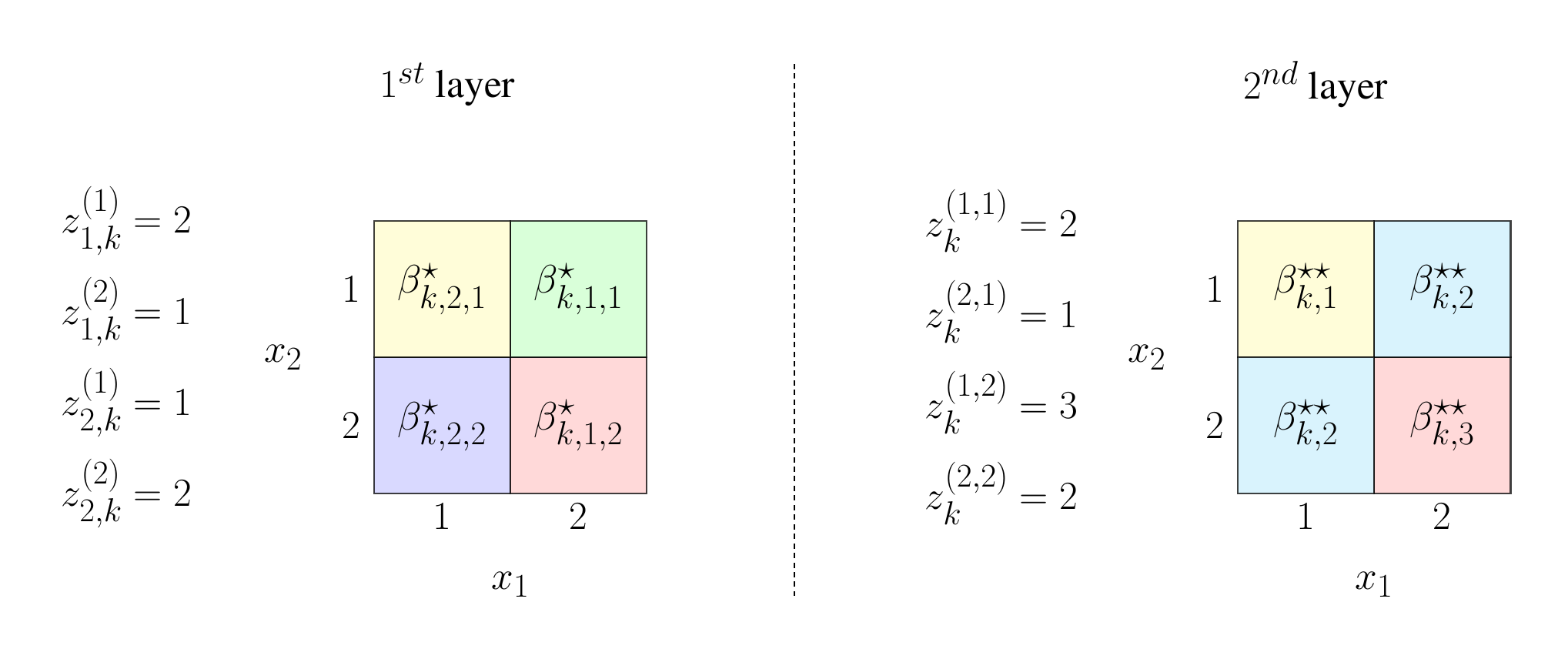} 
	\caption{
	Illustration of the two layers of latent variables that induce the partition of the covariate space at a fixed location $k$ in the case with two categorical predictors $x_{1},x_{2}$ with $x_{1,\max} = x_{2,\max} = 2$ levels each. 
In this example, $S_{k,1} = \{(1,1)\}$, $S_{k,2} = \{(1,2),(2,1)\}$, $S_{k,3} = \{(2,2)\}$, $\ell_{k} = 4$, $m_{k} = 2$.
}
	\label{fig: two_layers_latent}
\end{figure}

{Our two clustering layers thus play different roles in achieving parsimonious partitions of the covariate space in high dimensions - 
efficiently exploring and partitioning $\X$ in high dimensions is an extremely challenging task - 
we break this into two parts - 
the HOSVD first reduces the dimensions by creating a product of marginal partitions 
while also efficiently removing the unimportant covariates - 
the second layer then refines this smaller space to arrive at a fully flexible model.}

\vskip 5pt
So far, we have developed the HOSVD model separately for each knot location $k$. 
Next, we focus on introducing time-varying dependency structures between these building blocks appropriate for longitudinal settings.
We do this by assuming fHMM dynamics on the allocation variables $\bz_{j,k}$ that introduce dependencies between the local partitions at adjacent knot locations, 
and then assigning novel Markovian priors on the coefficients $\bbeta_{k}^{\star\star}$ that make these coefficients vary smoothly over time.

{\bf Dynamically Evolving Partition Structures:}
We first consider the problem of specifying probability models for the allocation variables $z_{j,k}^{(x_{j})}$
that allow them to be temporally dependent across $k$.
We model the temporal evolution of the $z_{j,k}^{(x_{j})}$'s using hidden Markov models (HMMs). 
For each predictor combination $(x_{1},\dots,x_{p})$, the collection $\bz^{(x_{1},\dots,x_{p})} = \{z_{j,k}^{(x_{j})}, k=1,\dots,K, j=1,\dots,p\}$ 
then defines a factorial HMM \citep{ghahramani1996factorial} (Figure \ref{fig: dag fHMM}). 
We characterize the dynamics of the fHMM component chains as 
\vspace{-5ex}\\
\bse
(z_{j,k}^{(x_{j})} \mid z_{j,k-1}^{(x_{j})} = z_{k-1}) \sim \Mult(\pi_{z_{k-1},1}^{(j)},\dots,\pi_{z_{k-1},z_{j,\max}}^{(j)})~~~~\text{for}~j = 1, \dots, p.
\ese
\vspace{-5ex}\\
We assign Dirichlet priors on the transition probabilities 
\vspace{-5ex}\\
\bse
&\bpi_{z}^{(j)} = (\pi_{z,1}^{(j)},\dots,\pi_{z,z_{j,\max}}^{(j)})\trans \sim \Dir(\alpha^{(j)}/z_{j,\max},\dots,\alpha^{(j)}/z_{j,\max})~~~\text{with}~~~\alpha^{(j)} \sim \Ga(a_{\alpha},b_{\alpha}).
\ese
\vspace{-5ex}\\
In general, the maximum number of distinct values of the $z_{j,k}^{(x_{j})}$'s is $x_{j,\max}$.
However, in most applications, $|\Z_{j,k}|$ will be much smaller than $x_{j,\max}$ uniformly for all $k$ and the restricted support $z_{j,k}^{(x_{j})} \in \{1, \dots, z_{j,\max}\}$, $z_{j,\max} < x_{j,\max}$ will suffice.
We impose parsimony by assigning exponentially decaying priors with finite support on the partition sizes $|\Z_{j,k}| = \ell_{j,k}$, favoring smaller partitions as  
\vspace{-5ex}\\
\bse
\ell_{j,k} \propto \exp(- \varphi_{j} \ell_{j,k}), ~~~\varphi_{j} \sim \Ga(a_{\varphi,j},b_{\varphi,j}),~~~j = 1, \dots, p, ~~~k = 1, \dots, K.
\ese 
\vspace{-5ex}\\
Larger values of $\varphi_{j}$ here induce faster decay and hence smaller model sizes. 
Gamma hyper-priors on the $\varphi_{j}$'s further make these shrinkage strengths data adaptive. 
Being shared across $k$, the $\varphi_{j}$'s also allow to share information on partition sizes across $k$ for each predictor $j$ separately. 
This is desirable since it is expected that for most predictors, especially the unimportant ones, the influence will be similar across all locations $k$.   

\begin{figure}[ht!]
	\centering
	\includegraphics[width=0.5\linewidth]{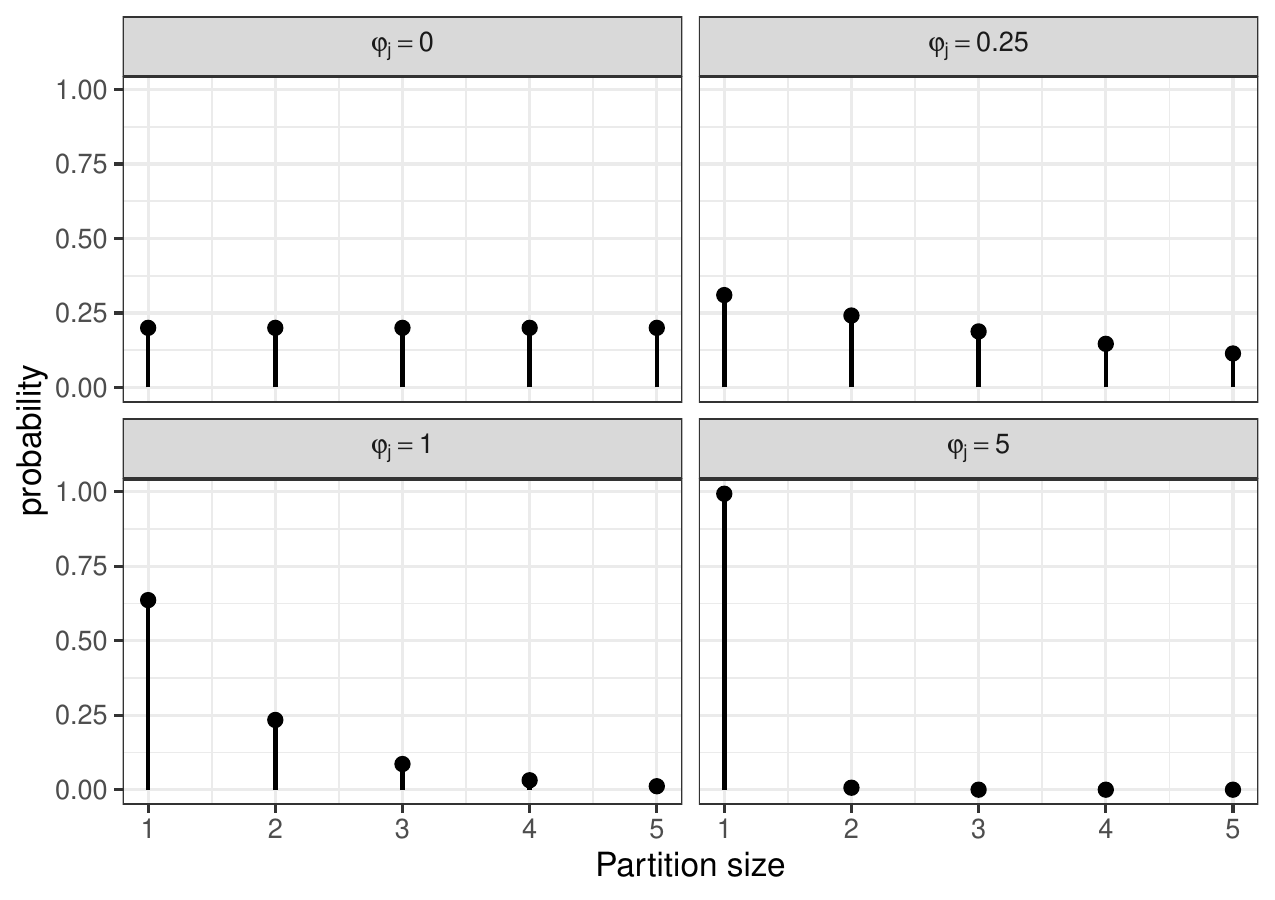} 
	\caption{
	Illustration of the prior distribution on the partition sizes for the categorical predictor $x_{j}$ with $x_{j,\max} = 5$ levels. Each panel corresponds to different \textit{fixed} values of the parameter $\varphi_{j}$. 
As $\varphi_{j}$ increases, the prior goes from discrete uniform ($\varphi_{j} = 0$) to a point mass at 1 ($\varphi_{j} \rightarrow +\infty$).
}
	\label{fig: prior_phi}
\end{figure}

\begin{figure}[ht!]
	\centering
	\hspace*{-0.00cm}\includegraphics[height=5.5cm, trim=2cm 1.25cm 1cm 1.25cm]{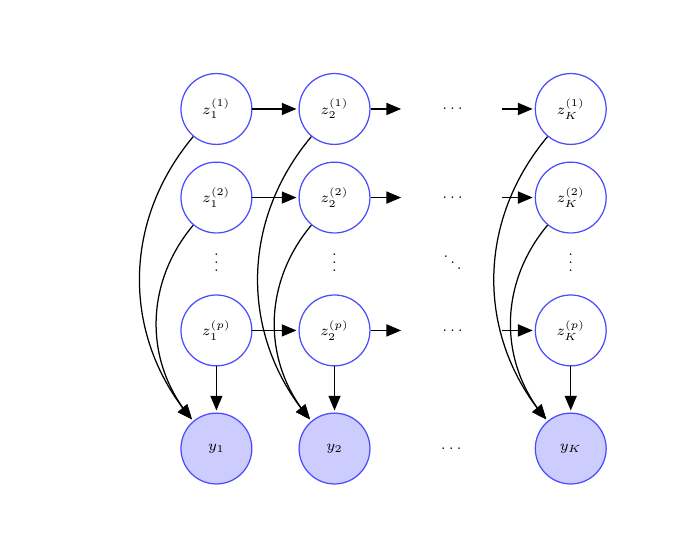} \quad\quad
	\hspace*{0.40cm}\includegraphics[height=5.5cm, trim=2cm 1.25cm 1cm 1.25cm]{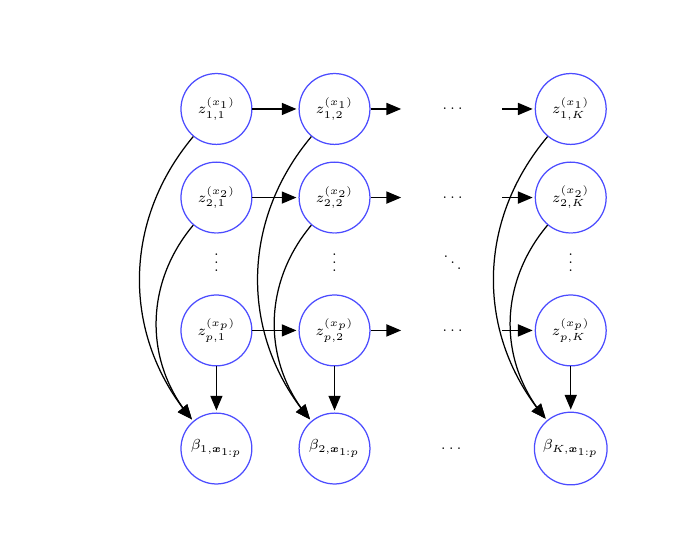}
	\caption{\baselineskip=10pt 
	Left panel: The directed acyclic graph (DAG) of a conventional fHMM with $p$ latent layers. 
	Right panel: DAG of our proposed fixed effects model \eqref{eq: TFM2B} with $p$ categorical predictors $\bx_{1:p} = (x_{1}, \dots, x_{p})$.
 	}
	\label{fig: dag fHMM}
\end{figure}

The second layer latent allocation variables $z_{k}^{(z_{1,k}, \dots, z_{p,k})}$ are assigned multinomial distributions with Dirichlet priors on the probability parameters as 
\vspace{-5ex}\\
\bse
&(z_{k}^{(z_{1,k}, \dots, z_{p,k})} \mid \bpi^{\star}_{k}) \sim \Mult(\pi_{k,1}^{\star},\dots,\pi_{k,\ell_{k}}^{\star}),
\\
&\bpi^{\star}_{k} = (\pi_{k,1}^{\star},\dots,\pi_{k,\ell_{k}}^{\star})\trans \sim \Dir(\alpha^{\star}/\ell_{k},\dots,\alpha^{\star}/\ell_{k})~~~\text{with}~~~\alpha^{\star} \sim \Ga(a_{\alpha^{\star}},b_{\alpha^{\star}}).
\ese
\vspace{-5ex}

When the $z_{j,k}^{(x_{j})}$'s corresponding to two different categories of $x_{j}$ are equal in a temporal region, 
the local support properties of B-splines then cause the underlying curves to be the same in that region. 
Conversely, if the $z_{j,k}^{(x_{j})}$'s corresponding to two different values of $x_{j}$ are different, 
the underlying curves will be distinct unless the second layer of latent variables maps them to the same joint partition element.

\begin{figure}[ht!]
	\centering
	\includegraphics[width=0.48\linewidth, trim=2cm 1.25cm 1cm 1.25cm]{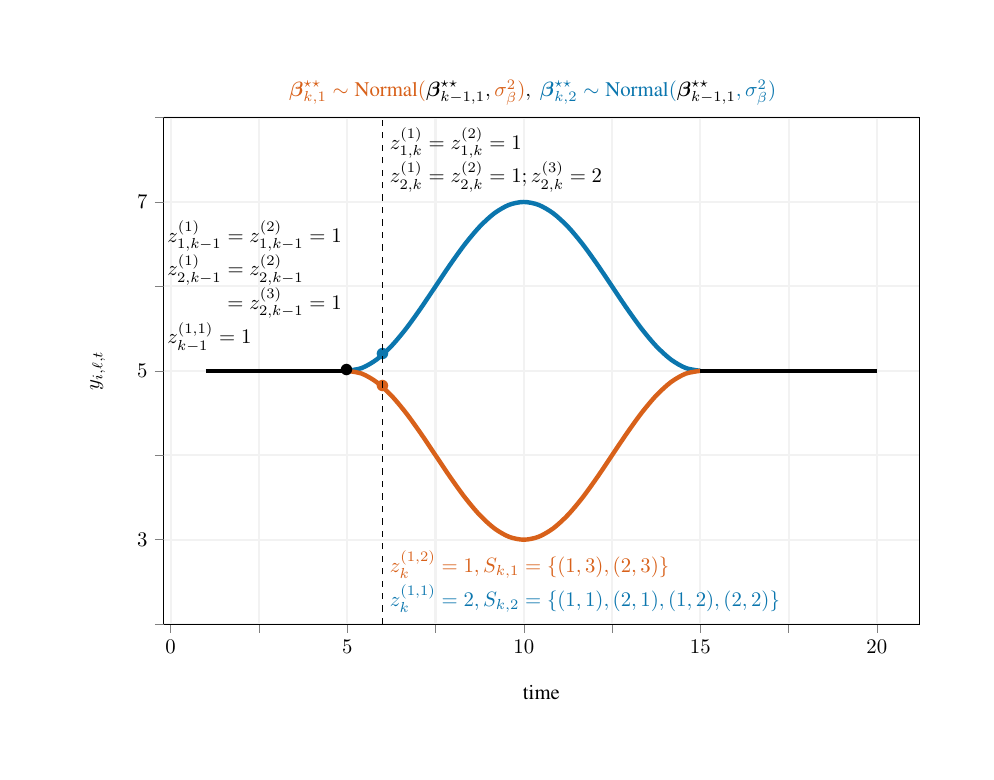}
	\hspace*{0.2cm}
	\includegraphics[width=0.48\linewidth, trim=2cm 1.25cm 1cm 1.25cm]{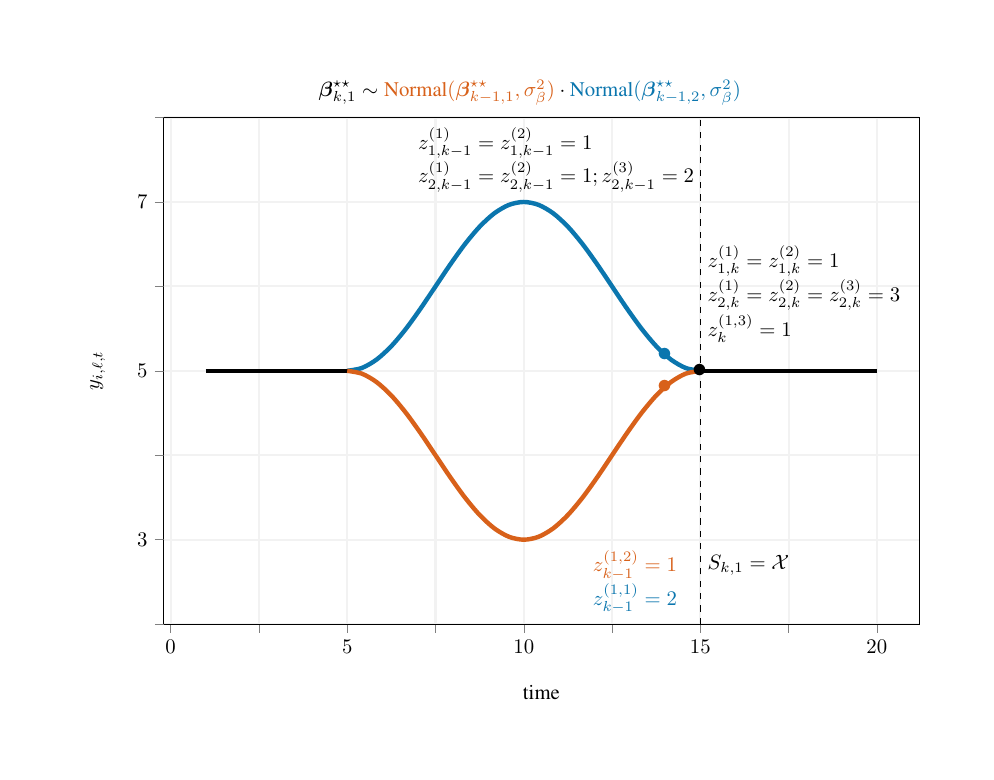}
	\caption{\baselineskip=10pt 
	An illustration of the prior on the spline core coefficients $\beta_{k,z_{k}}^{\star\star}$ at location $k$ (the dashed vertical lines) in 
	the fixed effects model developed in Section \ref{sec: fixed effects} 
	for a scenario with two categorical covariates $x_{1} \in \{1,2\}$ and $x_{2} \in \{1,2,3\}$,
	where the curves corresponding to all levels of $(x_{1}, x_{2})$ are initially equal, the curves for $x_{2}=1,2$ (in blue) and $x_{2}=3$ (in red) then diverge at $t=5$, merging back again at $t=15$. 
	}
	\label{fig: prior core beta}
\end{figure}

{\bf Conditionally Markov Core Coefficients:} 
We next consider priors for the unique core tensors $\beta_{k,z_{k}}^{\star\star}$. 
Conditional on the $z_{j,k}^{(x_{j})}$'s, $z_{k}^{(z_{1,k}, \dots, z_{p,k})}$'s, and the coefficients at the previous locations, for $k = 2, \dots, K$, we construct the priors sequentially as 
\vspace{-5ex}\\
\be
	\beta_{k,z_{k}}^{\star\star} \sim \prod_{h \in \Z_{k,z_{k}}^{-}} \Normal\left( \beta^{\star\star}_{k-1,h}, \sigma_{\beta}^{2}\right),
    \label{eq: prior_all_components}
\ee
\vspace{-4ex}\\
where $\Z_{k,z_{k}}^{-} = \big\{z_{k-1}: z_{k-1} = z_{k-1}^{(z_{1,k-1}, \dots, z_{p,k-1})}; (z_{1,k-1}, \dots, z_{p,k-1}) = (z_{1,k-1}^{(x_{1})}, \dots, z_{p,k-1}^{(x_{p})});$ $(x_{1},\dots,x_{p}) \in S_{k,z_{k}}\big\}$ and $S_{k,z_{k}}$ is the partition element comprising the covariates levels $(x_{1}, \dots, x_{p})$ that, at location $k$, are assigned the label $z_{k}$.
Simply put, 
we center the core coefficients around the ones that are `expressed' at the previous location (Figure \ref{fig: prior core beta}), 
thus effectively penalizing their differences.
The initial coefficients are assigned non-informative flat priors as $\beta_{1,z_{1}}^{\star\star} \sim 1$. 
The smoothness of the curves is thus controlled by the parameter $\sigma_{\beta}^{2}$ and is assigned a prior, allowing it to be informed by the data. 
We let 
\vspace{-5ex}\\
\bse
\sigma_{\beta} \sim \HC(0,s_{\sigma}),
\ese  
\vspace{-5ex}\\
where $\HC(a,b)$ denotes a half-Cauchy distribution with location parameter $a$ and scale parameter $b$. 
We chose a half-Cauchy prior over the more popular inverse gamma distributions since it has been shown that, 
with higher probability mass near zero, 
it is a more appropriate prior
for variance and smoothing parameters \citep{gelman2006prior, polson2012half}.

{\bf Characterization of Main and Interaction Effects:} 
As may be noted from our model description above, 
the HOSVD characterizes each $x_{j}$'s overall significance explicitly
and their joint influences implicitly but very compactly, 
efficiently eliminating the redundant variables and achieving significant reduction in dimensions, 
but avoids explicitly describing their main and lower-dimensional interaction effects 
which are often very useful to practitioners for their easy interpretation.  
These effects may, however, be meaningfully \emph{defined} (and easily estimated from the posterior samples) directly as 
\vspace{-5ex}\\
\be
\begin{split}
& \text{overall mean:~} f_{0}(t) = \frac{\sum_{\bx}f_{x_{1},\dots,x_{p}}(t)}  {\abs{\X}}, ~~~\text{main effects:~} f_{x_{j}}(t) = \frac{\sum_{\bx_{-j}}f_{x_{1},\dots,x_{p}}(t)}  {\abs{\X_{-j}}} - {f}_{0}(t), \\
& \text{interactions:~}  f_{x_{j_{1}},x_{j_{2}}}(t) = \frac{\sum_{\bx_{-j_{1},-j_{2}}}f_{x_{1},\dots,x_{p}}(t)}  {\abs{\X_{-j_{1},-j_{2}}}} - {f}_{x_{j_{1}}}(t) - {f}_{x_{j_{2}}}(t) - {f}_{0}(t), ~\text{etc.}, 
\end{split} \label{eq: main and interaction}
\ee
\vspace{-4ex}\\
where $\bx_{-j}=(x_{1},\dots,x_{j-1},x_{j+1},\dots, x_{p})\trans \in \X_{1} \times \cdots \times \X_{j-1} \times \X_{j+1} \times \cdots \times\X_{p} = \X_{-j}$, and so on. 
Section \ref{sec: sm main and interaction effects} in the supplementary materials provides additional details and plots, 
a general recipe for testing these effects, etc.

{\bf Single Predictor Special Cases:} 
The HOSVD approach is relevant particularly for the extremely challenging multivariate predictor problem $(x_{1},\dots,x_{p})$
but not for a single predictor $x$, in which case the fHMM (Figure \ref{fig: dag fHMM}, right panel) simplifies to an HMM with a single layer $z_{k}^{(x)}$, 
and the second layer clustering of the spline coefficients $\beta_{k,z_{k}^{(x)}}^{\star}$ (Figure \ref{fig: two_layers_latent}) is not needed. 
As discussed in the Introduction, the focus of the article is primarily on the multivariate case. 
Our implementation, however, is automated to adjust to both scenarios.

\subsection{Random Effects Model} \label{sec: random effects}

We model the random effects components $u_{i}(t)$ as 
\vspace{-5ex}\\
\be
    \begin{split}
        & \textstyle u_{i}(t) = \sum_{k=1}^{K} \beta_{k,i}^{(u)} b_{k}(t), ~~~~~
        \\
        & \bbeta_{i}^{(u)} \sim 
        \MVN_{K}\{\bzero,(\sigma_{u,a}^{-2}\bI_{K}+\sigma_{u,s}^{-2}\bP_{u})^{-1}\},~~~~~
        \\
        & \sigma_{u,s} \sim \HC (0, s_{\sigma}),~~~~~\sigma_{u,a} \sim \HC (0, s_{\sigma}),
    \end{split} 
    \label{eq: random effects}
\ee
\vspace{-4ex}\\
where $\bbeta_{i}^{(u)} = (\beta_{1,i}^{(u)},\dots,\beta_{K,i}^{(u)})\trans$ are subject specific spline coefficients, 
$\MVN_{K}(\bmu,\bSigma)$ denotes a $K$ dimensional multivariate normal distribution with mean $\bmu$ and covariance $\bSigma$.
The zero mean of the random effects distribution ensures that the random effects are separately nonparametrically identifiable \citep{guo2002functional,morris2006wavelet}.
We choose $\bP_{u} = \bD_{u}\trans \bD_{u}$, where the $(K-1) \times K$ matrix $\bD_{u}$ is such that $\bD_{u} \bbeta_{i}^{(u)}$ computes the first order differences in $\bbeta_{i}^{(u)}$.
The model thus penalizes $\sum_{k=1}^{K} (\nabla \beta_{k,i}^{(u)})^{2} = \bbeta_{i}^{(u)\hbox{\tiny T}}  \bP_{u} \bbeta_{i}^{(u)}$, the sum of squares of first order differences in $\bbeta_{i}^{(u)}$ \citep{eilers1996flexible}. 
This induces a first order Markov dynamics for the spline coefficients, 
evident from the {tridiagonal} structure of the precision matrix in (\ref{eq: random effects}) that encodes their conditional dependence relationships.
The random effects variance parameter $\sigma_{u,s}^{2}$ models the smoothness of the random effects curves, smaller $\sigma_{u,s}^{2}$ inducing smoother $u_{i}(t)$'s. 
Additional variations from the constant zero curve are explained by $\sigma_{u,a}^{2}$ (Figure \ref{fig: random_eff_ill}). 
The absence of random effects is signified by the limiting case $\sigma_{u,s}^{2} = \sigma_{u,a}^{2} = 0$.

\begin{figure}[!ht]
	\centering
	\begin{center}
	\includegraphics[width=0.7\linewidth, trim=0cm 0cm 0cm 0cm, clip=true]{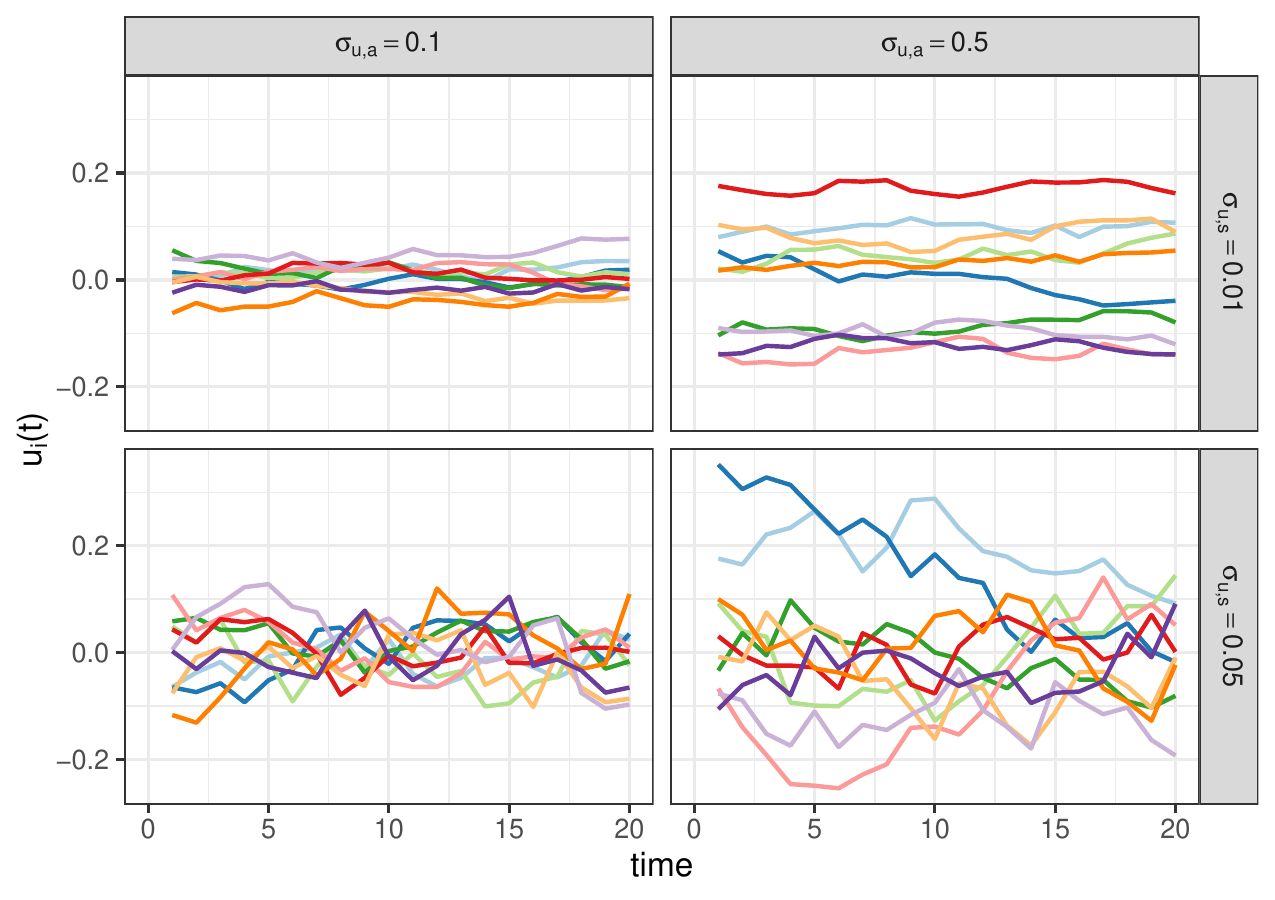}
	\end{center}
	\vskip -15pt
	\caption{An illustration of the functional random effects model proposed in Section \ref{sec: random effects}. 
	Each panel shows a collection of $10$ random draws from the random effects distribution for a combination of values of $(\sigma_{u,s}, \sigma_{u,a})$.  
	}
	\label{fig: random_eff_ill}
\end{figure}

A similar model for functional random effects but with additional assumptions on the covariance matrix has previously been developed in \cite{guo2002functional}.
If we ignore the sharing of information through model hierarchies, 
the data for estimating an individual level effect come from that individual alone 
whereas the data for estimating the fixed effects come from many individuals with shared predictor levels. 
In the literature on mixed models, the random effects are thus often kept much simpler compared to the associated fixed effects models.   
In similar vein, we have focused here on time-varying random intercept type models. 
When categorical covariates, say $x_{1}^{\prime},\dots,x_{p^{\prime}}^{\prime}$, are desired to be included in the random effects model, 
$u_{i}(t)$ can be modified as $u_{x_{1}^{\prime},\dots,x_{p^{\prime}}^{\prime},i}(t)$ 
and the modeling strategies for the fixed effects components described in Section \ref{sec: fixed effects} can potentially be used.

\section{Posterior Inference} \label{sec: post inference}

Inference for the proposed LFMM 
is based on samples drawn from the posterior using an MCMC algorithm. 
In our model, the values of $\ell_{j,k}$'s are crucial in controlling the model size since they act as local covariate importance indicators. 
Varying values of $\ell_{j,k}$'s, however, result in varying
dimensional models, posing daunting computational challenges.
Dynamic message passing algorithms, such as the forward-backward sampler, are popular strategies for inference in HMMs and fHMMs \citep{Rabiner:1989, Scott:2002}. 
However, it is not clear how message passing strategies can be adapted to include inferences about the $\ell_{j,k}$'s.

We address these challenges by designing an efficient trans-dimensional transition step 
which updates the partition structure and the corresponding local curves at every location.
First, for every location $k$, an update in the partition structure $\rho_{k}$ is proposed.
Second, conditional on $\rho_{k}$, samples of the spline coefficients $\bbeta_{k}^{\star \star} = \{\beta_{k,h}^{\star \star}\}_{h=1}^{m_{k}}$ are drawn from their Gaussian full conditional distributions.

Specifically, the first step involves updating, for every predictor $j$ at each location $k$, 
the first layer of latent variables $\bz_{j,k} = (z_{j,k}^{(1)}, \dots, z_{j,k}^{(x_{j,\max})})$, the implied partition sizes $(\ell_{1,k},\dots,\ell_{p,k})$, 
and the corresponding second layer of latent variables $\bz_{k} = z_{k}^{(z_{1,k}, \dots, z_{p,k})}$.
Designing an efficient such proposal is made challenging by the discrete and potentially high-dimensional support of the latent variables $\bz_{j,k}$ and $\bz_{k}$.
However, the proposal distribution can be defined sequentially as 
\vspace{-5ex}\\
\bse
	q(\ell_{j,k}^{\prime}, \bz_{j,k}^{\prime}, \bz_{k}^{\prime} \mid \ell_{j,k}, \bz_{j,k}, \bz_{k}) = q_{1}(\ell_{j,k}^{\prime}, \bz_{j,k}^{\prime} \mid \ell_{j,k}, \bz_{j,k}) q_{2}(\bz_{k}^{\prime} \mid \ell_{j,k}^{\prime}, \bz_{j,k}^{\prime}).
\ese
\vspace{-5ex}\\
First, we perturb the current state $\bz_{j,k}$ to a new configuration $\bz_{j,k}^{\prime}$ 
by sampling uniformly in a Hamming ball of radius $r$ around $\bz_{j,k}$ \citep{titsias2016hamming}, 
resulting in an efficient first layer proposal that shares many of the old components as 
\vspace{-5ex}\\
\bse
	q_{1}(\ell_{j,k}^{\prime}, \bz_{j,k}^{\prime} \mid \ell_{j,k}, \bz_{j,k}) = \Unif \{\bz_{j,k}^{\prime} \mid \H_{m}(\bz_{j,k})\} 1\{\ell_{j,k}^{\prime} = |\Z_{j,k}^{\prime}|\}.
\ese
\vspace{-5ex}\\
Conditioning on the first layer of latent variables, we update the second layer as 
\vspace{-5ex}\\
\bse
	q_{2}(\bz_{k}^{\prime} \mid \ell_{j,k}^{\prime}, \bz_{j,k}^{\prime}) = \Mult \left( 1 / \ell_{k}, \dots, 1 / \ell_{k} \right).
\ese
\vspace{-5ex}\\
In terms of the implied marginal partition structure $\Z_{j,k}$, 
when $r = 1$, 
this corresponds to (A) selecting a covariate level and 
either (Ba) merging it to one of the other existing partition elements or 
(Bb) creating a singleton by separating it from its partition element.
Since the first layer proposal distribution is symmetric, the resulting acceptance rate of the Metropolis-Hastings (M-H) step is 
\vspace{-5ex}\\
\be
	r_{acc} = \dfrac{p(\by_{k} \mid \rho_{k}^{\prime}, \sigma_{\varepsilon}^{2}, \sigma_{\beta}^{2}, \bzeta)}{p(\by_{k} \mid \rho_{k}, \sigma_{\varepsilon}^{2}, \sigma_{\beta}^{2}, \bzeta)} \cdot \dfrac{p(\bz_{j,k}^{\prime}) p(\ell_{j,k}^{\prime}) p(\bz_{k}^{\prime})}{p(\bz_{j,k}) p(\ell_{j,k}) p(\bz_{k})} \cdot \dfrac{q_{2}(\bz_{k} \mid \ell_{j,k}^{\prime}, \bz_{j,k}^{\prime})}{q_{2}(\bz_{k}^{\prime} \mid \ell_{j,k}^{\prime}, \bz_{j,k}^{\prime})},
	\label{eq: acc_rate}
\ee
\vspace{-4ex}\\
where $\bzeta$ denotes a generic variable that collects all other variables not explicitly mentioned here, including the data points, 
and $\by_{k} = \{y_{i,\ell,k}\}_{i,\ell}$.
Importantly, the spline coefficient parameters $\bbeta_{k}^{\star \star}$ at each location $k$ 
can be analytically integrated out of the posterior of the corresponding partition structure. 
This allows for an efficient scheme for sampling the random partition structures based on their marginal likelihood
\vspace{-5ex}\\
\bse
	p(\by_{k} \mid \rho_{k}, \sigma_{\varepsilon}^{2}, \sigma_{\beta}^{2}, \bzeta) = \prod_{h=1}^{m_{k}} \int p(\by_{k} \mid \beta_{k,h}^{\star \star}, S_{k,h}, \sigma_{\varepsilon}^{2}) p(\beta_{k,h}^{\star \star} \mid\sigma_{\beta}^{2}, \bzeta) d\beta_{k,h}^{\star \star}.
\ese
\vspace{-5ex}\\
The second term in the integral is the conditional smoothing prior for the spline coefficients
\vspace{-5ex}\\
\bse
	\begin{split}
		p(\beta_{k,h}^{\star \star} \mid \sigma_{\beta}^{2}, \bzeta) &\propto \prod_{h^{\prime} \in \Z_{k,h}^{-}} \Normal\left(\beta^{\star\star}_{k,h} \mid \beta^{\star\star}_{k-1,h^{\prime}}, \sigma_{\beta}^{2}\right) \prod_{h^{\prime \prime} \in \Z_{k,h}^{+}} \Normal\left( \beta^{\star\star}_{k+1,h^{\prime \prime}} \mid \beta^{\star\star}_{k,h} , \sigma_{\beta}^{2}\right)
		\\
		&= \Normal(\mu_{k,h}, \sigma_{k,h}^{2}),
	\end{split}
\ese
\vspace{-4ex}\\
where $\Z_{k,h}^{-} = \big\{z_{k-1}: z_{k-1} = z_{k-1}^{(z_{1,k-1}, \dots, z_{p,k-1})}; (z_{1,k-1}, \dots, z_{p,k-1}) = (z_{1,k-1}^{(x_{1})}, \dots, z_{p,k-1}^{(x_{p})});$ $(x_{1},\dots,x_{p}) \in S_{k,h}\big\}$ and
$\Z_{k,h}^{+} = \big\{z_{k+1}: z_{k+1} = z_{k+1}^{(z_{1,k+1}, \dots, z_{p,k+1})}; (z_{1,k+1}, \dots, z_{p,k+1}) = (z_{1,k+1}^{(x_{1})}, \dots, z_{p,k+1}^{(x_{p})});$ $(x_{1},\dots,x_{p}) \in S_{k,h}\big\}$ are the indexes of the coefficients expressed at the previous and following locations, respectively, $n^{-}_{k,h} = |\Z_{k,h}^{-}|$ and $n^{+}_{k,h} = |\Z_{k,h}^{+}|$ are the corresponding cardinalities, $\sigma_{k,h}^{2} = \sigma_{\beta}^{2} (n^{-}_{k,h} + n^{+}_{k,h})^{-1}$ and $\mu_{k,h} = \frac{\sum_{h^{\prime}} \beta_{k-1,h^{\prime }}^{\star \star} + \sum_{h^{\prime \prime}} \beta_{k+1,h^{\prime \prime}}^{\star \star}}{n^{-}_{k,h} + n^{+}_{k,h}}$ are the resulting smoothing prior variance and mean parameters. 
First order Markov priors we designed in (\ref{eq: prior_all_components}), 
as opposed to second order differences considered in \cite{eilers1996flexible} and elsewhere, 
make these calculations much more tractable here.
Using this, we get
\vspace{-5ex}\\
\bse
	\begin{split}
		& p(\by_{k} \mid \rho_{k}, \sigma_{\varepsilon}^{2}, \sigma_{\beta}^{2}, \bzeta) \\
		&= \prod_{h=1}^{m_{k}} \int \prod_{\substack{(i,\ell) \text{ s.t.} \\ \bx_{i,\ell,t} \in S_{k,h}}} \left\{ (2 \pi \sigma_{\varepsilon}^{2})^{-\frac{1}{2}} e^{-\frac{1}{2 \sigma_{\varepsilon}^{2}} (r_{i,\ell,k}^{(m)} - \beta_{k,h}^{\star \star})^{2}} \right\} (2 \pi \sigma_{k,h}^{2})^{-\frac{1}{2}} e^{-\frac{1}{2 \sigma_{k,h}^{2}} \left(\beta_{k,h}^{\star \star} - \mu_{k,h} \right)^{2}} d\beta_{k,h}^{\star \star}
		\\
		&= \prod_{h=1}^{m_{k}} (2 \pi \sigma_{\varepsilon}^{2})^{-\frac{n_{k,h}}{2}} (\sigma_{k,h}^{2})^{-\frac{1}{2}} (\sigma_{k,h}^{\star 2})^{\frac{1}{2}} e^{-\frac{1}{2} \left( \frac{\sum_{i,\ell} r_{i,\ell,k}^{(m) 2}}{\sigma_{\varepsilon}^{2}} + \frac{\mu_{k,h}^{2}}{\sigma_{k,h}^{2}}  - \frac{\mu_{k,h}^{\star 2}}{\sigma_{k,h}^{\star 2}}\right)},
	\end{split}
\ese
\vspace{-4ex}\\
where $\br^{(m)} = \{y_{i,\ell,t} - u_{i}(t) \}_{i,\ell,t}$ are the main effects residuals, $n_{k,h} = | \{(i, l) \text{ s.t. } \bx_{i,\ell,t} \in S_{k,h} \} |$ is the number of observations allocated to the spline coefficient $\beta_{k,h}^{\star \star}$, $\sigma_{k,h}^{\star 2} = \left(\sigma_{\varepsilon}^{-2} n_{k,h} + \sigma_{k,h}^{-2} \right)^{-1}$ and $\mu^{\star}_{k,h} = \sigma_{k,h}^{\star 2} \left( \sigma_{\varepsilon}^{-2} \sum_{i,\ell} r_{i,\ell,k}^{(m)} + \sigma_{k,h}^{-2} \mu_{k,h} \right)$.

Conditional on the partition structure $\rho_{k}$, the group specific curves are sampled from their Gaussian full conditional distribution
\vspace{-5ex}\\
\be
p(\beta^{\star \star}_{k,h} \mid \by_{k}, S_{k,h}, \sigma_{\varepsilon}^{2}, \sigma_{\beta}^{2}, \bzeta) \propto p(\by_{k} \mid \beta_{k,h}^{\star \star}, S_{k,h}, \sigma_{\varepsilon}^{2}) p(\beta_{k,h}^{\star \star} \mid\sigma_{\beta}^{2}, \bzeta)
=\Normal\left\{ \mu^{\star}_{k,h}, \sigma_{k,h}^{\star 2} \right\}. \label{eq: fullcond_beta_star}
\ee
\vspace{-5ex}

To simplify posterior sampling for the scale parameter $\sigma_{\beta}$, 
we used a hierarchical scale mixture representation of the half-Cauchy distribution \citep{makalic2016simple}. 
Introducing an auxiliary variable $\nu_{\beta}$, the $\HC(0,s_{\sigma})$ prior can be represented as 
\vspace{-5ex}\\
\bse
	\sigma_{\beta}^{2} \sim \IG (1/2, 1/\nu_{\beta}),~~~~~\nu_{\beta} \sim \IG(1/2,1/s_{\sigma}^{2}). 
\ese
\vspace{-5ex}\\
Posterior full conditionals for $\sigma_{\beta}^{2}$ and $\nu_{\beta}$ then belong to the inverse-Gamma family and can be easily sampled from. 
The same trick, however, does not yield tractable full conditionals for $\sigma_{u,s}^{2}$ and $\sigma_{u,a}^{2}$. 
M-H steps are used for these parameters.

The full MCMC sampler comprises the steps reported in Algorithm \ref{algo: MCMC} 
in Section \ref{sec: sm mcmc} in the supplementary materials. 
Our software implementation in \texttt{R} and \texttt{C++}, available as part of the online supplementary materials, is highly automated, 
requiring only the available data points and the values of a few prior hyper-parameters as inputs. 
These hyper-parameters appear deep inside the model hierarchy and inference is highly robust to their choices. 
Additional details on the default choices of the hyper-parameters, 
the runtime of the algorithm, etc. 
are provided in Sections \ref{sec: sm prior hyper-parameters} and \ref{sec: sm software} in the supplementary materials.

\section{Posterior Consistency} \label{sec: post con}

This section presents some convergence results for our proposed longitudinal functional mixed model. 
We focus on the case where $n \to \infty$ 
but $L$, the number of replicates per individual, 
and $T$, the number of data recording time points, 
are kept fixed, 
which constitutes an appropriate asymptotic regime for the applications discussed later. 
Under this framework, we focus mainly on the recovery of the fixed effects components. 
When $L \rightarrow \infty$, similar results can also be established for the individual specific effects.
We restrict ourselves to consistency at the knot points 
which coincide with the set of unique data observing time points in the setting of this article. 
The functional domain remaining fixed to a finite interval, say $[A,B]$, 
when 
the number of data recording time points inside the domain $T \rightarrow \infty$ 
and some additional mild smoothness assumptions are made on the true underlying functions, 
the results can also be extended to the entire domain.

Our proofs rely on some results and ideas from \cite{ghosal1999posterior} and \cite{suarez2016bayesian} 
and are presented in Section \ref{sec: sm proofs} in the supplementary materials. 
We first show consistency for the functional fixed effects. 
Using this result, we then show that our proposed
model can also recover the underlying true local partitions of the covariate space and hence perform consistent variable selection.

We let $\Pi(\cdot)$ denote the prior distribution induced by our model on the space of fixed effects functions $f_{\bx}(t)$ 
and $\Pi(\cdot \vert \data)$ denote the corresponding posterior.
We let $g(\bx)$ denote the
probability distribution of $\bx$. 
We consider the $g$-weighted local $L_{2}$-norm of the function $f_{\bx}(t)$, 
defined as $||f||_{2, g, loc}^{2} = \sum_{\bx \in \X} g(\bx) \sum_{k=1}^{K} f_{\bx}^{2}(k)$. 
For the linear B-spline mixtures used in this article, $f_{\bx}(k) = \beta_{k,\bx}$.

Integrating out the random effects distribution \eqref{eq: random effects} from model \eqref{eq: function 1}, we obtain
\vspace{-5ex}\\
\be
& \{y_{i,\ell,t} \mid x_{j,i,\ell,t} = x_{j}, j=1,\dots,p \} \sim \Normal \{f_{\bx}(t), \sigma_{\varepsilon}^{2} + \sigma_{u}^{2}(t)\},
\ee
\vspace{-5ex}\\
where $\sigma_{u}^{2}(t) = \{(\sigma_{u,a}^{-2}\bI_{K}+\sigma_{u,s}^{-2}\bP_{u})^{-1}\}_{t,t}$. 
In our proof, we deviate slightly from our stated model in assuming exponentially decaying tails for the priors on the variance parameters 
$\sigma_{u,a}^{2}$ and $\sigma_{u,s}^{2}$
instead of the more non-informative half-Cauchy priors we used in our implementation.

\begin{Thm}[function estimation] \label{thm: function estimation}
\label{thm: consistency}
	For any $\epsilon > 0$, $\Pi(||f - f_{0}||_{2,g,loc} < \epsilon \mid \data) \rightarrow 1$. 
\end{Thm}

Without any loss of generality, we assume that $g(\bx) > 0$ for all $\bx \in \X$. 
If not, we can simply restrict ourselves to the set on which $g(\bx)>0$. 
We then have $n_{\bx} \to \infty$ as $n \to \infty$ for all $\bx\in\X$.  
The asymptotic regime can then be understood as averaging over $n_{\bx}$ replications for each $\bx$, 
thus replacing $\sigma_{\varepsilon}^{2} + \sigma_{u}^{2}(t)$ by $\sigma_{n,\bx}^{2} = n_{\bx}^{-1}\{\sigma_{\varepsilon}^{2} + \sigma_{u}^{2}(t)\}$.
Theorem \ref{thm: consistency} then implies that, 
for any $\bx \in \X$ and $\epsilon > 0$, $\Pi(||f_{\bx} - f_{\bx,0}||_{2,loc} < \epsilon \mid \data) \rightarrow 1$, 
where $\b1f_{\bx} = (f_{1,\bx}, \dots, f_{K,\bx})\trans$ with $f_{k,\bx}=f_{\bx}(k)$ and $||f_{\bx}||_{2, loc}^{2} = \sum_{k=1}^{K} f_{\bx}^{2}(k)$.

For a given location $k$, let $\rho_{k} = \{S_{k,1}, \dots, S_{k,m_{k}}\}$ be a random partition of $\X$, 
the space of vectors of length $p$ whose individual entries have values in $\X_{j}$, respectively.
The partition $\rho_{k}$ is defined in the following way:
\vspace{-5ex}\\
\bse
\beta_{k,\bx}^{\star \star} = \beta_{k,\bx^{\prime}}^{\star \star} \Longleftrightarrow \bx, \bx^{\prime} \in S_{k,h} \text{ for some } h \in \{1, \dots, m_{k} \}.
\ese
\vspace{-5ex}\\
Our hierarchical prior for the random partitions assigns a positive probability to each possible configuration.
Let $\rho_{k,0}$ be the partition generated by the true values of the parameters at location $k$.
Then the following theorem holds.

\begin{Thm}[variable selection] \label{thm: variable selection}
\label{thm: partition} 
	$\Pi(\rho_{k} = \rho_{k,0} \mid \data) \rightarrow 1$.
\end{Thm}

The construction of our model in Section \ref{sec: fixed effects} is such that the influences of the predictors are encoded precisely by the model induced partition structures - 
the predictor $x_{j}$ is important at location $k$ if its levels belong to at least two different sets in the partition $\rho_{k}$. 
Consistency in recovering the local partitions 
thus immediately implies consistency in local variable selection.

\section{Simulation Studies} \label{sec: sim studies}

In synthetic experiments, the proposed longitudinal framework achieved excellent empirical performance 
in recovering the true fixed and random effect curves and associated local cluster configurations from noisy subject level data. 
Figure \ref{fig: eye} illustrates the scenario used in the simulation studies.
We considered $T=20$ time points $\{1, \dots, T\}$. 
We generated $p = 10$ predictors, $x_{1}, x_{2} \in \{1,2\}$ and $x_{3}, \dots, x_{10} \in \{1,2,3\}$. 
The total number of possible level combinations of $(x_{1},\dots,x_{10})$ across all time points to consider in a fully flexible but completely unstructured model 
would thus be $T\prod_{j=1}^{10}x_{j,\max} = 20 \times 2^{2} \times 3^{8} = 20 \times 26,244 = 524,880$.
The true data generating mechanism is such that $x_{1}$ and $x_{3}$ are locally important whereas all other covariates are redundant at all time points. 
The fixed effects curves corresponding to the levels $\{1,2\}$ and $\{3\}$ of $x_{3}$ are initially equal, then diverge at $t=5$ and finally merge back at $t=17$, conditional on $x_{1} = 1$. 
The fixed effects curves corresponding to the levels $\{1\}$ and $\{2\}$ of $x_{1}$ are initially equal and then diverge at $t=8$.
The true unique spline coefficients are 
\vspace{-5ex}\\
\bse
& \bbeta_{1}^{\star \star} = (5, 5, 5, 5, 6, 7.25, 8.5, 9, 9.25, 9.5, 9.5, 9.25, 9, 8.5, 7.25, 6, 5, 5, 5, 5)\trans,\\ 
& \bbeta_{2}^{\star \star} = (5, 5, 5, 5, 4, 2.75, 1.5, 1, 0.75, 0.5, 0.5, 0.75, 1, 1.5, 2.75, 4, 5, 5, 5, 5)\trans~\text{and}\\
& \bbeta_{3}^{\star \star} = (5, 5, 5, 5, 6, 7.25, 8.5, 10.5, 12, 13.25, 13.75, 13.75, 13.5, 13, 12.5, 12, 11.25, 10.5, 9.5, 8.5)\trans.
\ese
\vspace{-5ex}\\
We generated $n = 50$ individual specific curves with $L_{i,t} = 5$ repeated measurements at each time point. 	
The residual variance was set at $\sigma_{\varepsilon}^{2} = 1$, whereas the variance and the smoothness of the random effects were $\sigma_{u,s}^{2} = 0.1$ and $\sigma_{u,a}^{2} = 2$, respectively.

As shown in Figure \ref{fig: n_groups_sim}, our method correctly recovers $x_{1}$ and $x_{3}$ as the only significant predictors.
In fact, the estimated number of groups $\ell_{j,k}$ associated with the other predictors consistently equals to one.
The posterior probabilities also correctly estimate two groups for $x_{1}$ starting from $t = 8$ and two groups for $x_{3}$ starting from $t = 5$.
Estimates of the fixed effects curves and a few individual level curves obtained by our method are shown in Figure \ref{fig: eye}. 
Our model estimates the fixed (left panel) as well as the individual specific (right panel) effects very precisely 
by borrowing information whenever predictors are redundant or covariate levels are in the same cluster.

\begin{figure}[!ht]
	\centering
	\includegraphics[width=0.99\linewidth]{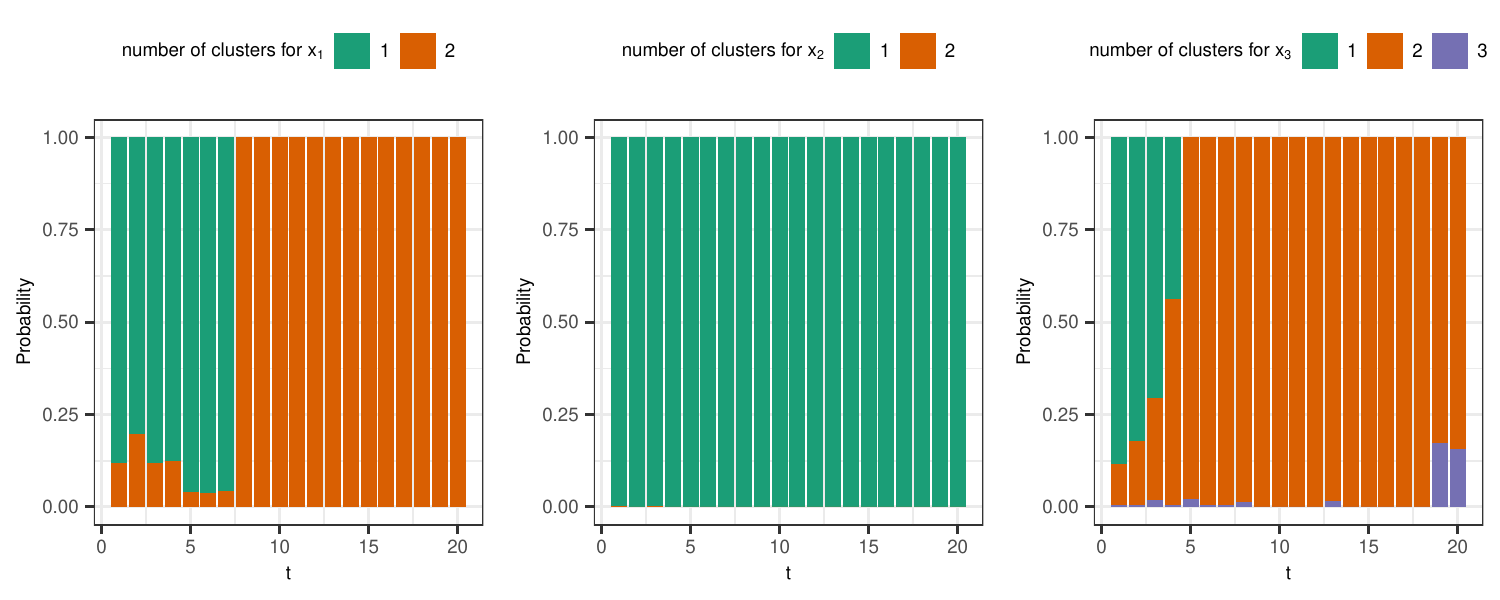}
	\caption{Results for synthetic data: 
	The estimated posterior probabilities for the number of clusters of the predictors' levels over time for $x_{1}, x_{2}$ and $x_{3}$. 
	The predictors $x_{1}$ and $x_{3}$ were locally important. 
	The remaining predictors, namely $(x_{2},x_{4},\dots,x_{10})$, including $x_{2}$ shown here, 
	were never included in the model - their levels always formed a single cluster.}
	\label{fig: n_groups_sim}
\end{figure}

\begin{figure}[!ht]
	\centering
	\begin{center}
		\includegraphics[width=0.99\linewidth]{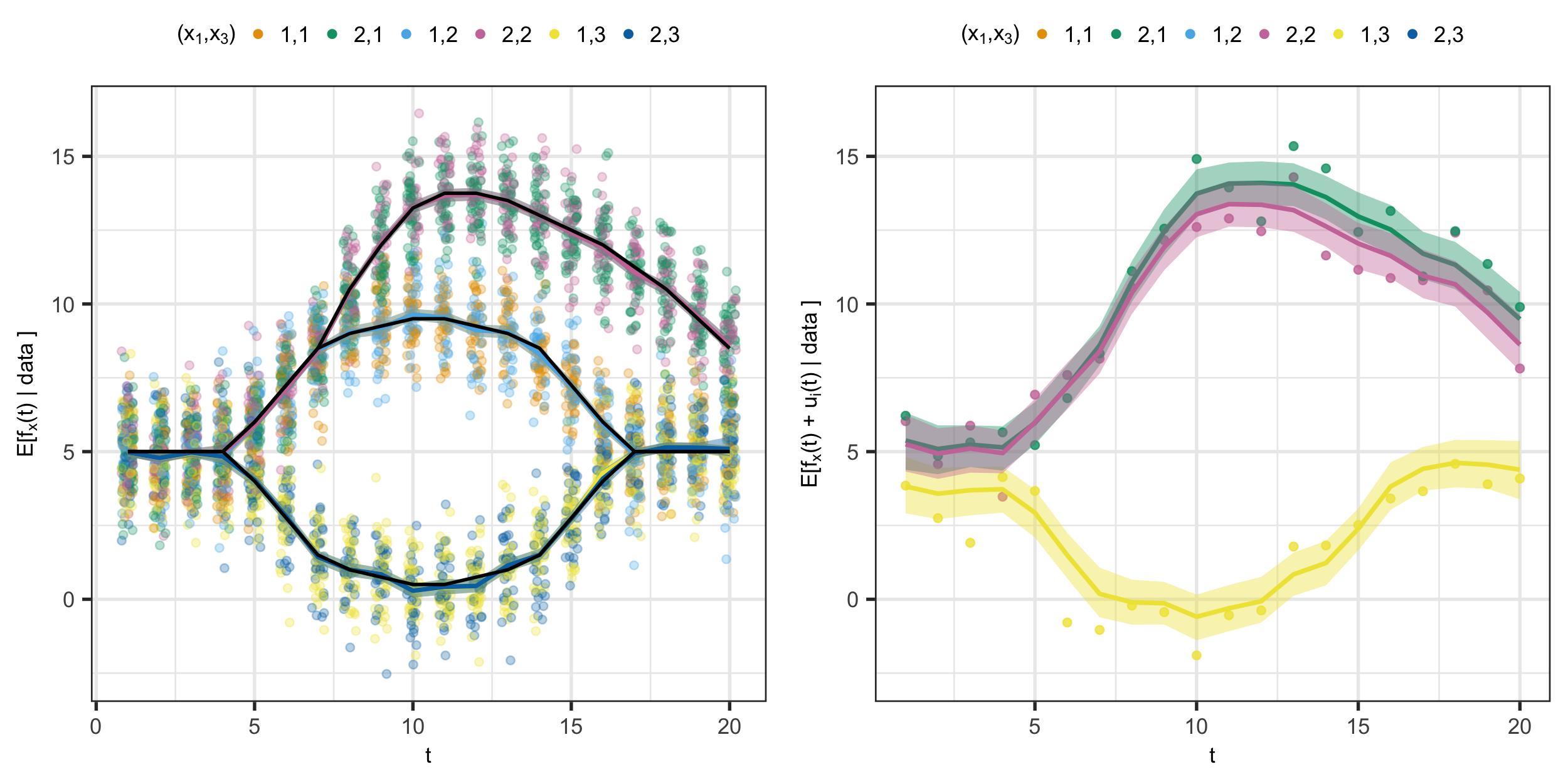}
	\end{center}
	\vskip-15pt
	\caption{Results for synthetic data: 
	Scenario with ten covariates $(x_{1},\dots,x_{10})$ where only $(x_{1}, x_{3})$ are locally important, as described in Section \ref{sec: sim studies}.
	Left panel: Estimated posterior means (colored lines) and $95\%$ point wise credible intervals for the fixed effects,  
	superimposed on slightly jittered response values $y_{i,\ell,t}$ for all combination of the levels of the significant predictors $(x_{1}, x_{3})$. 
	The true fixed effects are superimposed (black lines).
	Right panel: Estimated posterior means (colored lines) and $95\%$ point wise credible intervals for three individual specific curves, 
	superimposed on the associated observed individual response values $y_{i,\ell,t}$. 
	The figure here corresponds to the synthetic data set that produced the median root mean squared error. 	
	}
	\label{fig: eye}
\end{figure}

We compare the out-of-sample predictive performance of our proposed LFMM with state-of-the-art parametric and nonparametric regression alternatives.
We focus particularly on BART models by fitting both the original BART \citep{chipman2010bart} and the smooth BART \citep{linero2018bayesiansmooth} to the synthetic data sets.
In addition, we apply a LASSO regression model, implemented using the function \texttt{glmnet} in \texttt{R}, independently at each time point. 
\begin{figure}[ht!]
	\centering
	\includegraphics[width=0.325\linewidth]{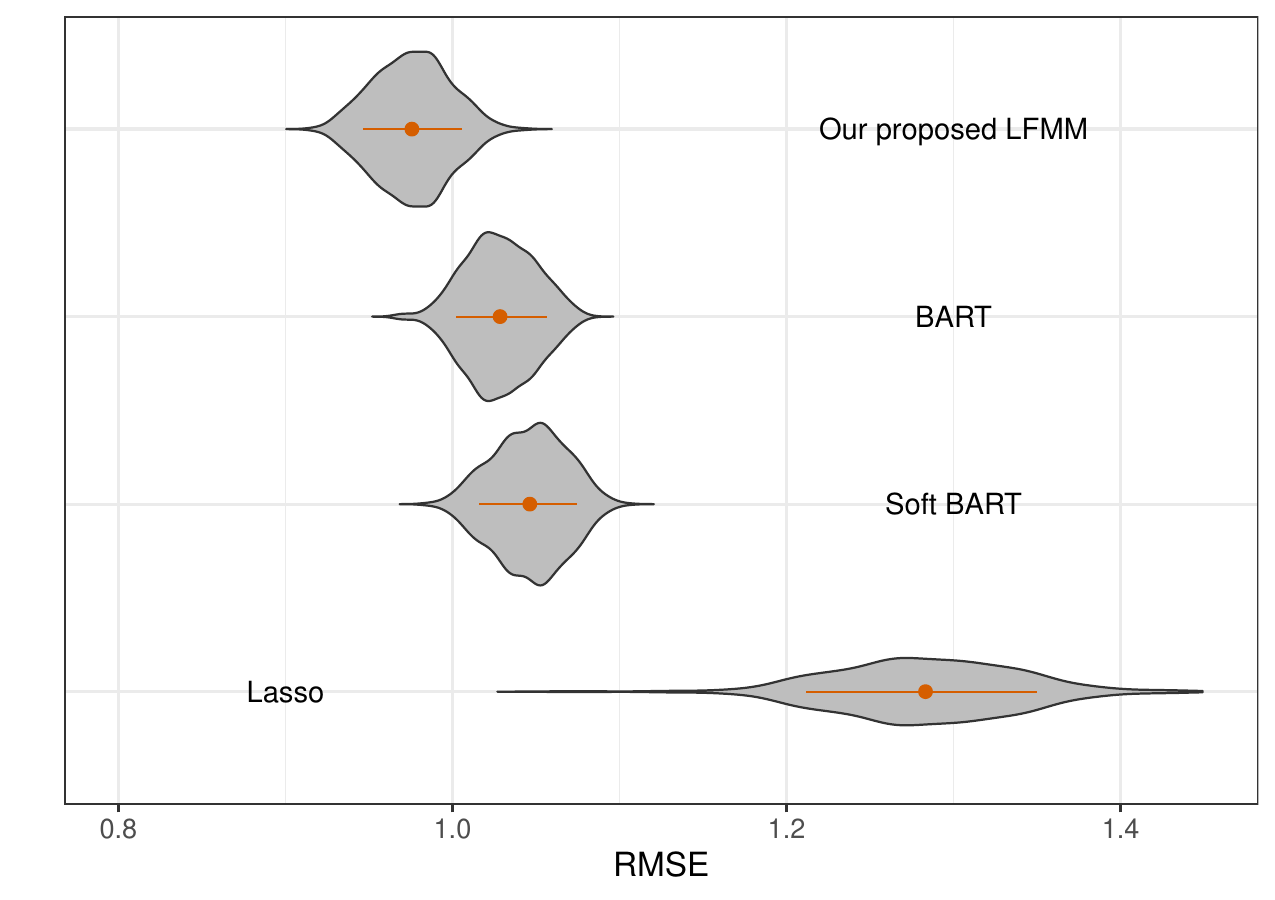}
	\includegraphics[width=0.325\linewidth]{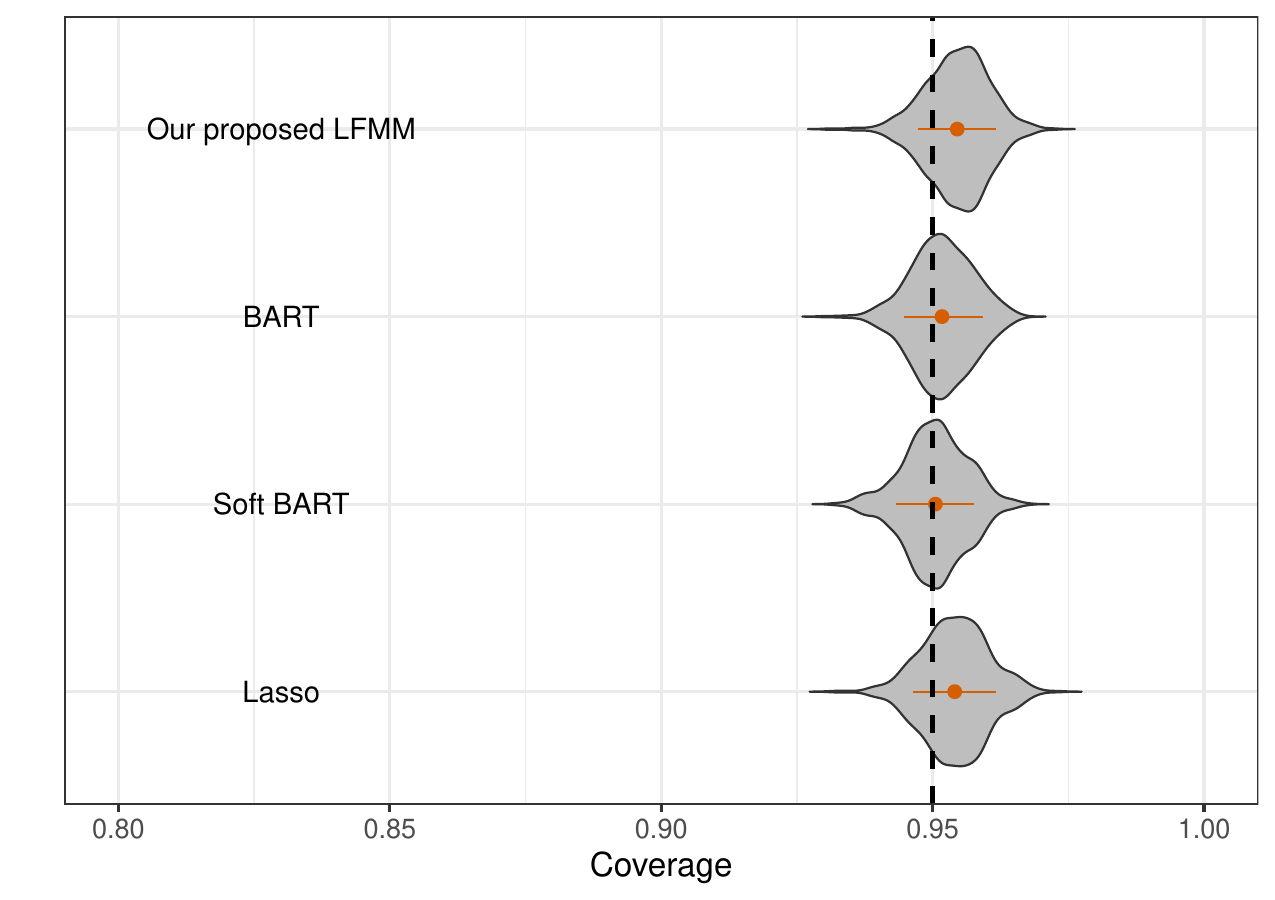}
	\includegraphics[width=0.325\linewidth]{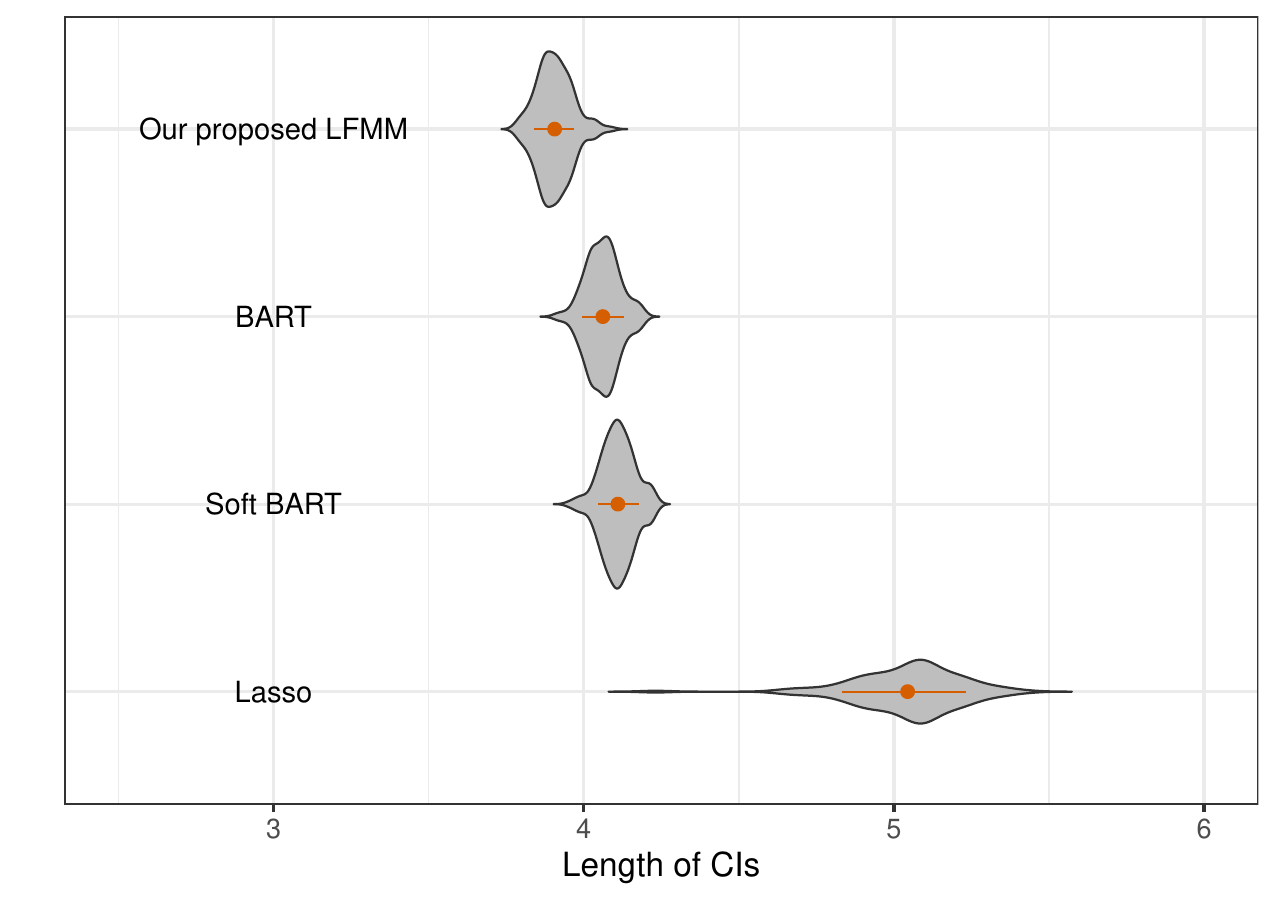}
	\caption{Results for synthetic data: The left panel shows the out-of-sample root mean squared error. 
	The middle panel shows the coverage of $95\%$ prediction intervals. 
	The right panel shows the widths of the prediction intervals.
	All measures reported are obtained over 500 $75\%$-$25\%$ training-test splits.
	The red points represent the averages across simulations, whereas the red intervals represent the interquartile ranges across simulations.}
	\label{fig: sim_RMSE}
\end{figure}
Figure \ref{fig: sim_RMSE} compares the out-of-sample predictive performance (left panel), the coverage of the $95\%$ prediction intervals (middle panel), and the lengths of these intervals (right panel),  
for the different methods for $500$ simulated data sets with $75\%$-$25\%$ training-test splits.
All methods produced prediction intervals with coverages probabilities close to the nominal rate. 
The coverage of Bayesian credible intervals is expected to exactly match with the corresponding nominal values in (frequentist) repeat simulations only in an asymptotic sense via Bernstein-von Mises theorems \citep{van2000asymptotic,ghosh2003}. 
Slight departures for complex models in high-dimensional finite sample settings is thus not completely unexpected. 
Remarkably, despite being very parsimonious, our proposed LFMM not only had substantially smaller out-of-sample RMSEs, it actually performed uniformly better than all other approaches in all simulated data sets. 
Furthermore, our method actually also achieved this with uniformly smaller interval widths.

{
We present the results of some additional simulation experiments in Section \ref{sec: sm add simulations} of the supplementary materials 
to assess the performance of our proposed model in the special but unrealistic case when no individual specific information is available 
as it  
provides a fairer comparison with our competitors that do not accommodate random effects. 
Our findings are, however, very similar to the scenario presented here. 
}

\section{Applications} \label{sec: applications}
In this section, we discuss the results of our method applied to two data sets. 
Three more examples, including one with time-varying predictors, are presented in Section \ref{sec: sm add data} of the supplementary materials.

\subsection{Progesterone Data}

We describe here an application of our proposed approach to modeling progesterone data \citep{brumback1998smoothing,nguyen2011dirichlet} 
that record the logarithm of the progesterone levels of women during the course of their menstrual cycles, measured by urinary hormone assay.
Measurements of 51 female subjects occur during a monthly cycle ranging from -8 to 15 (8 days pre-ovulation to 15 days post-ovulation). 
There are a total of 91 cycles: the first 70 cycles belong to the non-conceptive group, the remaining 21 cycles belong to the conceptive group. 
The type of cycle is the single categorical predictor used in the analysis.

\begin{figure}[!ht]
	\centering
	\includegraphics[width=0.985\linewidth]{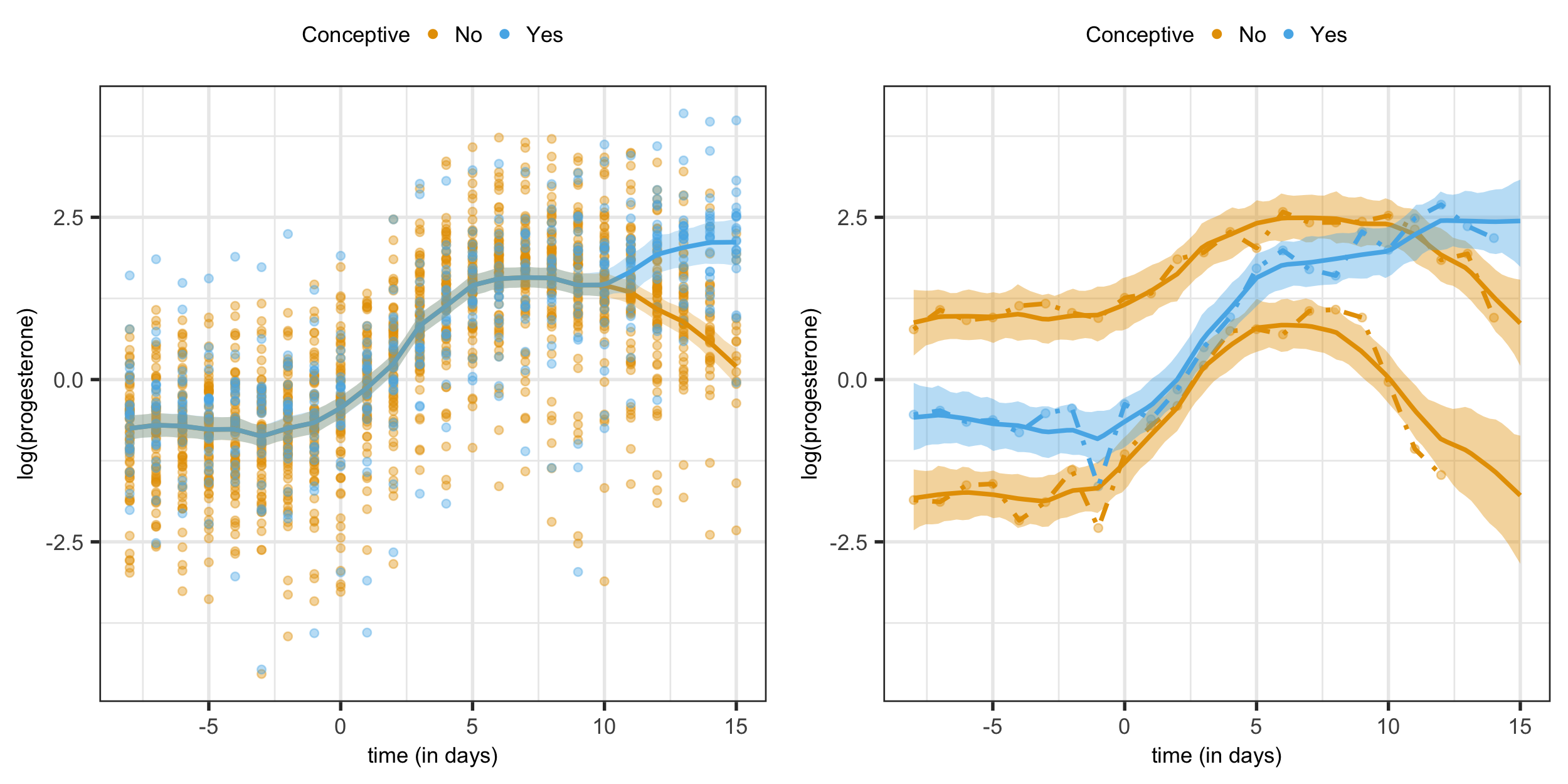}
	\caption{Results for the progesterone data: The left panel shows the estimated posterior means and $95\%$ point wise credible intervals for the fixed effects curves,  
	superimposed on slightly jittered response values $y_{i,\ell,t}$ for different levels of $x$. 
	Here $x$ is found to be only locally important: Its two levels have no effect on $y$ for $t\in[-8,10]$ but influence $y$ significantly differently for $t \in (10,15]$.
	The right panel shows {three} examples of individual specific curves, their estimated posterior means (solid lines) and $95\%$ point wise credible intervals, 
	superimposed on the associated observed individual response values (dashed lines) $y_{i,\ell,t}$.}
	\label{fig: prog_results}
\end{figure}

Figure \ref{fig: prog_results} (left) shows the estimated posterior means and associated $95\%$ point wise credible intervals for the group specific curves. 
The population level curves for conceptive and non-conceptive cycles are clustered together in the early part of the cycle 
but become different in the late post ovulation period.
In particular, the late conceptive cycles are associated with higher levels of progesterone. 
Global clustering methods would not allow clustering of the groups in the pre-ovulation period and would simply separate the two groups across all time points.  
Figure \ref{fig: prog_results} (right) shows the estimated posterior means and associated $95\%$ point wise credible intervals for the individual specific curves. 
These estimates show how our model can flexibly recover the individual level variations.

\subsection{Health and Retirement Study Data}

We analyze publicly available data from a longitudinal survey of US adults, the Health and Retirement Study (HRS). 
The HRS was established to assess the health implications of aging at both individual and population levels and has been fielded biennially years since 1992.
Three categories of data - public, sensitive and restricted - can be accessed on the \href{https://hrsonline.isr.umich.edu}{\texttt{HRS website}} or, alternatively, via the \href{https://www.rand.org/well-being/social-and-behavioral-policy/centers/aging/dataprod/hrs-data.html}{\texttt{RAND HRS longitudinal file}}.
The HRS is sponsored by the National Institute on Aging and the University of Michigan 
and has previously been analyzed in \citet{sonnega2014cohort} and most recently in \citet{deshpande2020vc}.

The goal of the study is to understand how life course processes influence the trajectories of cognitive health. 
Therefore, we focus on predicting each subject's later-life cognitive function over time using life course socio-economic position (SEP) indicators.
The $p=13$ covariates include measures of SEP in childhood (SEP index), early adulthood (educational attainment), and later-life (household wealth) as well as measures of later-life mental and physical health (binary indicators of physical activity, diabetes, heart problems, high blood pressure, loneliness and stroke as well as BMI and depression index) and socio-demographic factors (race, gender).
The size of the unstructured model $T\prod_{j=1}^{p}x_{j,\max} = 32 \times 580,608 = 18,579,456$ makes it impossible to estimate the parameters without adopting a dimensionality reduction approach.
The outcome is cognitive function as measured by a series of listening and memory tests that the HRS used to construct a score ranging from 0 to 35. 
We restricted our analysis to subjects aged between 65 and 96 years with at least two cognitive scores recorded between 2000 and 2016. 
This resulted in a sample of $n = 4,167$ subjects who were administered a total of $N = 27,820$ surveys, 
each individual being recorded either at even or at odd numbered ages but missing the intermediate values.

\begin{figure}[!ht]
	\centering
	\includegraphics[width=.99\linewidth]{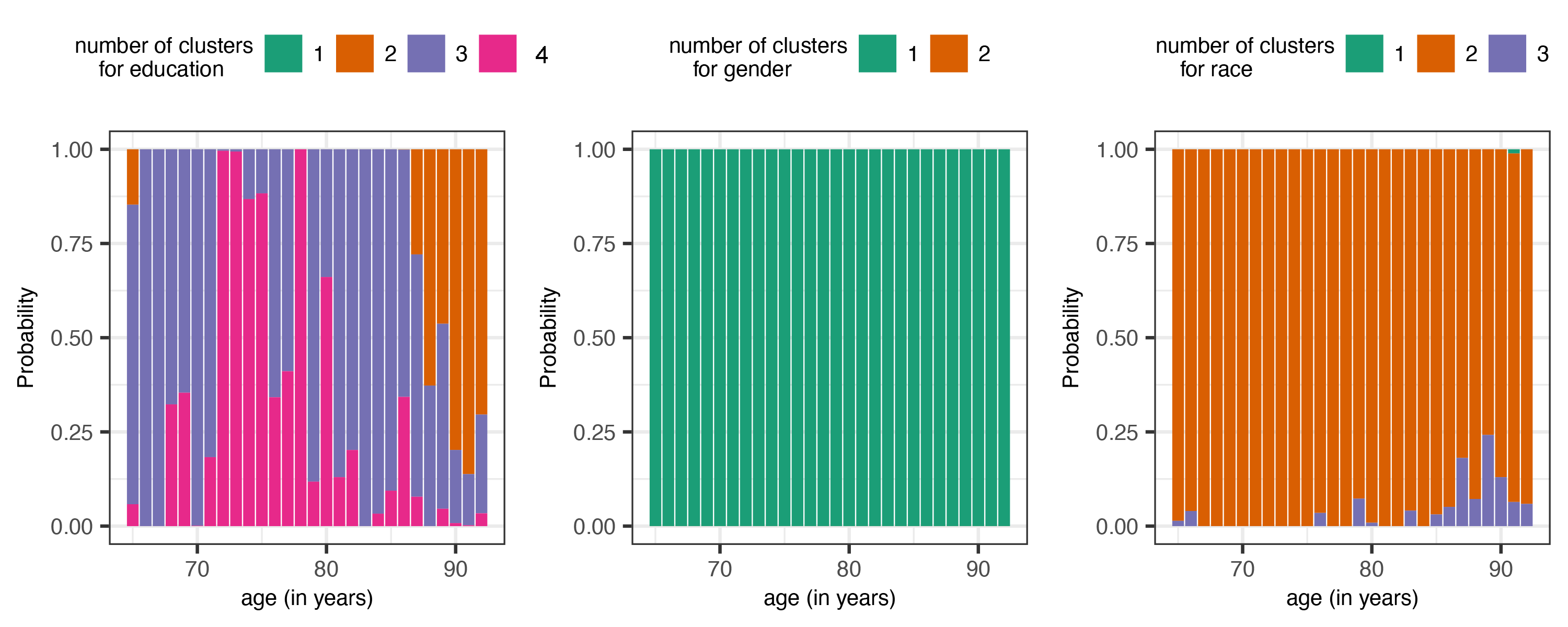}
	\caption{Results for the HRS: The estimated posterior probabilities for the number of clusters of the predictors' levels over time for $x_{1}=education$, $x_{2}=gender$, and $x_{3}=race$. 
	The predictors $x_{1}$ and $x_{3}$ were locally important. 
	The remaining predictors, including $x_{2}$ shown here, were never included in the model since the number of clusters of their levels was always $1$.}
	\label{fig: n_groups_cognitive}
\end{figure}

\begin{figure}[!ht]
	\centering
	\includegraphics[width=.99\linewidth]{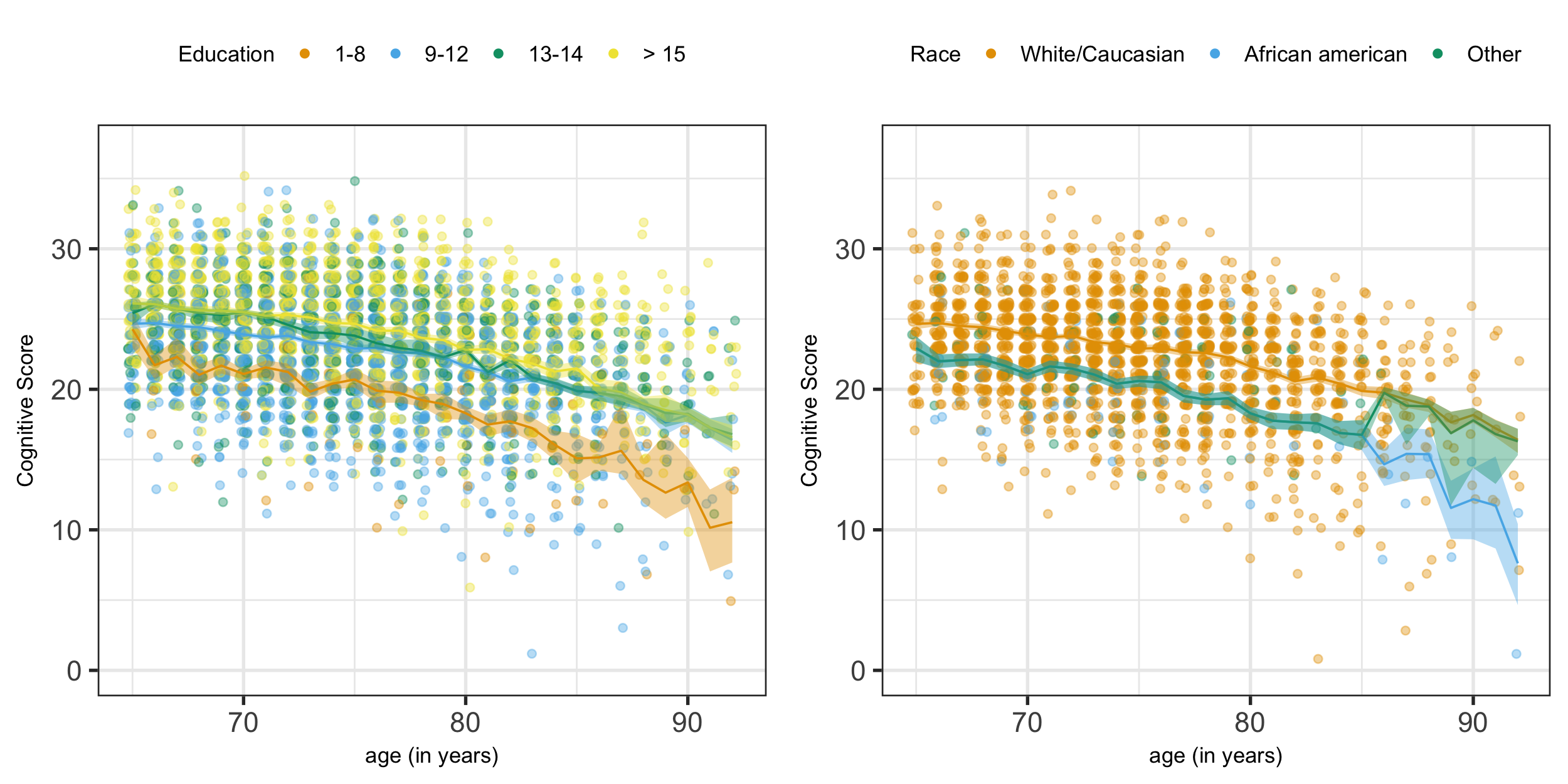}
	\caption{Results for the HRS: The left panel shows the estimated posterior means and $95\%$ point wise credible intervals for the fixed effects curves corresponding to different education levels,  
	superimposed on slightly jittered response values $y_{i,\ell,t}$. 
	The right panel shows the estimated posterior means and $95\%$ point wise credible intervals for the fixed effects curves corresponding to different races,  
	superimposed on slightly jittered response values $y_{i,\ell,t}$.}
	\label{fig: cognitive_results}
\end{figure}

Figure \ref{fig: n_groups_cognitive} shows the posterior probabilities for the number of groups $\ell_{j,k}$ associated to three of the predictors (education, gender, race). 
The other predictors' levels were grouped together at each location $k$ and therefore they did not affect the outcome.
Figure \ref{fig: cognitive_results} shows the effect of education and race, i.e., the two predictors that were selected by the model. 
These results highlight the importance of educational attainment due to its association with cognition.
It appears that higher levels educational attainment are associated with higher cognitive function across adulthood. 
This confirms that socioeconomic position in early adulthood as measured by education can have later life effects on cognition.
Conversely, it appears that the other SEP measures have no predictive effect on later-life cognition.
In middle aged invididuals, three groups of educational attainment seem to differently affect the outcome: 1-8, 9-12, 13+.
In old aged invididuals, instead, only two groups of educational attainment are significant: 1-8, and 9+.
As far as race is concerned, it appears that after controlling for the other covariates in this study, white and non-white individuals have significant differences in cognitive scores during later-life.
This finding also confirms the results in \citet{deshpande2020vc}, who estimated that white people's intercept parameter is larger than the one for other races, and is consistent with previous literature \citep{wilson2015cognitive, diaz2016racial}. 
This result indicates that other factors that are unaccounted for (i.e., quality of education or literacy) are affecting the estimated cognitive scores for each race/ethnic group.
Crucially, our model is able not only to flexibly estimate the cognitive score functions, but also to pool information across different covariate subgroups. 
Borrowing information across curves becomes especially important to estimate the cognitive score of older aged individuals due to the decrease in sample size.

\section{Discussion} \label{sec: discussion}

In this article, we developed a flexible Bayesian semiparametric approach to longitudinal functional mixed models in the presence of categorical covariates. 
Building on novel fHMM infused mixtures of locally supported B-splines, 
our proposed method allows the fixed effects components to vary flexibly with the associated covariates, 
allowing potentially different sets of important covariates to be included in the model at different time points. 
The mechanism not only allows different sets of covariates to be included in the model at different time points 
but also allows the selected predictors' influences to vary flexibly over time. 
Flexible time-varying additive random effects, modeled also by Markovian mixtures of B-splines, are used to capture subject specific heterogeneity. 
{We established theoretical results on posterior consistency of the proposed method for both function estimation and variable selection.} 
In simulation experiments, the method significantly outperformed the competitors. 
We illustrated the method's practical utility in real data applications.

The methodology presented here is highly generic and broadly adaptable to diverse other problems. 
While the focus of this article has been on dynamically varying longitudinal data models, 
the methodology could also be useful in static multiway mixed ANOVA designs. 
Methodological extensions we are pursuing as topics of separate research 
include 
dynamic partition models for observational units;
models for spatial and spatiotemporal settings;
models for multivariate responses; 
principled approaches to accommodate categorical and ordinal responses and continuous, ordinal and mixed type covariates; 
etc.

\section*{Supplementary Materials}
\baselineskip=14pt
The supplementary materials present brief reviews of B-splines, fHMMs, and tensor factorization methods for easy reference. 
The supplementary materials also include 
additional discussions on the characterization of overall, main and interaction effects and associated tests; 
proofs of the theoretical results; 
choice of the prior hyper-parameters; 
additional details of the MCMC algorithm used to sample from the posterior; 
MCMC diagnostics; 
results of some additional simulation experiments; 
additional real data applications; 
etc. 
\texttt{R} programs implementing the methods developed in this article and an accompanying `readme' file 
are also included as separate files in the supplementary materials. 

\vspace*{-10pt}
\section*{Funding}
This work was supported in part by the National Science Foundation grants DMS 1952679 to Mueller and DMS 1953712 to Sarkar.

\vspace*{-20pt}
\baselineskip=14pt
\bibliographystyle{natbib}
\bibliography{biblio}

\begin{thebibliography}{}

\bibitem[Berger(1985)Berger]{berger_book}
Berger, J.~O. (1985).
\newblock {\em {Statistical decision theory and Bayesian analysis}\/}.
\newblock Springer series in statistics. Springer-Verlag, New York, 2nd
  edition.

\bibitem[de~Boor(1978)de~Boor]{de1978practical}
de~Boor, C. (1978).
\newblock {\em A practical guide to splines\/}.
\newblock Springer-Verlag.

\bibitem[De~Lathauwer {\em et~al.}(2000)De~Lathauwer, De~Moore, and
  Vandewalle]{de_lathauwer_etal:2000}
De~Lathauwer, L., De~Moore, B., and Vandewalle, J. (2000).
\newblock A multilinear singular value decomposition.
\newblock {\em {SIAM} Journal on Matrix Analysis and Applications\/}, {\bf 21},
  1253--1278.

\bibitem[Eddelbuettel and Sanderson(2014)Eddelbuettel and
  Sanderson]{eddelbuettel2014rcpparmadillo}
Eddelbuettel, D. and Sanderson, C. (2014).
\newblock Rcpp{A}rmadillo: {A}ccelerating {R} with high-performance {C++}
  linear algebra.
\newblock {\em Computational Statistics \& Data Analysis\/}, {\bf 71},
  1054--1063.

\bibitem[Eddelbuettel {\em et~al.}(2011)Eddelbuettel, Fran{\c{c}}ois, Allaire,
  Ushey, Kou, Russel, Chambers, and Bates]{eddelbuettel2011rcpp}
Eddelbuettel, D., Fran{\c{c}}ois, R., Allaire, J., Ushey, K., Kou, Q., Russel,
  N., Chambers, J., and Bates, D. (2011).
\newblock Rcpp: {S}eamless {R} and {C++} integration.
\newblock {\em Journal of Statistical Software\/}, {\bf 40}, 1--18.

\bibitem[Escobar and West(1995)Escobar and West]{escobar1995bayesian}
Escobar, M.~D. and West, M. (1995).
\newblock Bayesian density estimation and inference using mixtures.
\newblock {\em Journal of the American Statistical Association\/}, {\bf 90},
  577--588.

\bibitem[Fr{\"u}hwirth-Schnatter(2006)Fr{\"u}hwirth-Schnatter]{fruhwirth2006finite}
Fr{\"u}hwirth-Schnatter, S. (2006).
\newblock {\em Finite mixture and {M}arkov switching models\/}.
\newblock Springer Science \& Business Media.

\bibitem[Gelman(2006)Gelman]{gelman2006prior}
Gelman, A. (2006).
\newblock Prior distributions for variance parameters in hierarchical models.
\newblock {\em Bayesian Analysis\/}, {\bf 1}, 515--534.

\bibitem[Geweke(1991)Geweke]{geweke1991evaluating}
Geweke, J. (1991).
\newblock Evaluating the accuracy of sampling-based approaches to the
  calculation of posterior moments.
\newblock In {\em Proceedings of the Fourth Valencia International Conference
  on Bayesian Statistics\/}, pages 169--193.

\bibitem[Ghahramani and Jordan(1997)Ghahramani and
  Jordan]{ghahramani1996factorial}
Ghahramani, Z. and Jordan, M.~I. (1997).
\newblock Factorial hidden {M}arkov models.
\newblock {\em Machine Learning\/}, {\bf 29}, 245--273.

\bibitem[Ghosal {\em et~al.}(1999)Ghosal, Ghosh, and
  Ramamoorthi]{ghosal1999posterior}
Ghosal, S., Ghosh, J.~K., and Ramamoorthi, R. (1999).
\newblock Posterior consistency of {D}irichlet mixtures in density estimation.
\newblock {\em Annals of Statistics\/}, {\bf 27}, 143--158.

\bibitem[Gilbody {\em et~al.}(2015)Gilbody, Littlewood, Hewitt, Brierley,
  Tharmanathan, Araya, Barkham, Bower, Cooper, Gask, {\em
  et~al.}]{gilbody2015computerised}
Gilbody, S., Littlewood, E., Hewitt, C., Brierley, G., Tharmanathan, P., Araya,
  R., Barkham, M., Bower, P., Cooper, C., Gask, L., {\em et~al.} (2015).
\newblock Computerised cognitive behaviour therapy (c{CBT}) as treatment for
  depression in primary care ({REEACT} trial): large scale pragmatic randomised
  controlled trial.
\newblock {\em BMJ\/}, {\bf 351}.

\bibitem[Group(2000)Group]{childhood2000long}
Group, C. A. M. P.~R. (2000).
\newblock Long-term effects of budesonide or nedocromil in children with
  asthma.
\newblock {\em New England Journal of Medicine\/}, {\bf 343}, 1054--1063.

\bibitem[Hitchcock(1927)Hitchcock]{hitchcock1927expression}
Hitchcock, F.~L. (1927).
\newblock The expression of a tensor or a polyadic as a sum of products.
\newblock {\em Journal of Mathematics and Physics\/}, {\bf 6}, 164--189.

\bibitem[Hothorn and Everitt(2014)Hothorn and Everitt]{hothorn2014handbook}
Hothorn, T. and Everitt, B.~S. (2014).
\newblock {\em A handbook of statistical analyses using {R}\/}.
\newblock CRC press.

\bibitem[Johndrow {\em et~al.}(2017)Johndrow, Bhattacharya, and
  Dunson]{johndrow2017tensor}
Johndrow, J.~E., Bhattacharya, A., and Dunson, D.~B. (2017).
\newblock Tensor decompositions and sparse log-linear models.
\newblock {\em Annals of Statistics\/}, {\bf 45}, 1--38.

\bibitem[Kolda and Bader(2009)Kolda and Bader]{kolda2009tensor}
Kolda, T.~G. and Bader, B.~W. (2009).
\newblock Tensor decompositions and applications.
\newblock {\em SIAM review\/}, {\bf 51}, 455--500.

\bibitem[McDonald and Zucchini(1997)McDonald and
  Zucchini]{mcdonald_zucchini:1997}
McDonald, S. and Zucchini, W. (1997).
\newblock {\em Hidden {M}arkov and other models for discrete-valued time
  series\/}.
\newblock Chapman \& Hall, London.

\bibitem[Moore {\em et~al.}(2000)Moore, Pedlow, Krishnamurty, and
  Wolter]{moore2000national}
Moore, W., Pedlow, S., Krishnamurty, P., and Wolter, K. (2000).
\newblock National longitudinal survey of youth 1997 ({NLSY97}).
\newblock Technical report, National Opinion Research Center.

\bibitem[Polson and Scott(2012)Polson and Scott]{polson2012half}
Polson, N.~G. and Scott, J.~G. (2012).
\newblock On the half-{C}auchy prior for a global scale parameter.
\newblock {\em Bayesian Analysis\/}, {\bf 7}, 887--902.

\bibitem[Proudfoot {\em et~al.}(2003)Proudfoot, Goldberg, Mann, Everitt, Marks,
  and Gray]{proudfoot2003computerized}
Proudfoot, J., Goldberg, D., Mann, A., Everitt, B., Marks, I., and Gray, J.
  (2003).
\newblock Computerized, interactive, multimedia cognitive-behavioural program
  for anxiety and depression in general practice.
\newblock {\em Psychological Medicine\/}, {\bf 33}, 217--227.

\bibitem[Tucker(1966)Tucker]{tucker:1966}
Tucker, L. (1966).
\newblock Some mathematical notes on three-mode factor analysis.
\newblock {\em Psychometrica\/}, {\bf 31}, 273--282.

\end{thebibliography}


\begin{thebibliography}{}

\bibitem[Albert and Chib(1993)Albert and Chib]{albert1993bayesian}
Albert, J.~H. and Chib, S. (1993).
\newblock Bayesian analysis of binary and polychotomous response data.
\newblock {\em Journal of the American Statistical Association\/}, {\bf 88},
  669--679.

\bibitem[Barry and Hartigan(1992)Barry and Hartigan]{barry1992product}
Barry, D. and Hartigan, J.~A. (1992).
\newblock Product partition models for change point problems.
\newblock {\em Annals of Statistics\/}, {\bf 20}, 260--279.

\bibitem[Breiman(2001)Breiman]{breiman2001statistical}
Breiman, L. (2001).
\newblock Statistical modeling: The two cultures.
\newblock {\em Statistical Science\/}, {\bf 16}, 199--231.

\bibitem[Brumback and Rice(1998)Brumback and Rice]{brumback1998smoothing}
Brumback, B.~A. and Rice, J.~A. (1998).
\newblock Smoothing spline models for the analysis of nested and crossed
  samples of curves.
\newblock {\em Journal of the American Statistical Association\/}, {\bf 93},
  961--976.

\bibitem[Chib and Hamilton(2002)Chib and Hamilton]{chib2002semiparametric}
Chib, S. and Hamilton, B.~H. (2002).
\newblock Semiparametric {B}ayes analysis of longitudinal data treatment
  models.
\newblock {\em Journal of Econometrics\/}, {\bf 110}, 67--89.

\bibitem[Chipman {\em et~al.}(2010)Chipman, George, McCulloch, {\em
  et~al.}]{chipman2010bart}
Chipman, H.~A., George, E.~I., McCulloch, R.~E., {\em et~al.} (2010).
\newblock {BART}: {B}ayesian additive regression trees.
\newblock {\em Annals of Applied Statistics\/}, {\bf 4}, 266--298.

\bibitem[Daniels and Pourahmadi(2002)Daniels and
  Pourahmadi]{daniels2002bayesian}
Daniels, M.~J. and Pourahmadi, M. (2002).
\newblock Bayesian analysis of covariance matrices and dynamic models for
  longitudinal data.
\newblock {\em Biometrika\/}, {\bf 89}, 553--566.

\bibitem[de~Boor(1978)de~Boor]{de1978practical}
de~Boor, C. (1978).
\newblock {\em A practical guide to splines\/}.
\newblock Springer-Verlag.

\bibitem[De~Iorio {\em et~al.}(2004)De~Iorio, M{\"u}ller, Rosner, and
  MacEachern]{de2004anova}
De~Iorio, M., M{\"u}ller, P., Rosner, G.~L., and MacEachern, S.~N. (2004).
\newblock An {ANOVA} model for dependent random measures.
\newblock {\em Journal of the American Statistical Association\/}, {\bf 99},
  205--215.

\bibitem[De~Lathauwer {\em et~al.}(2000)De~Lathauwer, De~Moore, and
  Vandewalle]{de_lathauwer_etal:2000}
De~Lathauwer, L., De~Moore, B., and Vandewalle, J. (2000).
\newblock A multilinear singular value decomposition.
\newblock {\em {SIAM} Journal on Matrix Analysis and Applications\/}, {\bf 21},
  1253--1278.

\bibitem[Denison {\em et~al.}(2002)Denison, Adams, Holmes, and
  Hand]{denison2002bayesian}
Denison, D., Adams, N., Holmes, C., and Hand, D. (2002).
\newblock Bayesian partition modelling.
\newblock {\em Computational Statistics \& Data Analysis\/}, {\bf 38},
  475--485.

\bibitem[Deshpande {\em et~al.}(2020)Deshpande, Bai, Balocchi, and
  Starling]{deshpande2020vc}
Deshpande, S.~K., Bai, R., Balocchi, C., and Starling, J.~E. (2020).
\newblock {VC}-{BART}: {B}ayesian trees for varying coefficients.
\newblock {\em arXiv preprint arXiv:2003.06416\/}.

\bibitem[D{\'\i}az-Venegas {\em et~al.}(2016)D{\'\i}az-Venegas, Downer, Langa,
  and Wong]{diaz2016racial}
D{\'\i}az-Venegas, C., Downer, B., Langa, K.~M., and Wong, R. (2016).
\newblock Racial and ethnic differences in cognitive function among older
  adults in the {USA}.
\newblock {\em International Journal of Geriatric Psychiatry\/}, {\bf 31},
  1004--1012.

\bibitem[Diggle {\em et~al.}(2002)Diggle, Diggle, Heagerty, Heagerty, Liang,
  Zeger, {\em et~al.}]{diggle2002analysis}
Diggle, P., Diggle, P.~J., Heagerty, P., Heagerty, P.~J., Liang, K.-Y., Zeger,
  S., {\em et~al.} (2002).
\newblock {\em Analysis of longitudinal data\/}.
\newblock Oxford University Press.

\bibitem[Duan {\em et~al.}(2007)Duan, Guindani, and
  Gelfand]{duan2007generalized}
Duan, J.~A., Guindani, M., and Gelfand, A.~E. (2007).
\newblock Generalized spatial {D}irichlet process models.
\newblock {\em Biometrika\/}, {\bf 94}, 809--825.

\bibitem[Dunson(2000)Dunson]{dunson2000bayesian}
Dunson, D.~B. (2000).
\newblock Bayesian latent variable models for clustered mixed outcomes.
\newblock {\em Journal of the Royal Statistical Society: Series B\/}, {\bf 62},
  355--366.

\bibitem[Efron(2020)Efron]{efron2020prediction}
Efron, B. (2020).
\newblock Prediction, estimation, and attribution.
\newblock {\em Journal of the American Statistical Association\/}, {\bf 115},
  636--655.

\bibitem[Eilers and Marx(1996)Eilers and Marx]{eilers1996flexible}
Eilers, P.~H. and Marx, B.~D. (1996).
\newblock Flexible smoothing with {B}-splines and penalties.
\newblock {\em Statistical Science\/}, {\bf 11}, 89--102.

\bibitem[Fitzmaurice {\em et~al.}(2008)Fitzmaurice, Davidian, Verbeke, and
  Molenberghs]{fitzmaurice2008longitudinal}
Fitzmaurice, G., Davidian, M., Verbeke, G., and Molenberghs, G. (2008).
\newblock {\em Longitudinal data analysis\/}.
\newblock CRC Press.

\bibitem[Fr{\"u}hwirth-Schnatter(2006)Fr{\"u}hwirth-Schnatter]{fruhwirth2006finite}
Fr{\"u}hwirth-Schnatter, S. (2006).
\newblock {\em Finite mixture and {M}arkov switching models\/}.
\newblock Springer Science \& Business Media.

\bibitem[Gelfand {\em et~al.}(2005)Gelfand, Kottas, and
  MacEachern]{gelfand2005bayesian}
Gelfand, A.~E., Kottas, A., and MacEachern, S.~N. (2005).
\newblock Bayesian nonparametric spatial modeling with {D}irichlet process
  mixing.
\newblock {\em Journal of the American Statistical Association\/}, {\bf 100},
  1021--1035.

\bibitem[Gelman(2006)Gelman]{gelman2006prior}
Gelman, A. (2006).
\newblock Prior distributions for variance parameters in hierarchical models.
\newblock {\em Bayesian Analysis\/}, {\bf 1}, 515--534.

\bibitem[Ghahramani and Jordan(1997)Ghahramani and
  Jordan]{ghahramani1996factorial}
Ghahramani, Z. and Jordan, M.~I. (1997).
\newblock Factorial hidden {M}arkov models.
\newblock {\em Machine Learning\/}, {\bf 29}, 245--273.

\bibitem[Ghosal {\em et~al.}(1999)Ghosal, Ghosh, and
  Ramamoorthi]{ghosal1999posterior}
Ghosal, S., Ghosh, J.~K., and Ramamoorthi, R. (1999).
\newblock Posterior consistency of {D}irichlet mixtures in density estimation.
\newblock {\em Annals of Statistics\/}, {\bf 27}, 143--158.

\bibitem[Ghosh and Ramamoorthi(2003)Ghosh and Ramamoorthi]{ghosh2003}
Ghosh, J. and Ramamoorthi, R. (2003).
\newblock {\em Bayesian Nonparametrics\/}.
\newblock Springer.

\bibitem[Gramacy {\em et~al.}(2013)Gramacy, Taddy, and
  Wild]{gramacy2013variable}
Gramacy, R.~B., Taddy, M., and Wild, S.~M. (2013).
\newblock Variable selection and sensitivity analysis using dynamic trees, with
  an application to computer code performance tuning.
\newblock {\em Annals of Applied Statistics\/}, {\bf 7}, 51--80.

\bibitem[Guhaniyogi {\em et~al.}(2017)Guhaniyogi, Qamar, and
  Dunson]{guhaniyogi2017bayesian}
Guhaniyogi, R., Qamar, S., and Dunson, D.~B. (2017).
\newblock Bayesian tensor regression.
\newblock {\em The Journal of Machine Learning Research\/}, {\bf 18},
  2733--2763.

\bibitem[Guo(2002)Guo]{guo2002functional}
Guo, W. (2002).
\newblock Functional mixed effects models.
\newblock {\em Biometrics\/}, {\bf 58}, 121--128.

\bibitem[Hartigan(1990)Hartigan]{hartigan1990partition}
Hartigan, J.~A. (1990).
\newblock Partition models.
\newblock {\em Communications in Statistics - Theory and Methods\/}, {\bf 19},
  2745--2756.

\bibitem[Hastie and Tibshirani(1993)Hastie and Tibshirani]{hastie1993varying}
Hastie, T. and Tibshirani, R. (1993).
\newblock Varying-coefficient models.
\newblock {\em Journal of the Royal Statistical Society: Series B\/}, {\bf 55},
  757--796.

\bibitem[Hoover {\em et~al.}(1998)Hoover, Rice, Wu, and
  Yang]{hoover1998nonparametric}
Hoover, D.~R., Rice, J.~A., Wu, C.~O., and Yang, L.-P. (1998).
\newblock Nonparametric smoothing estimates of time-varying coefficient models
  with longitudinal data.
\newblock {\em Biometrika\/}, {\bf 85}, 809--822.

\bibitem[Kolda and Bader(2009)Kolda and Bader]{kolda2009tensor}
Kolda, T.~G. and Bader, B.~W. (2009).
\newblock Tensor decompositions and applications.
\newblock {\em SIAM review\/}, {\bf 51}, 455--500.

\bibitem[Koslovsky {\em et~al.}(2020)Koslovsky, H{\'e}bert, Businelle,
  Vannucci, {\em et~al.}]{koslovsky2020bayesian}
Koslovsky, M.~D., H{\'e}bert, E.~T., Businelle, M.~S., Vannucci, M., {\em
  et~al.} (2020).
\newblock A {B}ayesian time-varying effect model for behavioral m{H}ealth data.
\newblock {\em Annals of Applied Statistics\/}, {\bf 14}, 1878--1902.

\bibitem[Li {\em et~al.}(2010)Li, Lin, and M{\"u}ller]{li2010bayesian}
Li, Y., Lin, X., and M{\"u}ller, P. (2010).
\newblock Bayesian inference in semiparametric mixed models for longitudinal
  data.
\newblock {\em Biometrics\/}, {\bf 66}, 70--78.

\bibitem[Linero and Yang(2018)Linero and Yang]{linero2018bayesiansmooth}
Linero, A.~R. and Yang, Y. (2018).
\newblock Bayesian regression tree ensembles that adapt to smoothness and
  sparsity.
\newblock {\em Journal of the Royal Statistical Society: Series B\/}, {\bf 80},
  1087--1110.

\bibitem[Little and Rubin(2019)Little and Rubin]{little2019statistical}
Little, R.~J. and Rubin, D.~B. (2019).
\newblock {\em Statistical analysis with missing data\/}.
\newblock John Wiley \& Sons.

\bibitem[MacLehose and Dunson(2009)MacLehose and
  Dunson]{maclehose2009nonparametric}
MacLehose, R.~F. and Dunson, D.~B. (2009).
\newblock Nonparametric {B}ayes kernel-based priors for functional data
  analysis.
\newblock {\em Statistica Sinica\/}, {\bf 19}, 611--629.

\bibitem[Makalic and Schmidt(2016)Makalic and Schmidt]{makalic2016simple}
Makalic, E. and Schmidt, D.~F. (2016).
\newblock A simple sampler for the horseshoe estimator.
\newblock {\em IEEE Signal Processing Letters\/}, {\bf 23}, 179--182.

\bibitem[Morris(2015)Morris]{morris2015functional}
Morris, J.~S. (2015).
\newblock Functional regression.
\newblock {\em Annual Review of Statistics and Its Application\/}, {\bf 2},
  321--359.

\bibitem[Morris and Carroll(2006)Morris and Carroll]{morris2006wavelet}
Morris, J.~S. and Carroll, R.~J. (2006).
\newblock Wavelet-based functional mixed models.
\newblock {\em Journal of the Royal Statistical Society: Series B\/}, {\bf 68},
  179--199.

\bibitem[M{\"u}ller {\em et~al.}(2013)M{\"u}ller, Quintana, Rosner, and
  Maitland]{muller2013bayesian}
M{\"u}ller, P., Quintana, F.~A., Rosner, G.~L., and Maitland, M.~L. (2013).
\newblock Bayesian inference for longitudinal data with non-parametric
  treatment effects.
\newblock {\em Biostatistics\/}, {\bf 15}, 341--352.

\bibitem[Nguyen(2010)Nguyen]{nguyen2010inference}
Nguyen, X. (2010).
\newblock Inference of global clusters from locally distributed data.
\newblock {\em Bayesian Analysis\/}, {\bf 5}, 817--845.

\bibitem[Nguyen and Gelfand(2011)Nguyen and Gelfand]{nguyen2011dirichlet}
Nguyen, X. and Gelfand, A.~E. (2011).
\newblock The {D}irichlet labeling process for clustering functional data.
\newblock {\em Statistica Sinica\/}, {\bf 21}, 1249--1289.

\bibitem[Page {\em et~al.}(2020)Page, Quintana, and Dahl]{page2020dependent}
Page, G.~L., Quintana, F.~A., and Dahl, D.~B. (2020).
\newblock Dependent random partition models.
\newblock {\em arXiv preprint arxiv:1912.11542\/}.

\bibitem[Papadogeorgou {\em et~al.}(2019)Papadogeorgou, Zhang, and
  Dunson]{papadogeorgou2019soft}
Papadogeorgou, G., Zhang, Z., and Dunson, D.~B. (2019).
\newblock Soft tensor regression.
\newblock {\em arXiv preprint arXiv:1910.09699\/}.

\bibitem[Paulon {\em et~al.}(2020)Paulon, Llanos, Chandrasekaran, and
  Sarkar]{paulon2020driftdiff}
Paulon, G., Llanos, F., Chandrasekaran, B., and Sarkar, A. (2020).
\newblock Bayesian semiparametric longitudinal drift-diffusion mixed models for
  tone learning in adults.
\newblock {\em Journal of the American Statistical Association\/}, pages 1--14.

\bibitem[Petrone {\em et~al.}(2009)Petrone, Guindani, and
  Gelfand]{petrone2009hybrid}
Petrone, S., Guindani, M., and Gelfand, A.~E. (2009).
\newblock Hybrid {D}irichlet mixture models for functional data.
\newblock {\em Journal of the Royal Statistical Society: Series B\/}, {\bf 71},
  755--782.

\bibitem[Polson and Scott(2012)Polson and Scott]{polson2012half}
Polson, N.~G. and Scott, J.~G. (2012).
\newblock On the half-{C}auchy prior for a global scale parameter.
\newblock {\em Bayesian Analysis\/}, {\bf 7}, 887--902.

\bibitem[Polson {\em et~al.}(2013)Polson, Scott, and
  Windle]{polson2013bayesian}
Polson, N.~G., Scott, J.~G., and Windle, J. (2013).
\newblock Bayesian inference for logistic models using {P}{\'o}lya--{G}amma
  latent variables.
\newblock {\em Journal of the American Statistical Association\/}, {\bf 108},
  1339--1349.

\bibitem[Quintana and Iglesias(2003)Quintana and
  Iglesias]{quintana2003bayesian}
Quintana, F.~A. and Iglesias, P.~L. (2003).
\newblock Bayesian clustering and product partition models.
\newblock {\em Journal of the Royal Statistical Society: Series B\/}, {\bf 65},
  557--574.

\bibitem[Quintana {\em et~al.}(2016)Quintana, Johnson, Waetjen, and
  B.~Gold]{quintana2016bayesian}
Quintana, F.~A., Johnson, W.~O., Waetjen, L.~E., and B.~Gold, E. (2016).
\newblock Bayesian nonparametric longitudinal data analysis.
\newblock {\em Journal of the American Statistical Association\/}, {\bf 111},
  1168--1181.

\bibitem[Rabiner(1989)Rabiner]{Rabiner:1989}
Rabiner, L. (1989).
\newblock A tutorial on hidden {M}arkov models and selected applications in
  speech recognition.
\newblock {\em \IEEE\/}, {\bf 77}, 257--286.

\bibitem[Ramsay and Silverman(2007)Ramsay and Silverman]{ramsay2007applied}
Ramsay, J.~O. and Silverman, B.~W. (2007).
\newblock {\em Applied functional data analysis: {M}ethods and case studies\/}.
\newblock Springer.

\bibitem[Ruppert(2002)Ruppert]{ruppert2002selecting}
Ruppert, D. (2002).
\newblock Selecting the number of knots for penalized splines.
\newblock {\em Journal of Computational and Graphical Statistics\/}, {\bf 11},
  735--757.

\bibitem[Sarkar and Dunson(2016)Sarkar and Dunson]{sarkar2016bayesian}
Sarkar, A. and Dunson, D.~B. (2016).
\newblock Bayesian nonparametric modeling of higher order {M}arkov chains.
\newblock {\em Journal of the American Statistical Association\/}, {\bf 111},
  1791--1803.

\bibitem[Scott(2002)Scott]{Scott:2002}
Scott, S.~L. (2002).
\newblock Bayesian methods for hidden {M}arkov models recursive computing in
  the 21st century.
\newblock {\em \JASA\/}, {\bf 97}, 337--351.

\bibitem[Singer {\em et~al.}(2003)Singer, Willett, Willett, {\em
  et~al.}]{singer2003applied}
Singer, J.~D., Willett, J.~B., Willett, J.~B., {\em et~al.} (2003).
\newblock {\em Applied longitudinal data analysis: {M}odeling change and event
  occurrence\/}.
\newblock Oxford University Press.

\bibitem[Sonnega {\em et~al.}(2014)Sonnega, Faul, Ofstedal, Langa, Phillips,
  and Weir]{sonnega2014cohort}
Sonnega, A., Faul, J.~D., Ofstedal, M.~B., Langa, K.~M., Phillips, J.~W., and
  Weir, D.~R. (2014).
\newblock Cohort profile: {T}he health and retirement study ({HRS}).
\newblock {\em International Journal of Epidemiology\/}, {\bf 43}, 576--585.

\bibitem[Sparapani {\em et~al.}(2016)Sparapani, Logan, McCulloch, and
  Laud]{sparapani2016nonparametric}
Sparapani, R.~A., Logan, B.~R., McCulloch, R.~E., and Laud, P.~W. (2016).
\newblock Nonparametric survival analysis using {B}ayesian additive regression
  trees ({BART}).
\newblock {\em Statistics in Medicine\/}, {\bf 35}, 2741--2753.

\bibitem[Starling {\em et~al.}(2020)Starling, Murray, Carvalho, Bukowski, and
  Scott]{starling2020bart}
Starling, J.~E., Murray, J.~S., Carvalho, C.~M., Bukowski, R.~K., and Scott,
  J.~G. (2020).
\newblock {BART} with targeted smoothing: {A}n analysis of patient-specific
  stillbirth risk.
\newblock {\em Annals of Applied Statistics\/}, {\bf 14}, 28--50.

\bibitem[Suarez and Ghosal(2016)Suarez and Ghosal]{suarez2016bayesian}
Suarez, A.~J. and Ghosal, S. (2016).
\newblock Bayesian clustering of functional data using local features.
\newblock {\em Bayesian Analysis\/}, {\bf 11}, 71--98.

\bibitem[Taddy {\em et~al.}(2011)Taddy, Gramacy, and Polson]{taddy2011dynamic}
Taddy, M.~A., Gramacy, R.~B., and Polson, N.~G. (2011).
\newblock Dynamic trees for learning and design.
\newblock {\em Journal of the American Statistical Association\/}, {\bf 106},
  109--123.

\bibitem[Titsias and Yau(2014)Titsias and Yau]{titsias2016hamming}
Titsias, M.~K. and Yau, C. (2014).
\newblock Hamming ball auxiliary sampling for factorial hidden {M}arkov models.
\newblock In {\em Advances in Neural Information Processing Systems\/}, pages
  2960--2968.

\bibitem[Tucker(1966)Tucker]{tucker:1966}
Tucker, L. (1966).
\newblock Some mathematical notes on three-mode factor analysis.
\newblock {\em Psychometrica\/}, {\bf 31}, 273--282.

\bibitem[Van~der Vaart(2000)Van~der Vaart]{van2000asymptotic}
Van~der Vaart, A.~W. (2000).
\newblock {\em Asymptotic statistics\/}.
\newblock Cambridge university press.

\bibitem[Wang {\em et~al.}(2016)Wang, Chiou, and
  M{\"u}ller]{wang2016functional}
Wang, J.-L., Chiou, J.-M., and M{\"u}ller, H.-G. (2016).
\newblock Functional data analysis.
\newblock {\em Annual Review of Statistics and Its Application\/}, {\bf 3},
  257--295.

\bibitem[Wilson {\em et~al.}(2015)Wilson, Capuano, Sytsma, Bennett, and
  Barnes]{wilson2015cognitive}
Wilson, R.~S., Capuano, A.~W., Sytsma, J., Bennett, D.~A., and Barnes, L.~L.
  (2015).
\newblock Cognitive aging in older black and white persons.
\newblock {\em Psychology and Aging\/}, {\bf 30}, 279--285.

\bibitem[Zucchini {\em et~al.}(2017)Zucchini, MacDonald, and
  Langrock]{zucchini2017hidden}
Zucchini, W., MacDonald, I.~L., and Langrock, R. (2017).
\newblock {\em Hidden Markov models for time series: {A}n introduction using
  {R}\/}.
\newblock CRC press.

\end{thebibliography}

%%%%%%%%%%%%%%%%%%%%%%%%%
%%% Supplementary Materials %%%
%%%%%%%%%%%%%%%%%%%%%%%%%

\clearpage\pagebreak\newpage
\newgeometry{textheight=9in, textwidth=6.5in,}
\pagestyle{fancy}
\fancyhf{}
\rhead{\bfseries\thepage}
\lhead{\bfseries SUPPLEMENTARY MATERIALS}

\baselineskip 20pt
\begin{center}
{\Large{\bf Bayesian Semiparametric\\ 
\vskip-10pt Hidden Markov Tensor Partition Models\\ 
\vskip-10pt for Longitudinal Data with\\
\vskip 5pt Local Variable Selection}}
\end{center}

\setcounter{equation}{0}
\setcounter{page}{1}
\setcounter{table}{1}
\setcounter{figure}{0}
\setcounter{section}{0}
\numberwithin{table}{section}
\renewcommand{\theequation}{S.\arabic{equation}}
\renewcommand{\thesubsection}{S.\arabic{section}.\arabic{subsection}}
\renewcommand{\thesection}{S.\arabic{section}}
\renewcommand{\thepage}{S.\arabic{page}}
\renewcommand{\thetable}{S.\arabic{table}}
\renewcommand{\thefigure}{S.\arabic{figure}}
\baselineskip=15pt

\vspace{0cm}

\begin{center}
Giorgio Paulon$^{a}$ (giorgio.paulon@utexas.edu)\\
Peter M\"uller$^{a,b}$ (pmueller@math.utexas.edu)\\
Abhra Sarkar$^{a}$ (abhra.sarkar@utexas.edu)\\

\vskip 7mm
$^{a}$Department of Statistics and Data Sciences, \\
The University of Texas at Austin,\\ 2317 Speedway D9800, Austin, TX 78712-1823, USA\\
\vskip 8pt 
$^{b}$Department of Mathematics, \\
The University of Texas at Austin,\\ 2515 Speedway C1200, Austin, TX 78712-1202, USA\\
\end{center}

\vskip 10mm
The supplementary materials present brief reviews of B-splines, fHMMs, and tensor factorization methods for easy reference. 
The supplementary materials also include 
additional discussions on the characterization of overall, main and interaction effects and associated tests; 
proofs of the theoretical results; 
choice of the prior hyper-parameters; 
additional details (software, run-time, etc.) of the MCMC algorithm used to sample from the posterior; 
MCMC diagnostics; 
results of some additional simulation experiments; 
additional real data applications; 
etc. 
\texttt{R} programs implementing the methods developed in this article and an accompanying `readme' file 
are also included as separate files in the supplementary materials.

\baselineskip=15pt
\newpage
\section{Linear B-splines} \label{sec: sm b-splines}
In the main article, we employed linear B-spline bases in the construction of functional factorial HMMs.  
The construction of linear B-spline bases is detailed below \citeplatex{de1978practical}. 
Consider knot points $t_{1} = t_{2} = A < t_{3} < \dots < B = t_{K+2} = t_{K+3}$ that divide $[A, B]$ into $K$ equal subintervals,
where $t_{2:(K+2)}$ are equidistant with $\delta = (t_{3} - t_{2})$.
For $j=2,3,\dots,K$, linear B-splines $b_{1,j}$ are then defined as 
\vskip-5ex
\bse
 b_{1,j}(t) &= \left\{\begin{array}{ll}
        (t - t_{j})/\delta  				& ~~~~\text{if } t_{j} \leq t < t_{j+1},  \\
        (t_{j+2} - t)/\delta     				& ~~~~\text{if } t_{j+1} \leq t < t_{j+2},  \\
        0  							& ~~~~ \text{otherwise}.
        \end{array}\right.
\ese
\vskip-3ex
\noindent The components at the ends are likewise defined as 
\vskip-5ex
\bse
b_{1,1}(t) &=&  \left\{\begin{array}{ll}
        (t_{3} - t)/\delta 		           	& ~~~~~~~~~~~~~~~~~~~~~~~\text{if } t_{2} \leq t < t_{3},  \\
        0  						& ~~~~~~~~~~~~~~~~~~~~~~~ \text{otherwise}.
        \end{array}\right.\\
b_{1,K+1}(t) &=&  \left\{\begin{array}{ll}
        (t - t_{K+2})/\delta 		  		& ~~~~~~~~~~~~~~~~~~~\text{if } t_{K+1} \leq t < t_{K+2},  \\
        0  					& ~~~~~~~~~~~~~~~~~~~ \text{otherwise}.
        \end{array}\right.
\ese
\vskip-3ex
\noindent Figure \ref{fig: b-splines} shows the B-spline bases used in the article.

\section{Factorial HMM (fHMM)} \label{sec: sm fHMM}

The basic HMM \citeplatex[][etc.]{fruhwirth2006finite,mcdonald_zucchini:1997}
consists of two processes: an \emph{observed} process $\{\by_{t}\}$ recorded sequentially over a set of discrete time points $t=1,2,\dots,T$
and an associated \emph{hidden} process $\{z_{t}\}$ which evolves according to a first order Markov chain with discrete state space.
Specifically, an HMM makes the following set of conditional independence assumptions to model the hidden and the observed processes
\vskip-5ex
\bse
& p(z_{t} \mid \bz_{1:(t-1)})  = p(z_{t}\mid z_{t-1}),	\\
& p(\by_{t} \mid \by_{1:(t-1)},\bz_{1:t}) = p(y_{t} \mid z_{t}).
\ese
\vskip-1.5ex
The distributions $p(z_{t} \mid z_{t-1})$ and $p(y_{t} \mid z_{t})$ are often referred to as the \emph{transition distribution} and the \emph{emission distribution}, respectively. 

In factorial HMMs \citeplatex{ghahramani1996factorial}, the latent states are represented by a collection of variables $\{\bz_{t}\} = \{(z_{t}^{(1)},\dots,z_{t}^{(m)})\}$ where each component $\{z_{t}^{(\ell)}\}$ now evolves according to a first order Markov chain with discrete state spaces, 
and the \emph{observed} process $\{y_{t}\}$ is observed sequentially as before over a set of discrete time points $t=1,2,\dots,T$.
An fHMM thus makes the following set of conditional independence assumptions to model the hidden and the observed processes
\vskip-5ex
\bse
& \textstyle p(\bz_{t} \mid \bz_{1:(t-1)})  = \prod_{\ell=1}^{m} p(z_{t}^{(\ell)} \mid z_{t-1}^{(\ell)}),	\\
& p(y_{t} \mid \by_{1:(t-1)},\bz_{1:t}) = p(y_{t} \mid \bz_{t}) = p(y_{t} \mid z_{t}^{(1)},\dots,z_{t}^{(m)}).
\ese

\vspace*{-0.5cm}
\begin{figure}[ht!]
	\centering
	\hspace*{-0.00cm}\includegraphics[height=2.85cm,width=6.75cm, trim=2cm 1.25cm 1cm 1.25cm]{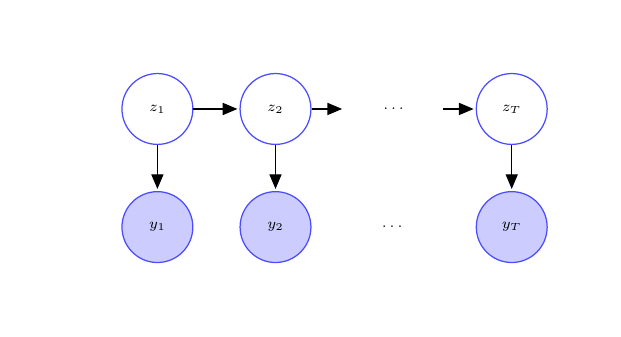} 
	\hspace*{-0.40cm}\includegraphics[width=8.5cm, trim=1cm 1.25cm 1cm 1.25cm]{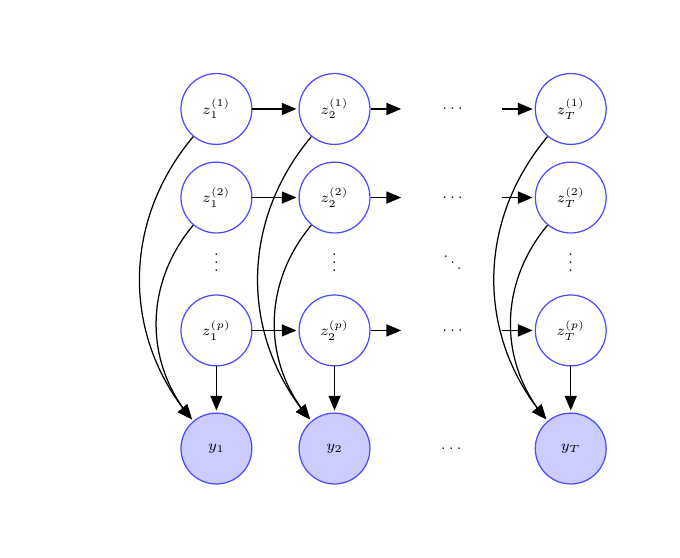}
	\caption{\baselineskip=10pt 
	Left panel: DAG of an HMM. 
	Right panel: DAG of an fHMM with $p$ layers. 
	}
	\label{fig: dag HMM and fHMM}
\end{figure}

In our work, we adapted the basic fHMM to characterize local influences of categorical predictors in longitudinal functional models.  
For each categorical predictor $x_{j} \in \{1,\dots,x_{j,\max}\}$, 
we introduced an fHMM $\{\bz_{j,t}=(z_{j,t}^{(1)},\dots,z_{j,t}^{(x_{j,\max})})\}$ with $x_{j,\max}$ layers, one for each level of $x_{j}$. 
Conditional on  $(z_{1,t}^{(x_{1})},\dots,z_{p,t}^{(x_{p})}) = (z_{1,t},\dots,z_{p,t})$, 
we then associated the coefficients $\beta_{t,x_{1},\dots,x_{p}}$ of a predictor dependent B-spline mixture model with atoms $\beta_{t,z_{1,t},\dots,z_{p,t}}^{\star}$. 
Specifically, we let
\vskip-5ex
\bse
& \textstyle p(\bz_{t} \mid \bz_{1:(t-1)})  = \prod_{j=1}^{p} p(\bz_{j,t} \mid \bz_{j,t-1}) = \prod_{j=1}^{p} \prod_{\ell=1}^{m} p(z_{j,t}^{(\ell)} \mid z_{j,t-1}^{(\ell)}),	\\
& \{\beta_{t,x_{1},\dots,x_{p}} \mid z_{j,t}^{(x_{j})}=z_{j,t}, j=1,\dots,p\} = \beta_{t,z_{1,t},\dots,z_{p,t}}^{\star}.
\ese

\section{Tensor Factorization Methods} \label{sec: sm tensor factorization}

In this section, we provide a brief review of the different main types of tensor factorizations \citeplatex{hitchcock1927expression,tucker:1966, de_lathauwer_etal:2000,kolda2009tensor}. 

A $d_{1} \times \dots \times d_{p}$ dimensional tensor $\bbeta = \{\beta_{h_{1},\dots,h_{p}}: h_{j}=1,\dots,d_{j}, j=1,\dots,p\}$ 
admits a parallel factor (PARAFAC) decomposition 
with rank $r$ (Figure \ref{fig: PARAFAC}) 
if it can be written as 
\vskip-5ex
\be
& \textstyle \beta_{h_{1},\dots,h_{p}} = \sum_{z=1}^{r} \prod_{j=1}^{p} a_{j,z}^{(h_{j})}~~~\text{for each}~(h_{1},\dots,h_{p}), \label{eq: sm parafac}
\ee
\vskip-2ex
\noindent where $\ba_{j,z} = \{a_{j,z}^{(h_{j})}: h_{j}=1,\dots,d_{j}\}, z=1,\dots, r, j=1,\dots,p$ are $d_{j} \times 1$ dimensional vectors. 

\begin{figure}[ht!]
	\centering
	\hspace*{-0.25cm}\includegraphics[width=0.6\linewidth, trim=1cm 1.25cm 1cm 1cm]{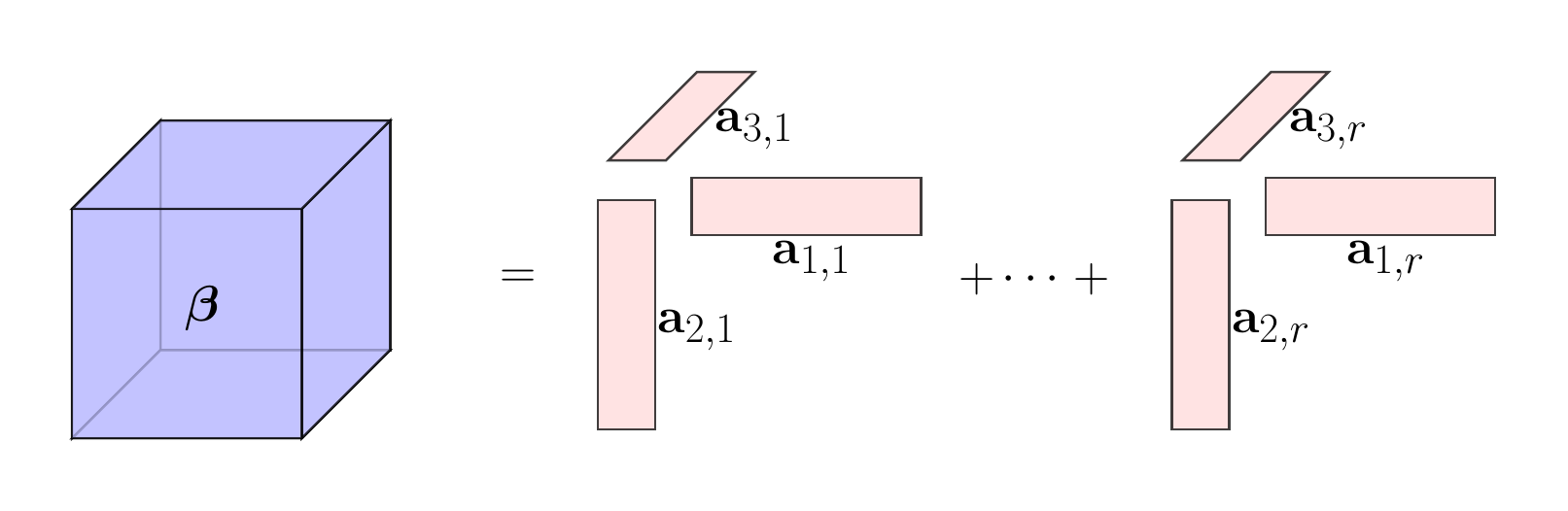}
	\caption{\baselineskip=10pt 
	Pictorial representation of PARAFAC of a three dimensional tensor. 
	}
	\label{fig: PARAFAC}
\end{figure}

A $d_{1} \times \dots \times d_{p}$ dimensional tensor $\bbeta = \{\beta_{h_{1},\dots,h_{p}}: h_{j}=1,\dots,d_{j}, j=1,\dots,p\}$ 
admits a Tucker decomposition 
with multi-linear rank $(r_{1}, \dots,r_{p})$ (Figure \ref{fig: HOSVD}) 
if it admits a representation 
\vskip-5ex
\be
& \textstyle \beta_{h_{1},\dots,h_{p}} = \sum_{z_{1}=1}^{r_{1}} \cdots\sum_{z_{p}=1}^{r_{p}} \beta_{z_{1},\dots,z_{p}}^{\star} \prod_{j=1}^{p} a_{j,z_{j}}^{(h_{j})}~~~\text{for each}~(h_{1},\dots,h_{p}), \label{eq: sm Tucker}
\ee
\vskip-2ex
\noindent where $\bbeta^{\star} = \{\beta_{z_{1},\dots,z_{p}}^{\star}: z_{j}=1,\dots,r_{j}, j=1,\dots,p\}$ is an $r_{1} \times \dots \times r_{p}$ dimensional `core tensor' 
with $1 \leq r_{j} \leq d_{j}$ for each $j$, 
and $\bA_{j} = \{a_{j,z_{j}}^{(h_{j})}: h_{j}=1,\dots,d_{j}, z_{j}=1,\dots,r_{j}\}, j=1,\dots,p$ are $d_{j} \times r_{j}$ dimensional `mode matrices' or `factor matrices' with full column rank $r_{j}$. 
The effective size of the model after the factorization is $\prod_{j=1}^{p} r_{j} + \sum_{j=1}^{p}r_{j}d_{j} \approx \prod_{j=1}^{p} r_{j}$. 
A significant reduction in dimensions is therefore achieved by the decomposition when $\prod_{j=1}^{p} r_{j} \ll \prod_{j=1}^{p} d_{j}$, 
that is, the size of the core tensor is much smaller than the size of the original tensor. 

\begin{figure}[ht!]
	\centering
	\hspace*{-0.25cm}\includegraphics[width=0.5\linewidth, trim=1cm 1.25cm 1cm 1cm]{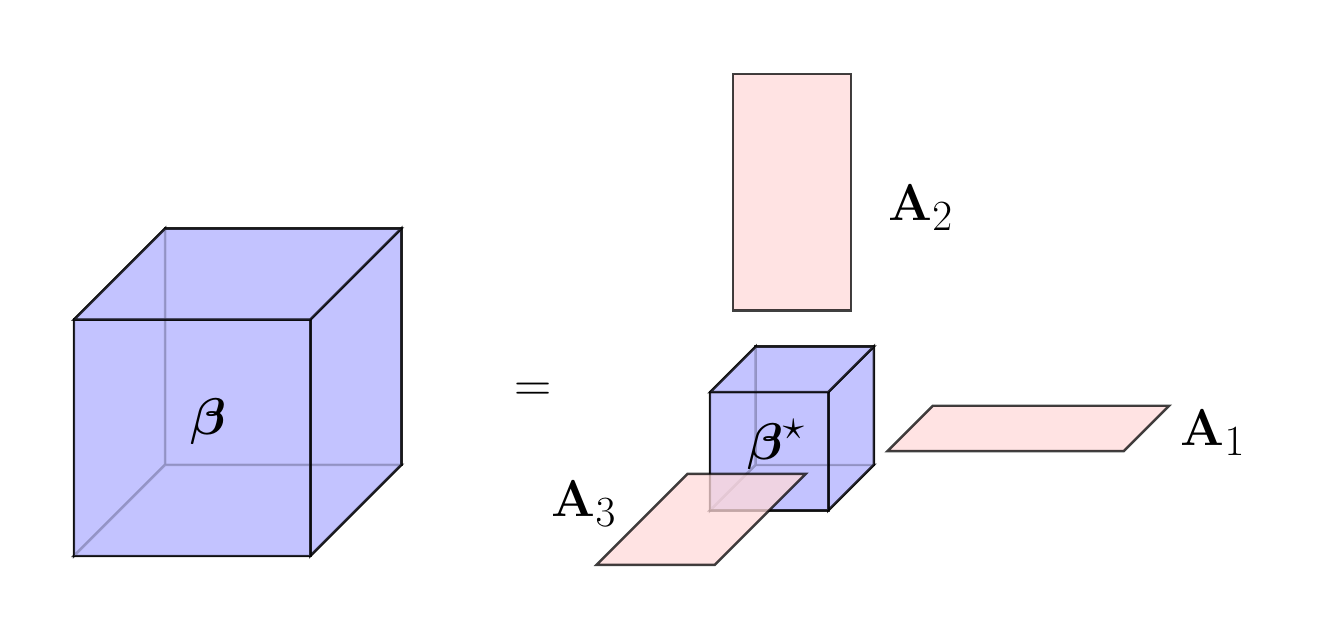}
	\caption{\baselineskip=10pt 
	Pictorial representation of HOSVD of a three dimensional tensor. 
	}
	\label{fig: HOSVD}
\end{figure}

The PARAFAC representation is obtained as a special case of the Tucker decomposition 
with $r_{1}=\dots=r_{p}=r$ and $\beta_{h_{1},\dots,h_{p}} = 1\{h_{1} = \dots = h_{p}\}$. 
Compared to the PARAFAC, the Tucker decomposition thus typically achieves a much greater reduction in the dimension of a tensor.  

The compact higher order singular value decomposition (compact HOSVD) of a tensor 
is a special case of the Tucker decomposition, 
where the mode matrices $\bA_{j}$'s are restricted to be semi-orthogonal, 
that is, they satisfy $\bA_{j}\trans \bA_{j} = \bI_{r_{j}}$ for all $j$. 

While none of these representations are fully identifiable, 
the compact HOSVD results in an equally flexible but much more interpretable form of the Tucker decomposition.

\vskip 15pt
In formulating our model for the fixed effects in Section \ref{sec: fixed effects} of the main paper, we structured the parameters for different predictor combinations 
as a $x_{1,\max} \times \dots \times x_{p,\max}$ dimensional tensor $\bbeta_{k} = \{\beta_{k, x_{1},\dots,x_{p}}: (x_{1},\dots,x_{p}) \in \X \}$ 
for different predictor combinations at each location $k$ 
and then applied a compact HOSVD-type factorization 
(Figure \ref{fig: dynamic HOSVD} in the main paper) as 
\vskip-5ex
\bse
& \textstyle \{\beta_{k, x_{1},\dots,x_{p}} \mid z_{j,k}^{(x_{j})}, j=1,\dots,p \} = \sum_{z_{1,k}} \cdots\sum_{z_{p,k}} \beta_{k,z_{1,k},\dots,z_{p,k}}^{\star} \prod_{j=1}^{p} 1\{z_{j,k}^{(x_{j})}=z_{j,k}\}, \label{eq: sm TFM2}
\ese
\vskip-2ex
\noindent where $\bbeta_{k}^{\star} = \{\beta_{k, z_{1,k},\dots,z_{p,k}}^{\star}: (z_{1,k},\dots,z_{p,k}) \in \Z_{k}\}$ is a $\ell_{1,k} \times \dots \times \ell_{p,k}$ dimensional core tensor 
and $\bz_{j,k}=\{{{\scriptstyle 1\{z_{j,k}^{(x_{j})}=z_{j,k}\} }}: x_{j} \in \X_{j}, z_{j,k} \in \Z_{j,k}\}$ are $x_{j,\max} \times \ell_{j,k}$ dimensional allocation matrices with binary entries. 
This is a compact HOSVD-type factorization since $\bz_{j,k}\trans\bz_{j,k}$ are diagonal matrices for all $j=1,\dots,p, k=1,\dots,K$.

\newpage
\section{Main and Interaction Effects} \label{sec: sm main and interaction effects}

As discussed in Section \ref{sec: lfmm} in the main paper, 
our proposed HOSVD based model for multiple predictor fixed effects 
achieves excellent dimension reduction properties 
by efficiently eliminating the redundant predictors while also characterizing the important predictors' joint influences implicitly but very compactly. 
It encodes each $x_{j}$'s overall significance explicitly
but does provide explicit description of the predictors' main and lower-dimensional interaction effects \citeplatex{johndrow2017tensor} 
which are often easy to interpret and hence appealing and useful to practitioners. 
This limitation can be sidestepped, however, 
by noting that these effects can be meaningfully \emph{defined} (and easily estimated from the posterior samples) directly. 
With some repetition from the main paper for easy reference, we have   
\vskip-5ex
\bse
& \text{overall mean:~} f_{0}(t) = \frac{\sum_{\bx}f_{x_{1},\dots,x_{p}}(t)}  {\abs{\X}}, ~~~\text{main effects:~} f_{x_{j}}(t) = \frac{\sum_{\bx_{-j}}f_{x_{1},\dots,x_{p}}(t)}  {\abs{\X_{-j}}} - {f}_{0}(t), \\
& \text{interactions:~}  f_{x_{j_{1}},x_{j_{2}}}(t) = \frac{\sum_{\bx_{-j_{1},-j_{2}}}f_{x_{1},\dots,x_{p}}(t)}  {\abs{\X_{-j_{1},-j_{2}}}} - {f}_{x_{j_{1}}}(t) - {f}_{x_{j_{2}}}(t) - {f}_{0}(t), ~\text{etc., where} \\
&\bx_{-j}=(x_{1},\dots,x_{j-1},x_{j+1},\dots, x_{p})\trans \in \X_{-j},\\ 
&\bx_{-j_{1},-j_{2}}=(x_{1},\dots,x_{j_{1}-1},x_{j_{1}+1},\dots, x_{j_{2}-1},x_{j_{2}+1},\dots,x_{p})\trans \in \X_{-j_{1},-j_{2}},~\text{and so on.}
\ese
\vskip-2ex

\begin{figure}[!h]
	\centering
	\begin{center}
		\includegraphics[width=0.85\linewidth,trim=0.1cm 0.1cm 0.1cm 0.1cm, clip=true]{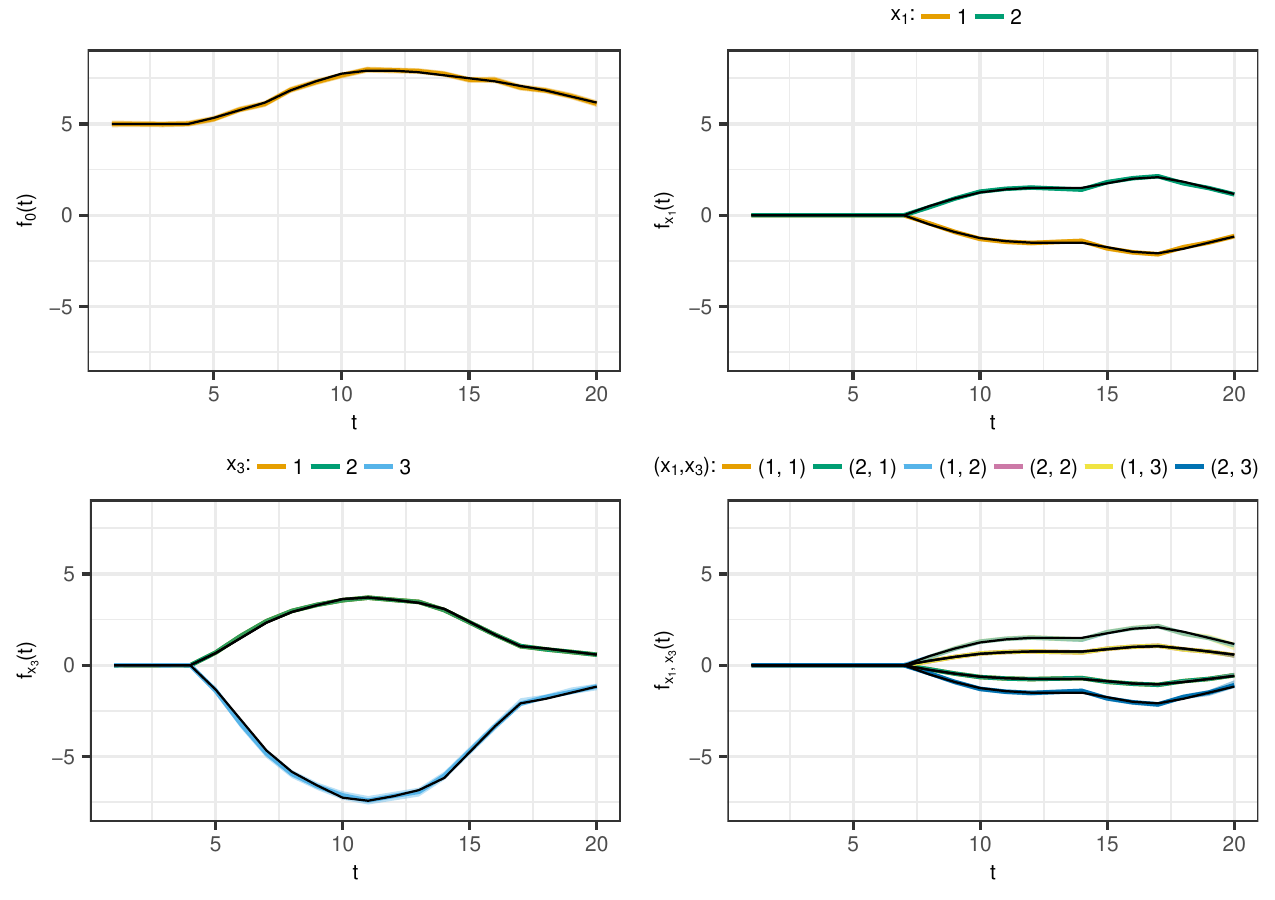}
	\end{center}
	\vskip-15pt
	\caption{Results for synthetic data: Scenario with 
	covariates $(x_{1},\dots,x_{10})$ with significant predictors $(x_{1}, x_{3})$ described in Section \ref{sec: sim studies}. 
	Showing their true effects (black lines) and their estimated posterior means (colored lines). 
	Clockwise from top left: overall mean; main effects of $x_{1}$; main effects of $x_{3}$; and interaction effects of $(x_{1},x_{3})$.
	The figure here corresponds to the synthetic data set that produced the median root mean squared error. 	
	}
	\label{fig: main and interactions}
\end{figure}
\setlength{\textfloatsep}{10pt}

For the simulation scenario described in Section \ref{sec: sim studies}, for instance, 
the true main and interaction effects for various levels and level combinations of $x_{1}$ and $x_{3}$, 
and the corresponding estimated posterior means and $95\%$ point wise credible intervals are shown in Figure \ref{fig: main and interactions}.
See also Figure \ref{fig: youth effects} in Section \ref{sec: sm youth data} in the supplementary materials. 

Our proposed HOSVD based methodology provides a straightforward way to test for the presence of local overall effects of different predictors $x_{j}$ 
using their marginal posterior inclusion probabilities (Figure \ref{fig: eye} in the main paper),
with consistency proven in Theorem \ref{thm: variable selection}.

When a predictor is found important overall, 
interest may additionally lie in testing the significance of its main and lower order interaction effects. 
We do not pursue the problem in more detail in this article 
but provide a general recipe for doing this by testing their pairwise differences here. 
For the main effects of predictor $x_{j} \in \X_{j} = \{1,\dots,x_{j,\max}\}$, for example, 
we may consider, at each time point $t$, ${x_{j,\max}}\choose{2}$ pairwise tests of the type  
\vskip-5ex
\bse
	H_{0,j,\ell_{1},\ell_{2}}(t): \abs{f_{x_{j_{\ell_{1}}}}(t) - f_{x_{j_{\ell_{2}}}}(t)} \leq \Delta_{j}(t) ~~~ \text{ vs. } ~~~ H_{1,j,\ell_{1},\ell_{2}}(t): \abs{f_{x_{j_{\ell_{1}}}}(t) - f_{x_{j_{\ell_{2}}}}(t)}> \Delta_{j}(t), 
\ese
\vskip-2ex
\noindent where we have followed \citelatex[Chapter 4, pp. 148]{berger_book} in replacing conventional point nulls 
by reasonable interval nulls. 
If $H_{0,j,\ell_{1},\ell_{2}}(t)$ is rejected in favor of $H_{1,j,\ell_{1},\ell_{2}}(t)$ for some $\ell_{1} \neq \ell_{2}$, 
we can conclude that the main effects of $x_{j}$ are significant at time $t$. 
The interaction effects can be similarly tested.

\clearpage\newpage
\section{Proofs of Theoretical Results}\label{sec: sm proofs}

\subsection{Proof of Theorem 1}

For notational convenience, we let $f_{k,\bx,0} = f_{\bx,0}(k)$ be denoted, without loss of generality, by $\beta_{k,\bx,0}$. 
As our focus is on the parameters $f$ of a conditional probability distribution of the type $p(y \vert f,\bx)$, 
we can fix the marginal distribution $g(\bx)$ of $\bx$ at its true value, say $g_{0}(\bx)$, 
and model the unknown conditional distribution $p(y \vert f,\bx)$ independently of $g(\bx)$.
We can simply restrict ourselves 
to the set of joint distributions such that $p(y,\bx \vert f) = p(y \vert f, \bx) g_{0}(\bx)$.
With some abuse of notation, we have thus not distinguished between $g(\bx)$ and $g_{0}(\bx)$, 
nor between the joint distribution of $(y,\bx \vert f)$ 
and the conditional distribution of $(y \vert f,\bx)$ 
but denote them both by $p_{f}$.

We start by proving that the true data generating density $p_{f_{0}}$ is in the Kullback-Leibler support of the prior $\Pi$, i.e., $\Pi \left(d_{KL}(p_{f}, p_{f_{0}}) < \epsilon \right) > 0 \ \forall \epsilon > 0$. 
The KL support property shows the theoretical flexibility of the proposed model in encompassing a large class of true data generating scenarios. 
We note that
\vskip-5ex
\bse
		&& \textstyle  \Pi \left(\norm{f - f_{0}}^{2}_{2,g, loc} < \epsilon^{2} \right) 
		= \Pi \left( \sum_{\bx \in \X} g(\bx)\sum_{k=1}^{K} |\beta_{k,\bx} - \beta_{k,\bx,0}|^{2} < \epsilon^{2} \right)
		\\
		&& \geq \Pi \left\{|\beta_{k,\bx} - \beta_{k,\bx,0}|^{2} < \epsilon^{2}/(K\abs{\X}\delta_{g,\max}), \ \forall  \bx, \forall k \right\}
		\\
		&& \geq \Pi \left\{|\beta_{k,z}^{\star \star} - \beta_{k,z,0}^{\star \star}|^{2} < \epsilon^{2}/(K\abs{\X}\delta_{g,\max}), \ \forall  \bx, \forall k \mid A \right\} \Pi(A),
\ese		
\vskip-2ex
\noindent where $\delta_{g,\max} = \max_{\bx}g(\bx)$, 
the event $A = \{z_{j,k}^{(x_{j})} = x_{j} \ \forall j,k; z_{k}^{(x_{1}, \dots, x_{p})} = z; \ m_{k} = \ell_{k} \}$ denotes a special case when no clustering occurs at any time point, 
or, in other words, the different possible level combinations of $\bx$ all form their own separate clusters at all time points. 
By the construction of our partition model, the event $A$ has a positive prior probability.
It is possible to explicitly calculate the prior on the spline coefficients, conditional on the event $A$. 
Specifically, we get 
\vskip-5ex
\bse
\left(\begin{array}{c}
\beta_{1,\bx}^{\star \star} \\ 
\beta_{2,\bx}^{\star \star} \\ 
\vdots \\ 
\beta_{K-1,\bx}^{\star \star} \\ 
\beta_{K,\bx}^{\star \star}
\end{array}  \right)
\sim \MVN_{K} \left\{
\left( \begin{array}{c}
\mu_{0} \\ 
\mu_{0} \\ 
\vdots \\ 
\mu_{0} \\ 
\mu_{0}
\end{array} \right); \left(
\begin{array}{ccccc}
\sigma_{\beta}^{-2} + \sigma_{0}^{-2} & -\sigma_{\beta}^{-2} & &  &  \\ 
-\sigma_{\beta}^{-2} & 2 \sigma_{\beta}^{-2} & -\sigma_{\beta}^{-2} &  &  \\ 
 & -\sigma_{\beta}^{-2} & \ddots & \ddots &  \\ 
 &  & \ddots & \ddots & -\sigma_{\beta}^{-2} \\ 
 &  & & -\sigma_{\beta}^{-2} & \sigma_{\beta}^{-2}
\end{array} \right)^{-1}
\right\}.
\ese
\vskip-2ex
\noindent The above result follows by assuming that $\beta_{1,\bx}^{\star \star} \sim \Normal(\mu_{0}, \sigma_{0}^{2})$, 
which in the limit $\sigma_{0}^{2} \rightarrow \infty$ approximates the setting of the article.
When $\sigma_{0}^{2}$ is finite, the precision matrix of the prior on the spline coefficients is symmetric positive definite. 
Since the joint distribution of the spline coefficients has full support on $\mathbb{R}^{K}$, it follows that $\Pi \left(\norm{f - f_{0}}_{2,g,loc} < \epsilon \right) > 0$.
This shows the positivity of any Kullback-Leibler neighborhood 
since in the case of a Gaussian likelihood $d_{KL}(p_{f}, p_{f_{0}}) = \norm{f - f_{0}}_{2,g,loc}^{2} / (2 \sigma_{n}^{2})$.

To establish strong consistency for the posterior distribution of $f(t)$, we apply Theorem 2 of \citeplatex{ghosal1999posterior} stated below for easy reference.

\begin{Thm}[Ghosal et al., 1999]
Let $\Pi$ be a prior on $\F$. 
Suppose $p_{f_{0}} \in \F$ is in the KL support of $\Pi$ and let $U = \{p_{f}: \norm{p_{f} - p_{f_{0}}} < \epsilon \}$, 
where $\norm{\cdot}$ is the $L_{1}$-norm for the densities. 
If there is a $\delta < \epsilon/4, c_{1}, c_{2} > 0, \alpha < \epsilon^{2}/8$ and $\F_{n} \subset \F$ such that, for all $n$ large:
\begin{enumerate}[label=(\roman*)]
\item $\Pi(\F_{n}^{c}) < c_{1} \exp(-n c_{2})$, and
\item $J(\delta,\F_{n}, \norm{\cdot}) < n \alpha$,
\end{enumerate}
then $\Pi(U \mid \text{data}) \rightarrow 1$.
\end{Thm}

For $\delta > 0$, the metric entropy $J(\delta, \F_{n}, \norm{\cdot})$ is the logarithm of the minimum of all $k$ such that there exist $p_{1}, p_{2}, \dots , p_{k} \in \F$ with the property $\F_{n} \subset \cup_{i=1}^{k} \{p : \norm{p - p_{i}} < \delta \}$.
We can construct a sieve in the parameter space
$\H_{n} = \{\btheta: ||\bbeta^{\star \star}||_{\infty} < M_{1n}, m_{2n} < \sigma_{n}^{2} < M_{2n} \}$ and $\F_{n} = \{ p_{f} : \btheta \in \H_{n} \}$.

We have already verified the KL support condition.
Similarly, we can use results for the Gaussian likelihood to bound the $L_{1}$ distance as 
$\norm{p_{f} - p_{f_{0}}} \leq C_{1} \frac{\norm{f - f_{0}}_{\infty}}{\sigma_{n}}$ for some constant $C_{1}$.
The logarithm of the minimum number of brackets of size $\delta$ required to cover $\F_{n}$ is bounded as 
\vskip-5ex
\bse
    \begin{split}
    J(\delta, \F_{n}, \norm{\cdot}) &< J(m_{2n} \delta / C_{1},\{\bbeta^{\star \star}, \sigma_{n}^{2}: \norm{\bbeta^{\star \star}}_{\infty} < M_{1n}, m_{2n} < \sigma_{n}^{2} < M_{2n} \}, \norm{\cdot}_{\infty})
    \\
    & < K \log \{3C_{1} K M_{1n} / (\delta m_{2n})\}.
    \end{split} 
\ese
\vskip-3ex
\noindent We need to analyze the tail behavior of $\sigma_{n}^{2}(t) = \sigma_{\varepsilon}^{2} + \sigma_{u}^{2}(t)$ to find a bound for the sieve complement.
Using Cramer's rule to calculate the inverse matrix, 
it is easy to see that $\sigma_{u}^{2}(t) = \{(\sigma_{u,a}^{-2}\bI_{K}+\sigma_{u,s}^{-2}\bP_{u})^{-1}\}_{t,t} = O(\sigma_{u,a}^{2})$.
Thus, $\sigma_{n}^{2} = O(\sigma_{\varepsilon}^{2}) + O(\sigma_{u,a}^{2})$.
Assuming exponentially decaying tails for the prior on the variances, the prior probability of the sieve-complement can then be bounded as
\vskip-5ex
\be
    \begin{split}
    \Pi(\H_{n}^{c}) & < \Pi(\bbeta^{\star \star} \notin [-M_{1n}, M_{1n}]^{K} ) + \Pi(\sigma_{n}^{2} \notin [m_{2n}, M_{2n}]) 
    \\
    & < K \exp(-R_{1} M_{1n}^{t_{1}}) + \exp(-R_{2} M_{2n}^{t_{2}})
    \end{split} \label{eq: sieve 1}
\ee
\vskip-3ex
\noindent for some constants $R_{1}, t_{1}, R_{2}, t_{2}$ and some sequences $M_{1n}, m_{2n}, M_{2n}$. 
In order to apply Theorem 2, we then need, for $\delta < \epsilon/4$ and $\alpha < \epsilon^{2}/8$, that  
\vskip-6ex
\be
    \begin{split}
    &K ~ \log \{3C_{1} K M_{1n} / (\delta m_{2n})\} < n \alpha ~~~\text{and}
    \\
    &K \exp(-R_{1} M_{1n}^{t_{1}}) + \exp(-R_{2} M_{2n}^{t_{2}}) < c_{1} \exp(-n c_{2}).
    \end{split} \label{eq: sieve 2}
\ee
\vskip-2ex
\noindent Conditions \ref{eq: sieve 1} and \ref{eq: sieve 2} are satisfied by choosing $M_{1n}$ to be a positive polynomial of $n$, 
$m_{2n}$ be a negative polynomial of $n$, 
and $M_{2n}$ be an exponential function of $n$, 
depending appropriately on the constants in these equations.
Hence, we have $\Pi(||p_{f} - p_{f_{0}}|| < \epsilon \mid \text{data}) \rightarrow 1$.
We conclude that the posterior distribution is consistent relative to the $L_{1}$ distance. 
Finally, since $\norm{f - f_{0}}_{2,g,loc} \lesssim \norm{p_{f} - p_{f_{0}}}$ we get $\Pi(\norm{f - f_{0}}_{2,g,loc} < \epsilon \mid \text{data}) \rightarrow 1$.

The property of exponentially decaying tails holds for common classes of priors such as the inverse gamma but not for the half-Cauchy used in our analysis.
Form a practical point-of-view, half-Cauchy is indeed preferable to inverse-Gamma \citeplatex{gelman2006prior, polson2012half} 
since, while the former assigns significant probability mass in a neighborhood of zero which corresponds to smooth curves in our models, the latter assigns a vanishing mass there. 
To satisfy both the practical and the proof requirements, one could easily specify a mixture prior 
combining a half-Cauchy truncated to an interval, say, $[0,C]$ and an inverse-Gamma restricted to $[C,\infty)$, 
thereby mimicking the behavior of the half-Cauchy around zero and that of the inverse gamma in the tails. 
This would come at the cost of a slightly more complicated MCMC step for the variance parameters. 
In small scale simulations not reported here, the performance of such methods remained practically indistinguishable from our the half-Cauchy based approach.

\subsection{Proof of Theorem 2}

We define a compatible model be a collection of all parameter values corresponding to a partition which is finer than $\rho_{k,0}$.
An incompatible model is then any collection of parameters that does not result in a compatible model.

Theorem \ref{thm: consistency} implies that the posterior probability for any neighborhood of the true value $\beta_{k,\bx,0}$ for the spline coefficients converges to $1$.
For incompatible models, there exists an incorrect assignment for a pair $(\bx, \bx^{\prime})$, i.e., $\beta_{k,\bx} = \beta_{k,\bx^{\prime}}$ when $\beta_{k,\bx,0} \neq \beta_{k,\bx^{\prime},0}$.
Therefore, is it possible to find a neighborhood of $\beta_{k,\bx,0}$ that does not contain $\beta_{k,\bx}$, which contradicts Theorem 1.

Hence, we can focus on the set of compatible models only. 
Let $\rho_{k} = \{S_{k,1}, \dots, S_{k,m_{k}}\}$ be a compatible model and $\rho_{k,0} = \{S_{k,1,0}, \dots, S_{k,m_{k,0},0}\}$ be the true model. 
Since $\rho_{k}$ is a finer partition, we can assume without loss of generality that $m_{k} > m_{k,0}$ and that $S_{k,h} \subseteq S_{k,h,0}$ for $h = 1, \dots, m_{k,0}$.
In Section \ref{sec: post inference}, we calculated the marginal likelihood of the model as 
\vskip-5ex
\bse
	p(\by_{k} \mid \rho_{k}, \bzeta) &=& \prod_{h=1}^{m_{k}} (2 \pi \sigma_{n}^{2})^{-\frac{n_{k,h}}{2}} (\sigma_{k,h}^{2})^{-\frac{1}{2}} (\sigma_{k,h}^{\star 2})^{\frac{1}{2}} e^{-\frac{1}{2} \left( \frac{\sum_{i,\ell} y_{i,\ell,k}^{2}}{\sigma_{n}^{2}} + \frac{\mu_{k,h}^{2}}{\sigma_{k,h}^{2}}  - \frac{\mu_{k,h}^{\star 2}}{\sigma_{k,h}^{\star 2}}\right)}
	\\
	&=& \prod_{h=1}^{m_{k}} (2 \pi \sigma_{n}^{2})^{-\frac{n_{k,h}}{2}} \left(1 + n_{k,h} \sigma_{n}^{- 2} \sigma_{k,h}^{2} \right)^{-\frac{1}{2}} e^{-\frac{1}{2} \left( \frac{\sum_{i,\ell} y_{i,\ell,k}^{2}}{\sigma_{n}^{2}} + \frac{\mu_{k,h}^{2}}{\sigma_{k,h}^{2}}  - \frac{\mu_{k,h}^{\star 2}}{\sigma_{k,h}^{\star 2}}\right)}.
\ese
\vskip-3ex
Now, 
\vskip-5ex
\bse
	&& \frac{p(\by_{k} \mid \rho_{k}, \bzeta)}{p(\by_{k} \mid \rho_{k,0}, \bzeta)}
	\\
	&& =\quad \sqrt{ \frac{\prod_{h=1}^{m_{k,0}} (1 + n_{k,h,0} \sigma_{n}^{- 2} \sigma_{k,h,0}^{2})}{ \prod_{h=1}^{m_{k}} (1 + n_{k,h} \sigma_{n}^{- 2} \sigma_{k,h}^{2})}}  \exp\left\{ -\frac{1}{2} \sum_{h=1}^{m_{k}} \left(\frac{\mu_{k,h}^{2}}{\sigma_{k,h}^{2}}  - \frac{\mu_{k,h}^{\star 2}}{\sigma_{k,h}^{\star 2}}\right) +\frac{1}{2} \sum_{h=1}^{m_{k,0}} \left(\frac{\mu_{k,h,0}^{2}}{\sigma_{k,h,0}^{2}}  - \frac{\mu_{k,h,0}^{\star 2}}{\sigma_{k,h,0}^{\star 2}}\right) \right\}.
\ese
\vskip-2ex
\noindent To examine the behavior of this expression, note that $m_{k} > m_{k,0}$, and $n_{k,h} < n_{k,h,0}$, $h = 1, \dots ,m_{k,0}$. 
Now, the expression under the square root converges to $0$ as $n \rightarrow \infty$ since $m_{k} > m_{k,0}$ implies that the denominator is of a higher order.

Let us now focus on the exponential term.
Notice that $\mu_{k,h}^{2} / \sigma_{k,h}^{2}$ and $\mu_{k,h,0}^{2} / \sigma_{k,h,0}^{2}$ are $O_{n}(1)$.
Moreover, the only terms in the exponential that are not $O_{n}(1)$ can be rewritten as 
\vskip-5ex
\bse
- \frac{1}{2 \sigma_{n}^{2}} \left[ \sum_{h=1}^{m_{k}} \frac{\left(\sum_{\substack{(i,\ell) \text{ s.t.} \\ \bx_{i,\ell,t} \in S_{k,h}}} y_{i,\ell,k}\right)^{2}}{n_{k,h}} - \sum_{h=1}^{m_{k,0}} \frac{\left(\sum_{\substack{(i,\ell) \text{ s.t.} \\ \bx_{i,\ell,t} \in S_{k,h,0}}} y_{i,\ell,k}\right)^{2}}{n_{k,h,0}} \right].
\ese
\vskip-3ex
The expression in the square brackets can be rewritten as a quadratic form $\by^{\intercal} \bA \by$ for some matrix $\bA$ which is $O_{n}(1)$.
Since the dispersion matrix of $\by$ is $\sigma_{n}^{2} \bI_{n}$, the variance of $\by^{\intercal} \bA \by$ is $O_{n}(1/n)$ as $n \rightarrow \infty$. 
Therefore, the exponential term is bounded in probability as $n \rightarrow \infty$.

Thus, we have shown that the probability of any incompatible model goes to zero.
Along with the fact that for any compatible model the marginal likelihood ratio tends to $0$ implies that the only model that can possibly retain positive probability is the truth. 
Since for any fixed $k$ there are only finitely many models for $\{\beta_{k,\bx}\}_{\bx \in \X}$, the probability of the true model must tend to $1$.

\section{Prior Hyper-parameters} \label{sec: sm prior hyper-parameters}

The fixed effects parameters of the longitudinal mixed effects model (\ref{eq: function 1})
are initialized at the maximum likelihood estimate for the spline coefficients of the simplified model with no smoothing and no predictors included. 
The random effects are initialized at zero.

The hyper-parameter for the half-Cauchy prior on the smoothing parameters is $s_{\sigma} = 1$.
The $\HC(0, 1)$ distribution has its mode at zero and hence is capable of capturing strong smoothness but also has heavy tails and is thus also capable of capturing wiggly functions.
The hyper-parameters for the inverse-Gamma prior on the residual variance are set at $a_{\sigma} = b_{\sigma} = 1$.
The hyper-parameters on the Gamma prior for the mass of the Dirichlet distributions on the transition dynamics are 
$a_{\alpha} = b_{\alpha} = a_{\alpha^{\star}} = b_{\alpha^{\star}} = 1$, as recommended in \citetlatex{escobar1995bayesian}.
Finally, the hyper-parameters for the Gamma prior on $\varphi_{j,k}$ are $a_{\varphi,j} = 5, b_{\varphi,j} = 1$.

\newpage
\section{Posterior Inference}

\subsection{MCMC Algorithm} \label{sec: sm mcmc}
We summarize here the steps of the MCMC algorithm used to sample from the posterior of our model. 
See also Section \ref{sec: post inference} in the main paper for additional details.  

\vspace*{1cm}
\begin{algorithm}[ht!] 
\caption{}
\label{algo: MCMC}
\begin{algorithmic}[1]
\vspace{0.2cm}

\Algphase{Updating the core tensor sizes $\ell_{j,k}$ and the latent variables $z_{j,k}^{(x_{j})}, z_{k}^{(z_{1,k}, \dots, z_{p,k})}$}
\State
For $k = 1, \dots, K$, $j = 1, \dots, p$, sample $\ell_{j,k}$, $\bz_{j,k}$ and $z_{k}^{(z_{1,k}, \dots, z_{p,k})}$ using the M-H step with acceptance rate \eqref{eq: acc_rate}.

\Algphase{Updating the cluster specific parameters $\beta_{k,h}^{\star \star}$ and the smoothness $\sigma_{\beta}^{2}$}
\State
For $k = 1, \dots, K$, $h = 1, \dots, m_{k}$ sample the group specific curves $\beta_{k,h}^{\star \star}$ from their Gaussian full conditionals \eqref{eq: fullcond_beta_star}
\vspace{-0.3cm}
\begin{equation*}
	p(\beta^{\star \star}_{k,h} \mid \by_{k}, S_{k,h}, \sigma_{\varepsilon}^{2}, \sigma_{\beta}^{2}, \bzeta) = \Normal\left\{ \mu^{\star}_{k,h}, \sigma_{k,h}^{\star 2} \right\}.
\end{equation*}

\State
Sample the smoothness parameter $\sigma_{\beta}^{2}$ from its inverse-Gamma full conditional
\vspace{-0.3cm}
\begin{equation*}
	p(\sigma_{\beta}^{2} \mid \bzeta) \propto p_{0}(\sigma_{\beta}^{2}) \prod_{k} \prod_{h} p(\beta_{k,h}^{\star \star} \mid \sigma_{\beta}^{2}).
\end{equation*}

\Algphase{Updating the initial distributions $\bpi_{0}^{(j)}$ and the transition dynamics $\bpi_{h}^{(j)}$} 
\State
Let $n_{0,h}^{(j)} = \sum_{x_{j}}1\{z_{j,1}^{(x_{j})} = h\}$. For $j = 1, \dots, p$, sample $\bpi_{0}^{(j)}$ as
\vspace{-0.3cm}
\bse
	\{\pi_{0}^{(j)}(1),\dots,\pi_{0}^{(j)}(x_{j,\max})\} \mid \bzeta \sim \Dir\{\alpha^{(j)}/x_{j,\max} + n_{0,1}^{(j)},\dots,\alpha^{(j)}/x_{j,\max}+ n_{0,x_{j,\max}}^{(j)} \}.
\ese

\State
Let $n_{h,h^{\prime}}^{(j)} = \sum_{k = 2}^{K} \sum_{x_{j}}1\{z_{j,k}^{(x_{j})} = h,z_{j,k-1}^{(x_{j})} = h^{\prime}\}$. 
For $j = 1, \dots, p$, $h = 1, \dots, x_{j,\max}$, sample $\bpi_{h}^{(j)}$ as 
\vspace{-0.3cm}
\bse
\{\pi_{h}^{(j)}(1),\dots,\pi_{h}^{(j)}(x_{j,\max})\} \mid \bzeta \sim \Dir\{\alpha^{(j)}/x_{j,\max} + n_{h,1}^{(j)},\dots,\alpha^{(j)}/x_{j,\max}+ n_{h,x_{j,\max}}^{(j)} \}.
\ese
\State
For $j = 1, \dots, p$, sample $\alpha^{(j)}$ using an M-H step from its full conditional
\vspace{-0.3cm}
\bse
\textstyle p(\alpha^{(j)} \mid \bzeta) \propto p_{0}(\alpha^{(j)}) p(\bpi_{0}^{(j)} \mid \alpha^{(j)}) \prod_{h=1}^{x_{j,\max}} p(\bpi_{h}^{(j)} \mid \alpha^{(j)}).
\ese

\Algphase{Updating the second layer probabilities $\bpi^{\star}_{k}$} 

\State
Let $n_{k,h}^{\star} = \sum_{z_{1,k}, \dots, z_{p,k}}1\{z_{k}^{(z_{1,k}, \dots, z_{p,k})} = h\}$. Sample $\bpi^{\star}_{k}$ as 
\vspace{-0.4cm}
\bse
\{\pi^{\star}_{k}(1),\dots,\pi^{\star}_{k}(\ell_{k})\} \mid \bzeta \sim \Dir\{\alpha^{\star}/\ell_{k} + n_{k,1}^{\star},\dots,\alpha^{\star}/\ell_{k}+ n_{k,\ell_{k}}^{\star} \}.
\ese

\algstore{myalg}
\end{algorithmic}
\end{algorithm}
\setlength{\textfloatsep}{0pt}

\begin{algorithm}[ht!]
\begin{algorithmic}[1]
\vspace{0.2cm}
\algrestore{myalg}

\State
Sample $\alpha^{\star}$ using an M-H step from its full conditional
\vspace{-0.4cm}
\bse
\textstyle p(\alpha^{\star} \mid \bzeta) \propto p_{0}(\alpha^{\star}) \prod_{k=1}^{K} p(\bpi^{\star}_{k} \mid \alpha^{\star}).
\ese

\Algphase{Updating the random effects parameters}
\State
Let $\br^{(r)} = \{y_{i,\ell,t} - f_{x_{1}, \dots, x_{p}}(t) \}_{i,\ell,t}$ be the random effects residuals.
For $i = 1, \dots, n$, sample the random effects curve parameters $\bbeta_{i}^{(u)}$ 
\vspace{-0.4cm}
\bse
\textstyle p(\bbeta_{i}^{(u)} \mid \bzeta) \sim \MVN_{K}\left( \bSigma^{\star}_{i} \bB_{i}^{\intercal} \frac{\br_{i}^{(r)}}{\sigma_{\varepsilon}^{2}}, \bSigma^{\star}_{i} \right),
\ese

\vspace{-0.4cm}
\noindent where $\bSigma^{\star}_{i} = \left( \sigma_{\varepsilon}^{-2} \bB_{i}^{\intercal} \bB_{i} + \sigma_{u,s}^{-2} \bP_{u} + \sigma_{u,a}^{-2} \bI_{K} \right)^{-1}$, $\bB_{i} = \{b_{k}(t_{i})\}_{t_{i}, k}$.

\State
Sample the random effects smoothness parameter $\sigma_{u,s}$ using an M-H step from its full conditional 
\vspace{-0.4cm}
\bse
\textstyle p(\sigma_{u,s} \mid \bzeta) \propto p_{0}(\sigma_{u,s}) \prod_{i=1}^{n} p(\bbeta_{i}^{(u)} \mid \sigma_{u,s}, \sigma_{u,a}).
\ese

\State
Sample the random effects scale parameter $\sigma_{u,a}$ using an M-H step from its full conditional
\vspace{-0.4cm}
\bse
\textstyle p(\sigma_{u,a} \mid \bzeta) \propto p_{0}(\sigma_{u,a}) \prod_{i=1}^{n} p(\bbeta_{i}^{(u)} \mid \sigma_{u,s}, \sigma_{u,a}).
\ese

\Algphase{Updating the global variance parameter}
\State
Let $\br = \{y_{i,\ell,t} - f_{x_{1}, \dots, x_{p}}(t) - u_{i}(t) \}_{i,\ell,t}$ be the residuals. Sample the error variance $\sigma_{\varepsilon}^{2}$ as
\vspace{-0.4cm}
\bse
\textstyle \sigma_{\varepsilon}^{2} \mid \bzeta \sim \IG \{ a_{\sigma} + nLT/2, b_{\sigma} + \br^\intercal \br / 2 \}.
\ese

\end{algorithmic}
\end{algorithm}
\setlength{\textfloatsep}{35pt}

\newpage
\vspace*{0cm} 
\subsection{Software, Runtime, etc.} \label{sec: sm software}

The results reported in this article are all based on $7,500$ MCMC iterations with the initial $2,500$ iterations discarded as burn-in.
The remaining samples were further thinned by an interval of $5$. 
We programmed in \texttt{R} interfaced with \texttt{C++} through Rcpp \citeplatex{eddelbuettel2011rcpp} and RcppArmadillo \citeplatex{eddelbuettel2014rcpparmadillo}.
The codes are available as part of the supplementary materials. 
The MCMC algorithm takes 10 minutes on a Macbook laptop with 8 Gb RAM for the synthetic examples.
The execution time complexity is compared with the other methods in Table \ref{tab: comp_time}.
A `readme' file, providing additional details for a practitioner, is also included in the supplementary material.
In all experiments, the posterior samples produced very stable estimates of the population and individual level parameters of interest. 
MCMC diagnostic checks were not indicative of any convergence or mixing issues. 
While the general methodology does not rely on the hypothesis of equidistant time points, all data examples presented in the manuscript have this feature. 
Therefore, the current implementation of the algorithm assumes that the observations are collected on a regular time grid. 
We plan to release future software updates relaxing this assumption.

\begin{table}[!ht]
\centering
\begin{tabular}{l|c}
\multicolumn{1}{c|}{\textbf{Method}} & \textbf{\begin{tabular}[c]{@{}c@{}}Execution Time {[}seconds{]}\\ Median (95\% CI)\end{tabular}} \\ \hline
		Our Proposed LFMM & 1967.3 (1285.1, 2169.6) \\
		BART              & 268.3 (192.8, 279.0) 
		\\
		Soft BART         & 644.0 (351.2, 789.7) 
		\\
		Lasso             & 2.07 (1.12, 2.68)                                                                                   
\end{tabular}
\caption{Execution time of the different methods compared in Section \ref{sec: sim studies}, in seconds. 
The competing methods used the packages \texttt{BART}, \texttt{SoftBart} and \texttt{glmnet}, respectively.}
\label{tab: comp_time}
\end{table}

\newpage
\subsection{MCMC Diagnostics}

This section presents some MCMC convergence diagnostics for the Gibbs sampler described in Section \ref{sec: sm mcmc}.
The results presented here are obtained on one of the $500$ replications of the synthetic experiment described in Section \ref{sec: sim studies}, 
namely the one achieving the median RMSE.

\begin{table}[!ht]
	\centering
	\begin{tabular}{|l|ccccc|} \hline
		& $k=1$  & $k=5$  & $k=10$  & $k=15$  & $k=20$ \\ \hline
		\multicolumn{1}{|c|}{\multirow{2}{*}{$\beta_{k,x_{1}, \dots, x_{p}}$}}	& -0.236 & 0.640  & 0.295  & -1.074  & -1.330 \\
		\multicolumn{1}{|c|}{}	& (0.81) & (0.52)  & (0.77)  & (0.28)  & (0.18) \\ \hline
		\multirow{2}{*}{$\beta_{k,x_{1}^{\prime}, \dots, x_{p}^{\prime}}$}	& -0.251 & -0.690 & -1.040 & -0.982  & -0.930  \\
		& (0.80) & (0.49) & (0.30) & (0.33) & (0.35)  \\ \hline
	\end{tabular}
	\caption{Geweke statistics and associated p-values assessing convergence of the covariate specific spline parameters $\beta_{k,x_{1},\dots, x_{p}}$ evaluated at several locations $k \in \{1, 5, 10, 15, 20\}$.
	}
	\label{tab: param_geweke}
\end{table}

We analyze convergence for the covariate specific parameters, i.e. the $\bbeta_{x_{1}, \dots, x_{p}}$'s, as opposed to the mixture atoms $\bbeta_{h}^{\star \star}$'s as the latter are not affected by label switching.
Figure \ref{fig: traceplots} and Figure \ref{fig: densityplots} respectively show the trace plots and the estimated marginal posterior distributions
of these parameters at different time points for two combinations of the categorical predictors.
Similar results can be obtained for other level combinations of the categorical predictors. 
These results are based on the MCMC thinned samples. 
As these figures show, the running means are very stable and there seems to be no convergence issues. 
Additionally, the Geweke test \citeplatex{geweke1991evaluating} for stationarity of the chains, which formally compares the means of the first and last part of a Markov chain, was also performed. 
If the samples are drawn from the stationary distribution of the chain, the two means are equal and Geweke's statistic has an asymptotically standard normal distribution. 
The results of the test, reported in Table \ref{tab: param_geweke}, indicate that convergence was satisfactory for the parameters considered since they fail to reject the null hypothesis of stationarity of the corresponding chains.
Only one parameter in the first row of Table \ref{tab: param_geweke}, had a significant p-value. 
Some chance rejections are expected in multiple hypothesis testing scenarios. 
A visual inspection of the corresponding trace plot, however, does not indicate any serious issue.

Likewise, Figure \ref{fig: reff traceplots} and Figure \ref{fig: reff densityplots} respectively show the trace plots and the marginal posterior distributions
of the random effects parameters $u_{i}(t)$ at several time points $t$ for two randomly selected  different individuals $i$.

\vskip -10pt
\begin{figure}[!h]
	\centering
	\includegraphics[width=0.65\linewidth]{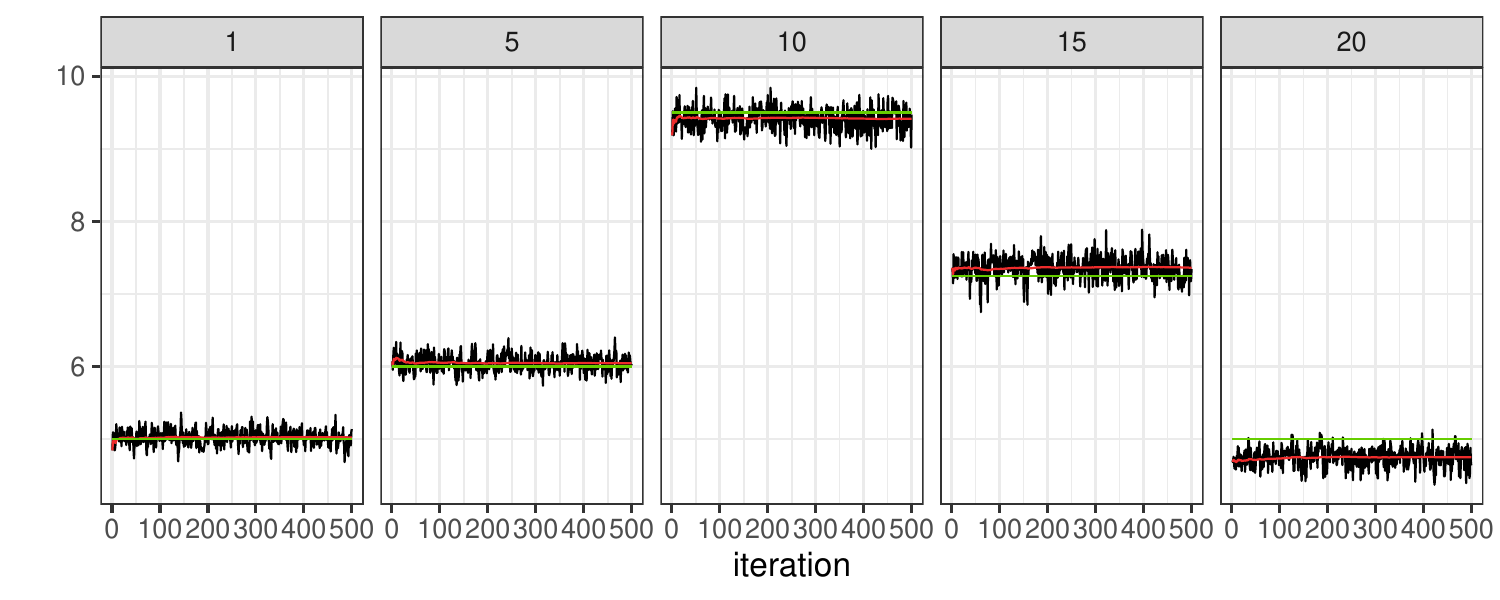}
	\includegraphics[width=0.65\linewidth]{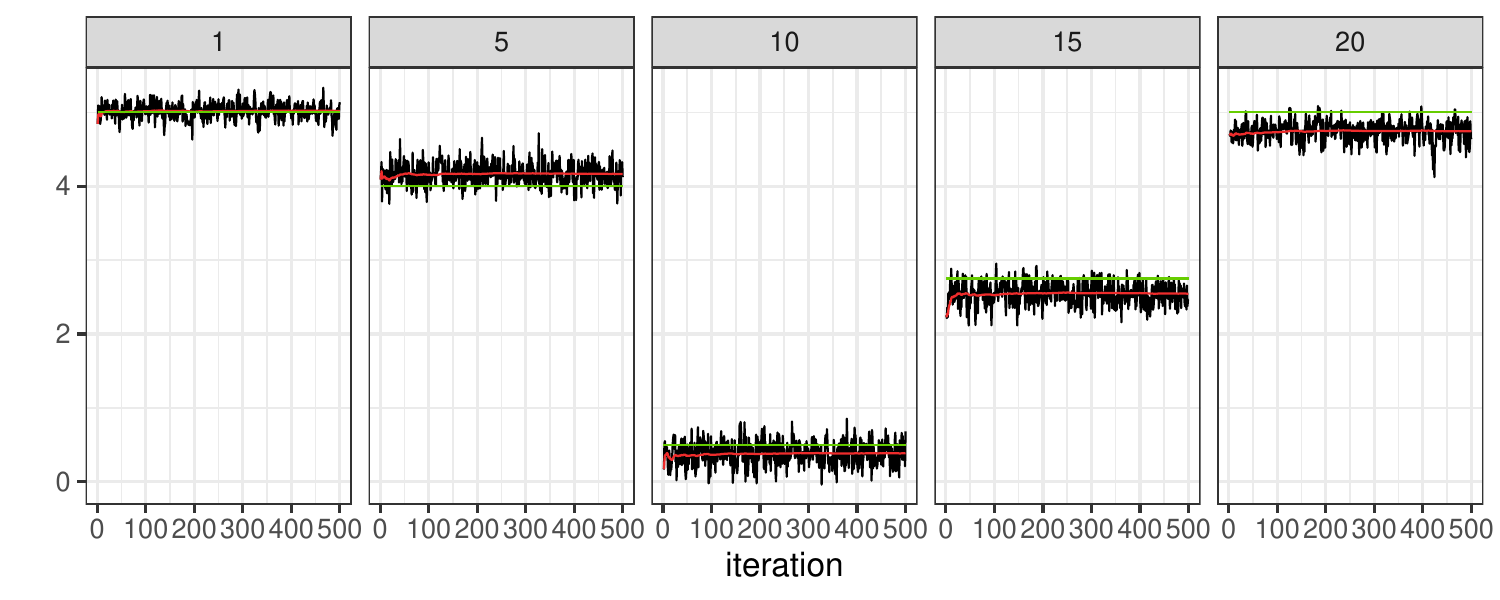}
	\vskip-05pt
	\caption{
	Results for synthetic data: Trace plots of the covariate specific spline parameters $\beta_{k,x_{1},\dots, x_{p}}$ evaluated at several time points $k \in \{1, 5, 10, 15, 20\}$ (along the columns) for two different combinations of the categorical predictors (along the rows). 
	In the first row, $(x_{1}, \dots, x_{p}) = (1,1,2,1,1,1,1,1,1,1)$.
	In the second row, $(x_{1}^{\prime}, \dots, x_{p}^{\prime}) = (1,1,3,1,1,1,1,1,1,1)$.
	In each panel, the solid red line shows the running mean, 
	and the solid green line the corresponding simulation truth.
	}
	\label{fig: traceplots}
\end{figure}

\vskip -30pt
\begin{figure}[!ht]
	\centering
	\includegraphics[width=0.65\linewidth]{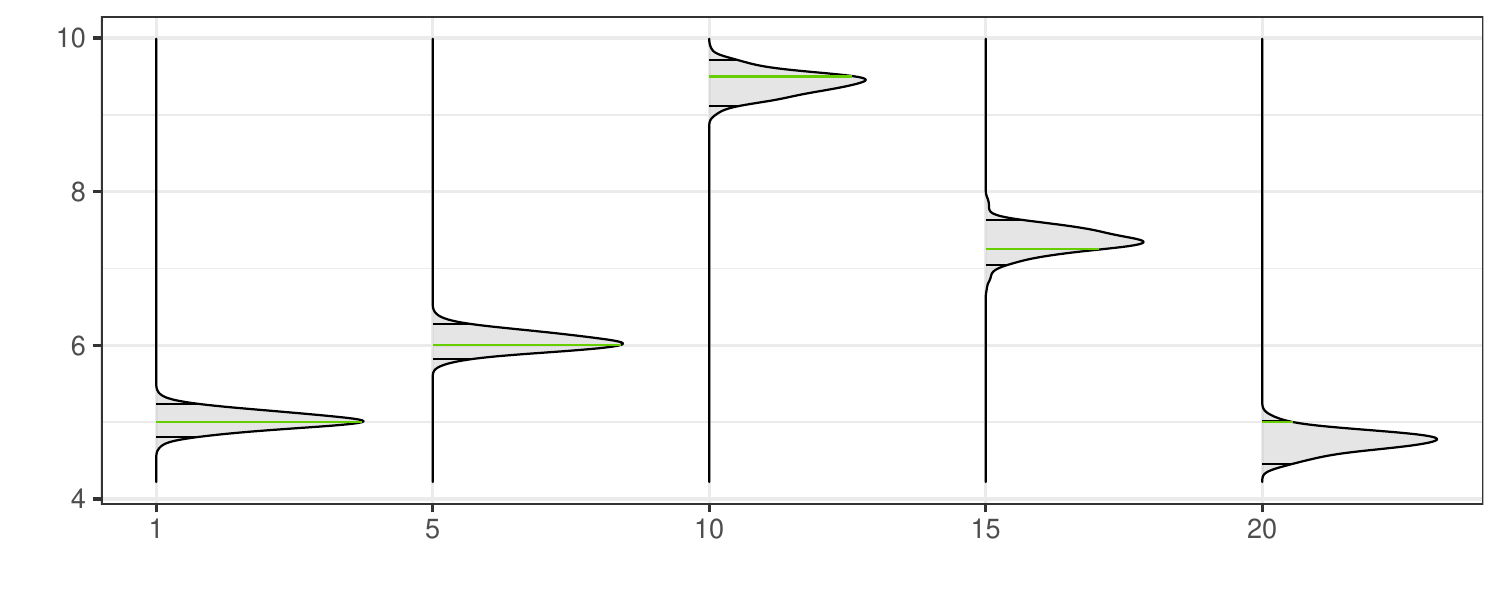}
	\includegraphics[width=0.65\linewidth]{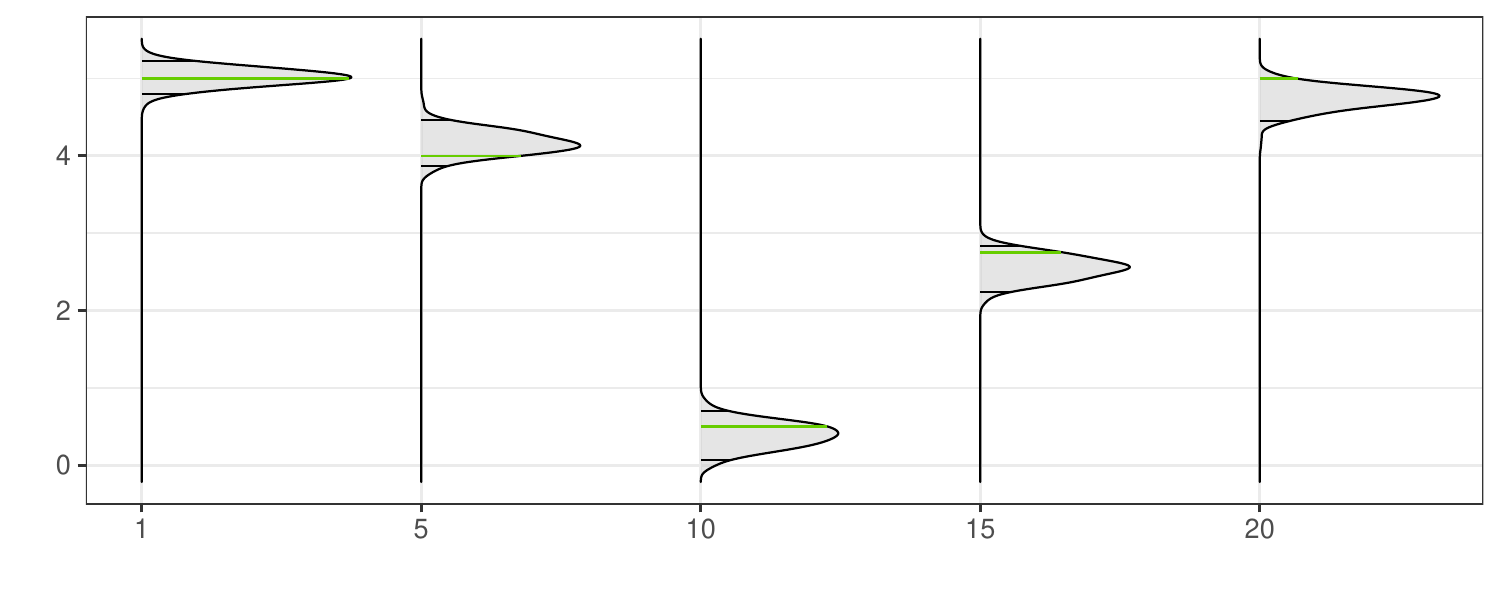}
	\vskip-05pt
	\caption{
	Results for synthetic data: Estimated posterior densities of the covariate specific spline parameters $\beta_{k,x_{1},\dots, x_{p}}$ evaluated at several time points $k \in \{1, 5, 10, 15, 20\}$ (along the columns) for two different combinations of the categorical predictors (along the rows). 
	In the top row, $(x_{1}, \dots, x_{p}) = (1,1,2,1,1,1,1,1,1,1)$.
	In the bottom row, $(x_{1}^{\prime}, \dots, x_{p}^{\prime}) = (1,1,3,1,1,1,1,1,1,1)$.
	In each panel, the black vertical lines show the corresponding $95\%$ posterior credible intervals, 
	and the solid green line the corresponding simulation truth.
	}
	\label{fig: densityplots}
\end{figure}

\begin{figure}[!ht]
	\centering
	\includegraphics[width=0.65\linewidth]{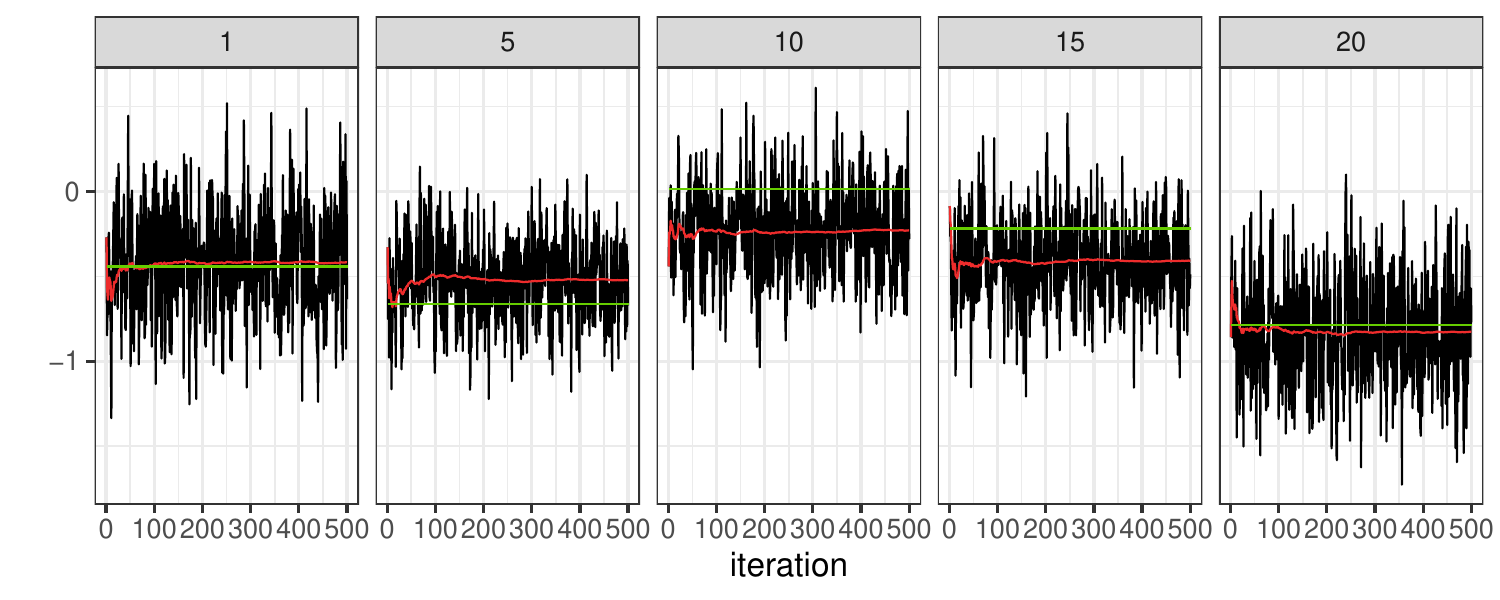}
	\includegraphics[width=0.65\linewidth]{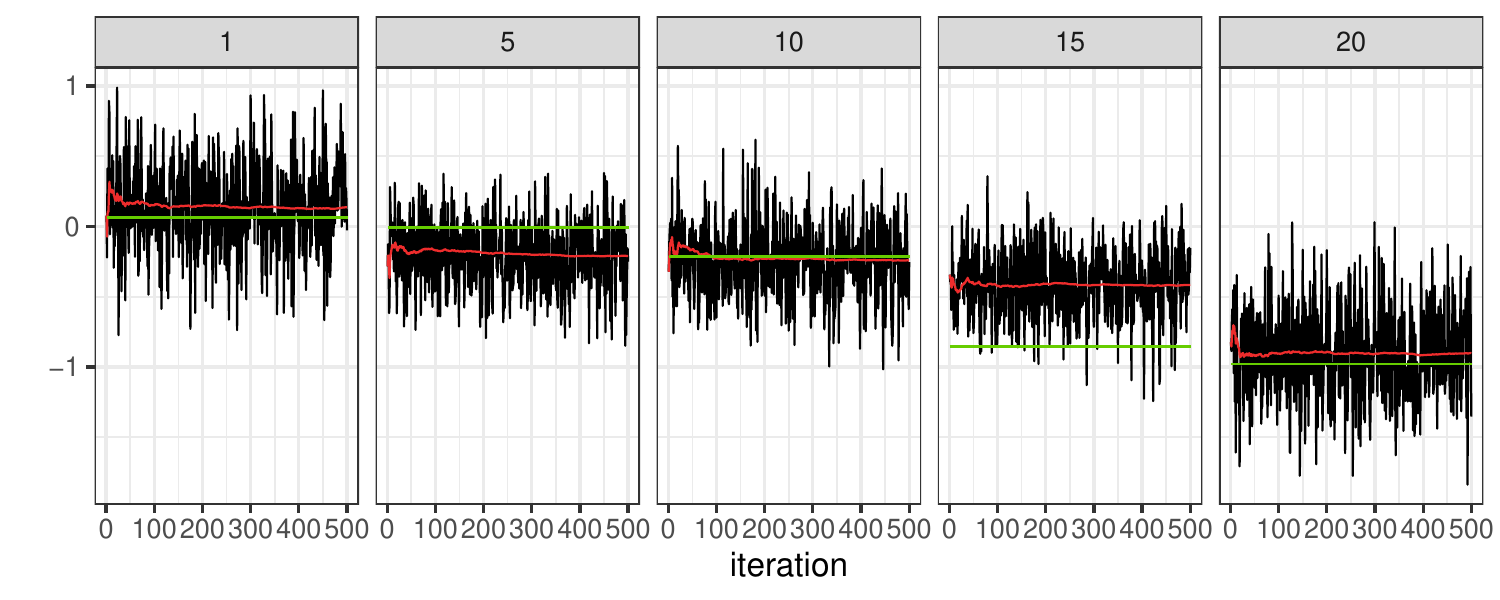}
	\vskip-05pt
	\caption{
	Results for synthetic data: Trace plots of the individual specific random effects parameters $u_{i}(t)$ evaluated at several time points $t \in \{1, 5, 10, 15, 20\}$ (along the columns) for two different individuals $i$ (along the rows). 
	In each panel, the solid red line shows the running mean, 
	and the solid green line the corresponding simulation truth.
	}
	\label{fig: reff traceplots}
\end{figure}

\begin{figure}[!hb]
	\centering
	\includegraphics[width=0.65\linewidth]{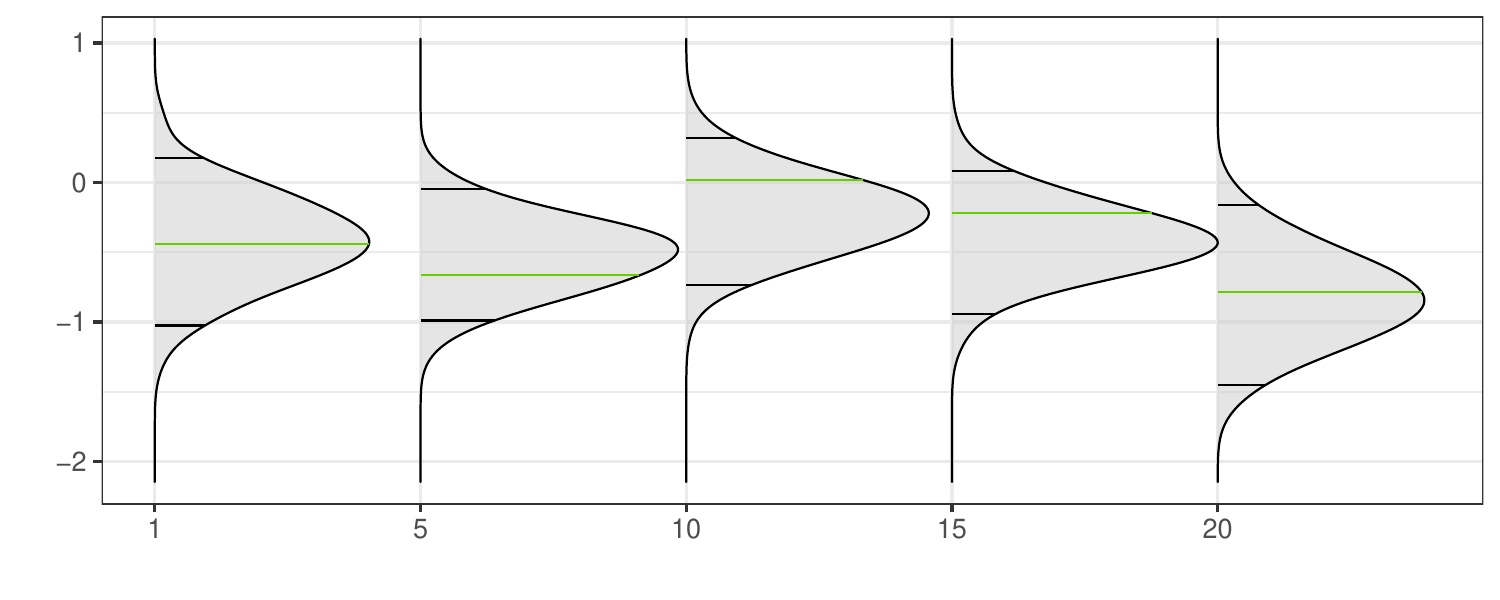}
	\includegraphics[width=0.65\linewidth]{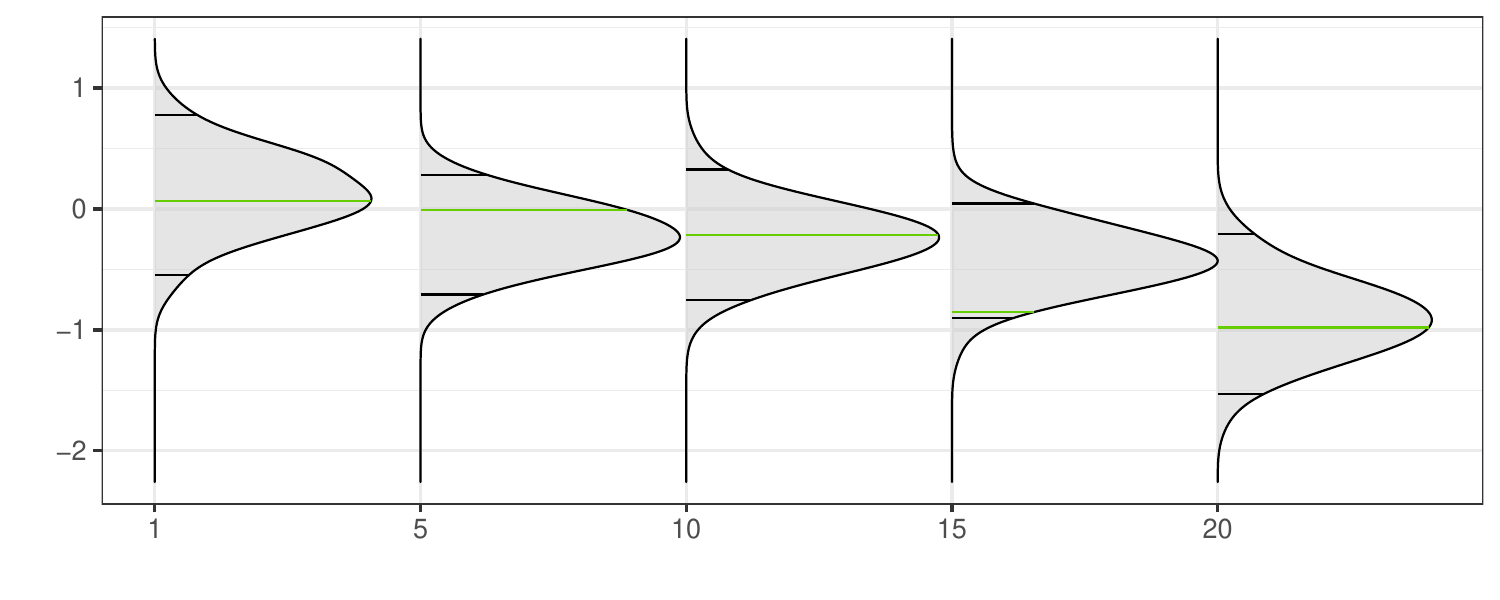}
	\vskip-05pt
	\caption{
	Results for synthetic data: Estimated posterior densities of the individual specific random effects parameters $u_{i}(t)$ evaluated at several time points $t \in \{1, 5, 10, 15, 20\}$ (along the columns) for two different individuals $i$ (along the rows). 
	In each panel, the black vertical lines show the corresponding $95\%$ posterior credible intervals, 
	and the solid green line the corresponding simulation truth.
	}
	\label{fig: reff densityplots}
\end{figure}

\clearpage
\newpage
\section{Additional Simulations} \label{sec: sm add simulations}

We consider here additional simulations to assess the performance of our proposed model in the special case when no individual specific information is available. 
Although not very realistic for longitudinal data applications, this scenario serves as a fairer comparison with nonparametric regression techniques that do not accommodate random effects.
We use the same true fixed effects employed in Section \ref{sec: sim studies}, which are also shown in Figure \ref{fig: eye}.
We consider $p = 10$ predictors, $T=20$ time points and $L_{t} = 200$ repeated measurements at each time point, yielding the same total sample size $\prod_{t=1}^{T} L_{t} = 4000$ of the experiments in Section \ref{sec: sim studies}.
The residual variance is set at $\sigma_{\varepsilon}^{2} = 1$.

\begin{figure}[!ht]
	\centering
	\includegraphics[width=0.99\linewidth]{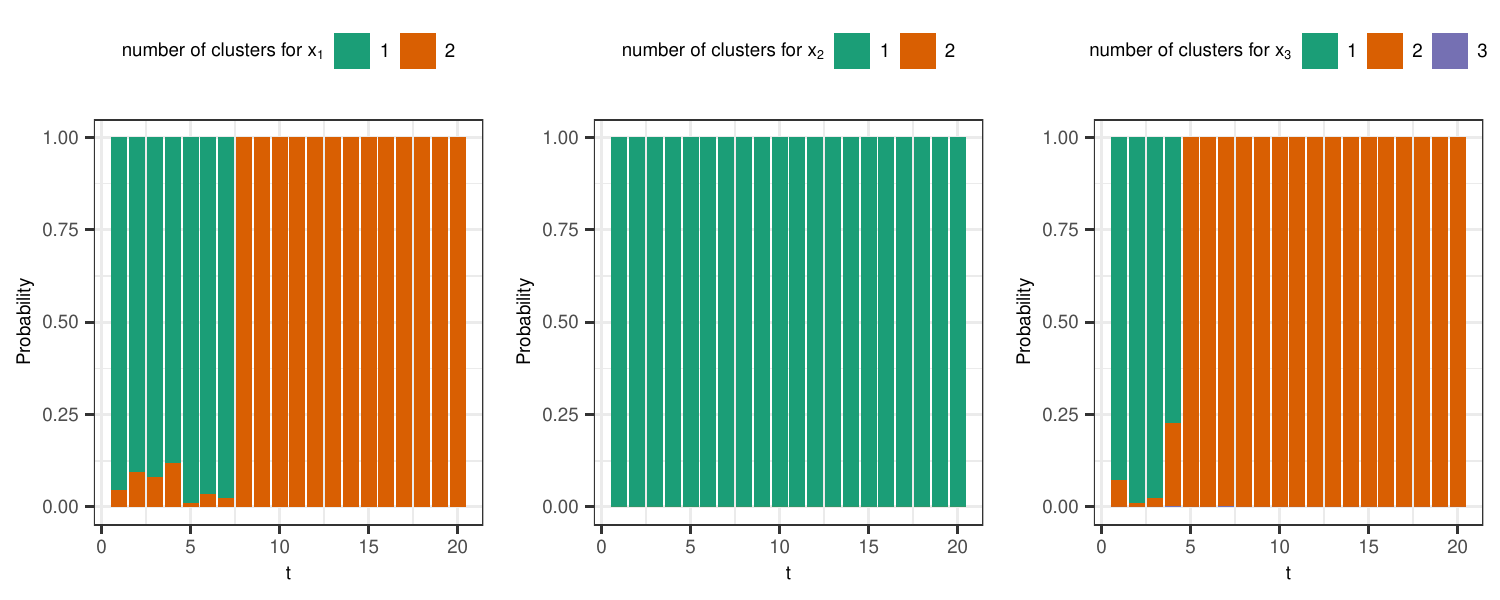}
	\caption{Results for synthetic data: 
	The estimated posterior probabilities for the number of clusters of the predictors' levels over time for $x_{1}, x_{2}$ and $x_{3}$. 
	The predictors $x_{1}$ and $x_{3}$ were locally important. 
	The remaining predictors, namely $(x_{2},x_{4},\dots,x_{10})$, including $x_{2}$ shown here, 
	were never included in the model - their levels always formed a single cluster.}
	\label{fig: n_groups_sim_no_reff}
\end{figure}

\begin{figure}[!ht]
	\centering
	\begin{center}
		\includegraphics[width=0.49\linewidth]{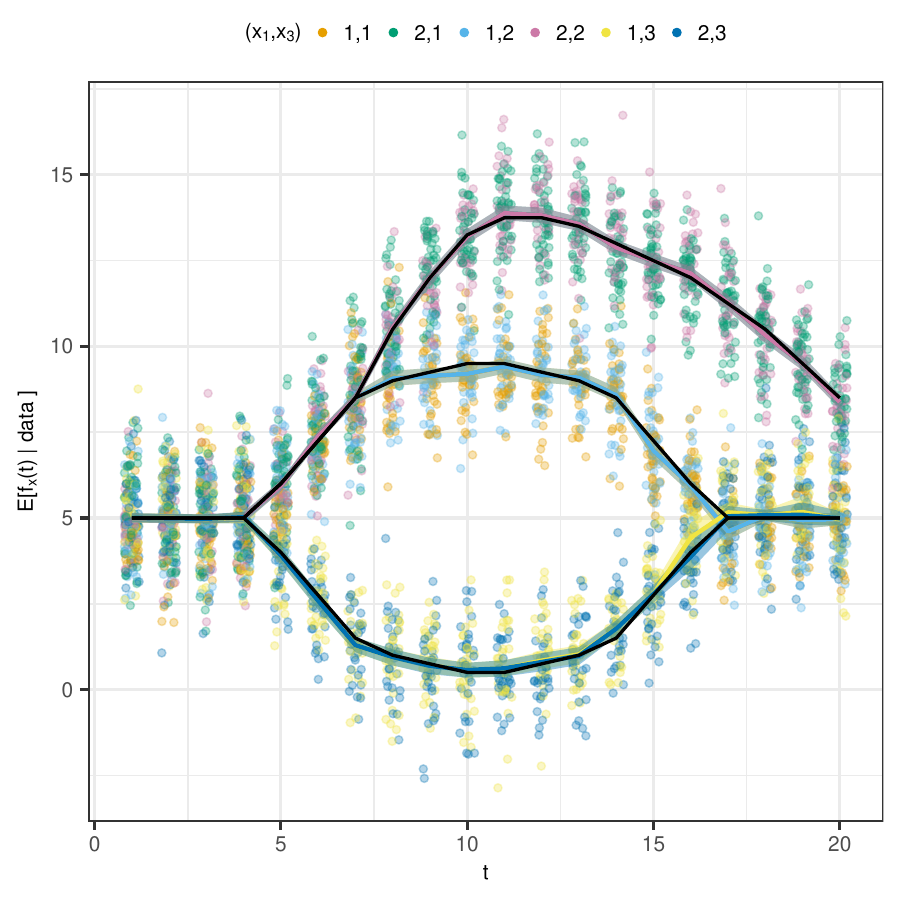}
	\end{center}
	\vskip-15pt
	\caption{Results for synthetic data: 
	Scenario with ten covariates $(x_{1},\dots,x_{10})$ where only $(x_{1}, x_{3})$ are locally important, as described in Section \ref{sec: sim studies}.
	The colored lines represent the estimated posterior means and $95\%$ point wise credible intervals for the fixed effects,  
	superimposed on slightly jittered response values $y_{i,\ell,t}$ for all combination of the levels of the significant predictors $(x_{1}, x_{3})$. 
	The true fixed effects are superimposed (black lines).
	The figure here corresponds to the synthetic data set that produced the median root mean squared error. 	
	}
	\label{fig: eye_no_reff}
\end{figure}

As shown in Figure \ref{fig: n_groups_sim_no_reff}, our method correctly recovers the significant predictors $x_{1}$ and $x_{3}$.
A $0.5$ posterior probability cutoff also correctly estimates two groups for $x_{1}$ starting from $t = 8$ and two groups for $x_{3}$ starting from $t = 5$.
Estimates of the fixed effects curves obtained by our method are shown in Figure \ref{fig: eye_no_reff}. 
Our model estimates the fixed effects very precisely by borrowing information whenever predictors are redundant or covariate levels are in the same cluster.

Figure \ref{fig: sim_RMSE_no_reff} compares the out-of-sample predictive performance (left panel), 
the coverage of the $95\%$ prediction intervals (middle panel) 
and the widths of these intervals (right panel)
for different methods for $500$ simulated data sets with $75\%$-$25\%$ training-test splits.

\begin{figure}[ht!]
	\centering
	\includegraphics[width=0.32\linewidth]{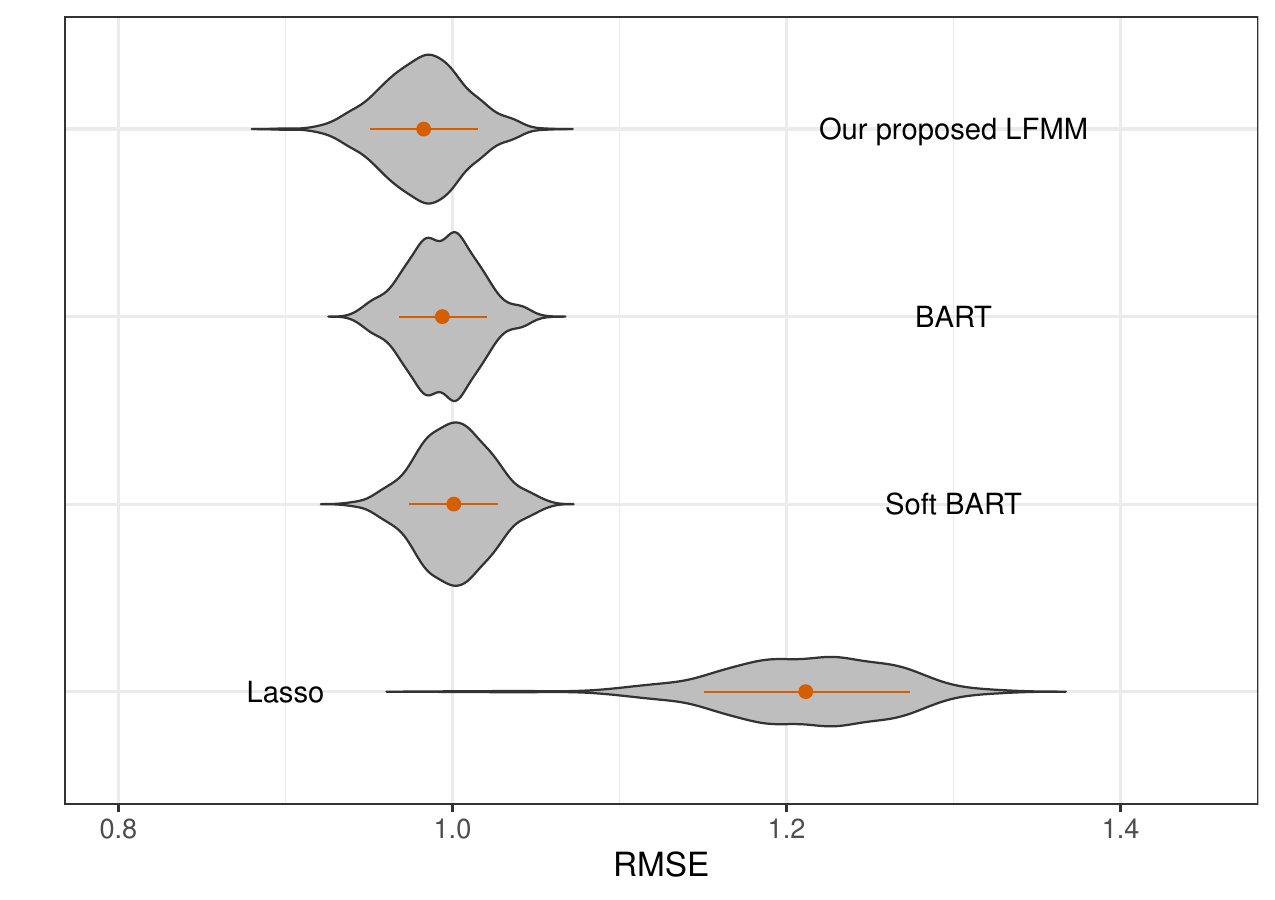}
	\includegraphics[width=0.32\linewidth]{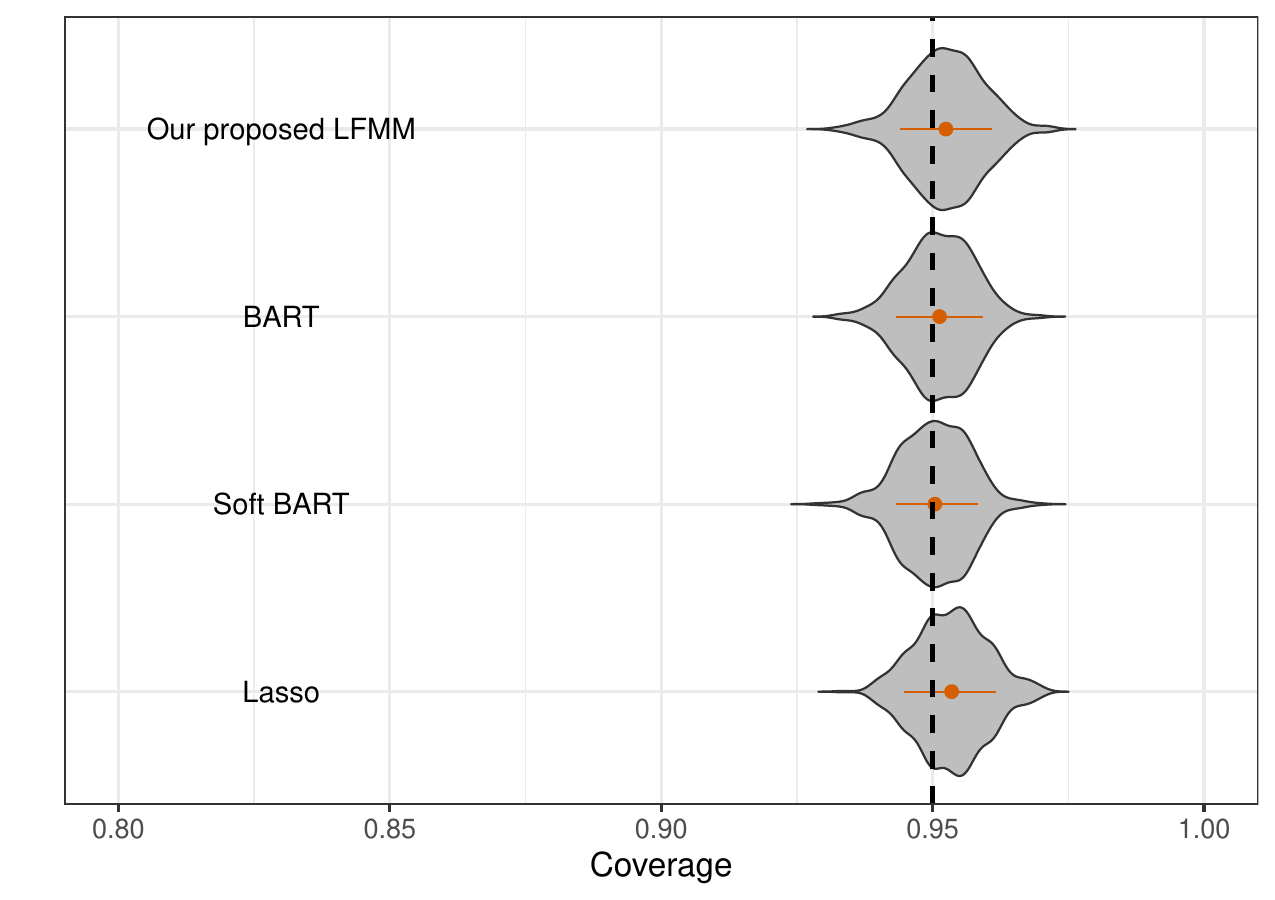}
	\includegraphics[width=0.32\linewidth]{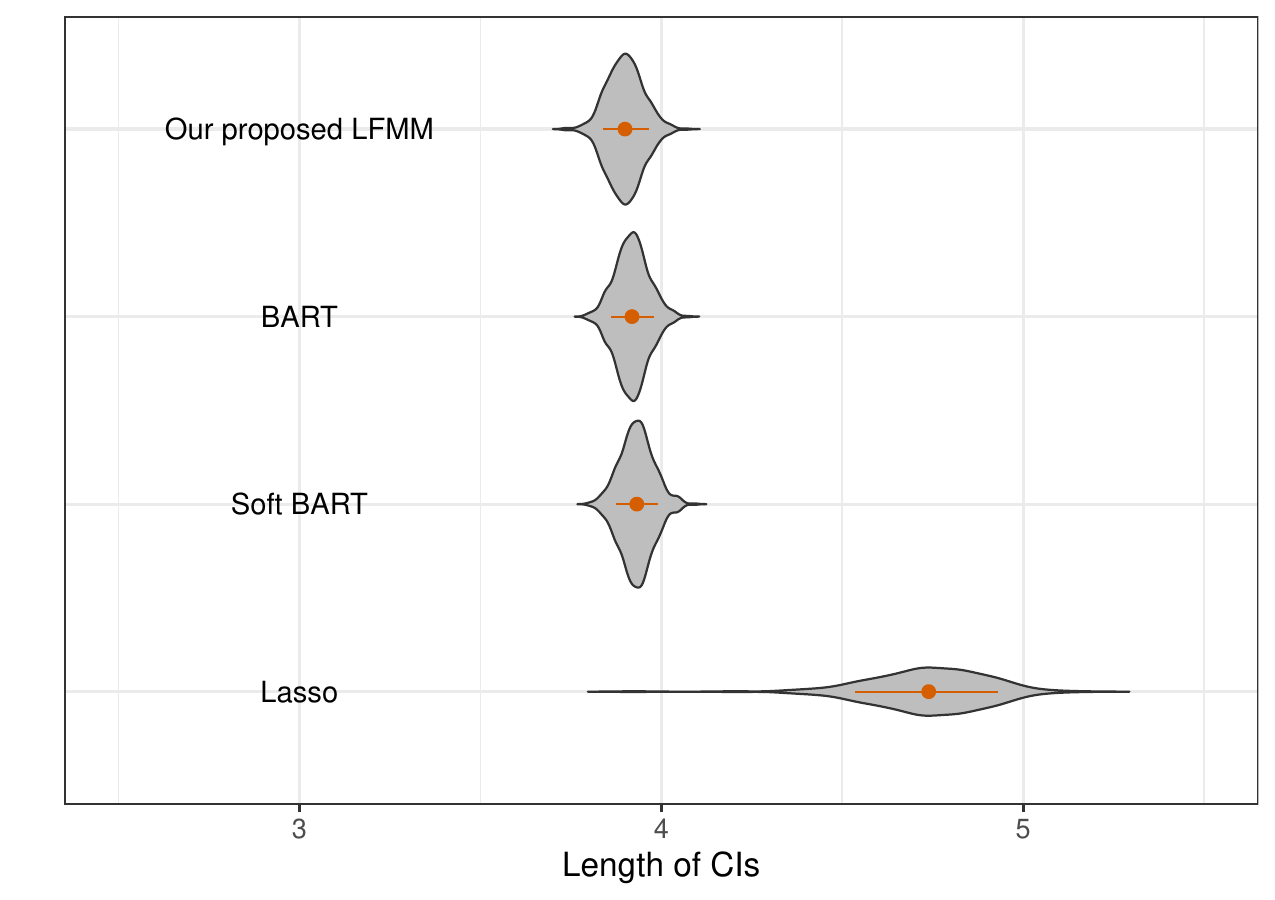}
	\caption{Results for synthetic data: The left panel shows the out-of-sample root mean squared error. 
	The middle panel shows the coverage of $95\%$ prediction intervals. All measures reported are obtained over 500 $75\%$-$25\%$ training-test splits.
	The right panel shows the widths of the prediction intervals.
	The red points represent the averages across simulations, whereas the red intervals represent the interquartile ranges across simulations.}
	\label{fig: sim_RMSE_no_reff}
\end{figure}

\clearpage\newpage
\section{Additional Applications} \label{sec: sm add data}

We describe here some additional real data applications of our proposed method.

\subsection{Beat the Blues Data} \label{sec: sm btheb data}

We consider longitudinal data from a randomized clinical trial of an interactive multimedia program known as ``Beat the Blues'' 
which was designed to deliver cognitive behavioral therapy to depressed patients via a computer terminal. 
Patients with depression recruited in primary care were randomized to either the Beating the Blues program, or to ``Treatment as Usual'' (TAU), and they were followed up for a maximum of 4 visits.
Other than the treatment indicator, the two additional predictors include 
dummy variables indicating if patients take anti-depressant drugs and if the length of the current episode of depression is less or more than six months.
Thus, the size of the unstructured model $T \prod_{j=1}^{p} x_{j,\max} = 5 \times 2^3 = 40$ makes it hard to estimate 
the parameters with the small sample size $n = 380$, typical of a clinical trial. 
The measured outcome is the Beck Depression Inventory II (BDI), a popular depression screening instrument.
The data is publicly available, for instance, via the \texttt{R} package \texttt{HSAUR2} \citeplatex{hothorn2014handbook}.
The efficacy of computerized cognitive behavioral therapy was first detected in \citetlatex{proudfoot2003computerized} via a linear mixed effects model 
but was not replicated in the randomized clinical trial of \citetlatex{gilbody2015computerised}.

\begin{figure}[!ht]
	\centering
	\includegraphics[width=.99\linewidth]{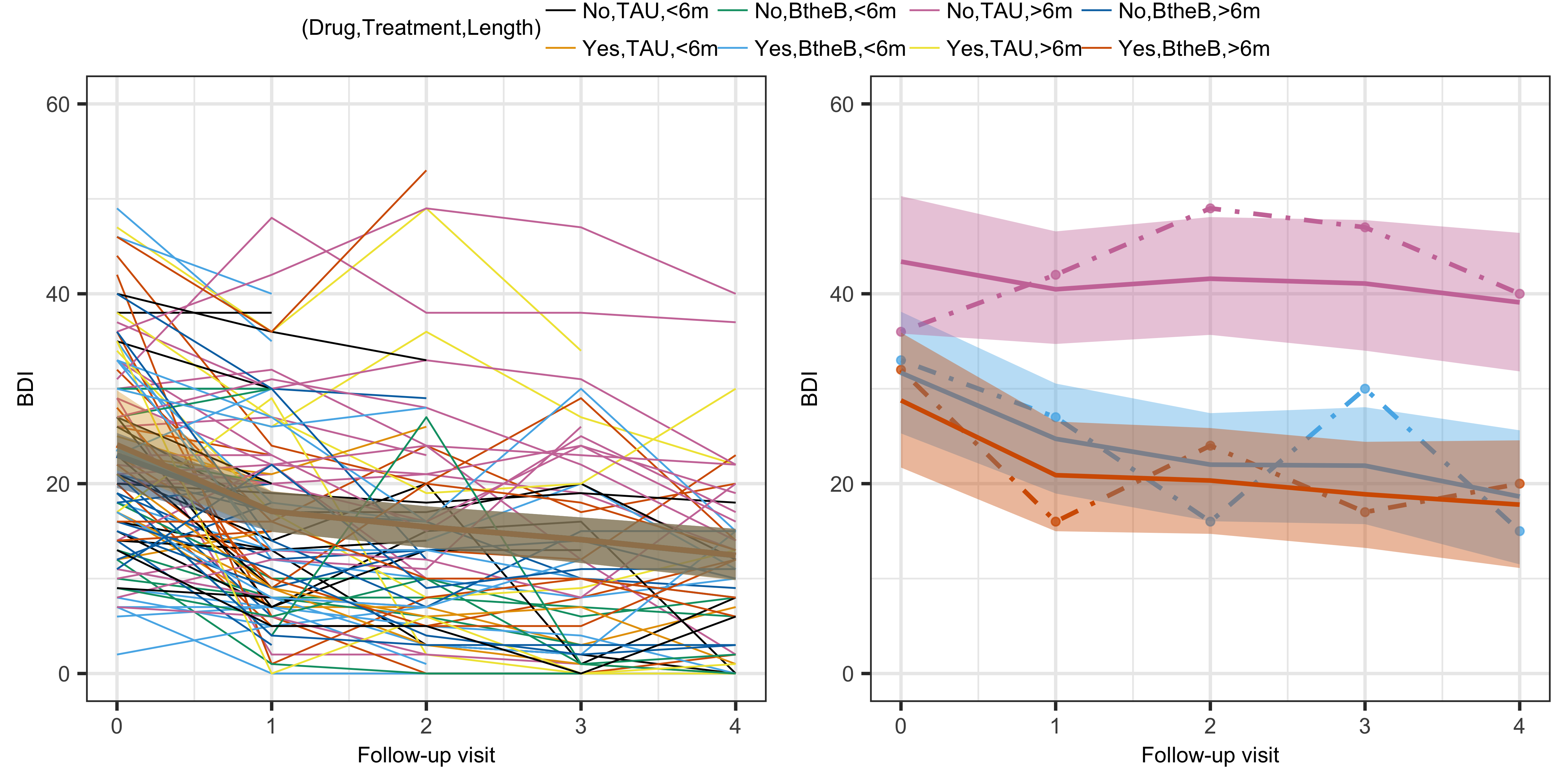}
	\caption{Results for the Beat the Blues data: The left panel shows the estimated posterior means and $95\%$ point wise credible intervals for the fixed effects,
	superimposed on the observed trajectories $y_{i,\ell,t}$. 
	The right panel shows the estimated posterior means and $95\%$ point wise credible intervals for three individual specific curves,
	superimposed on the associated individual responses $y_{i,\ell,t}$.}
	\label{fig: depression_results}
\end{figure}

Figure \ref{fig: depression_results} (left panel) shows the estimated posterior means and associated $95\%$ point wise credible intervals for the group specific curves for each of the eight possible combinations of the three categorical predictors.
As illustrated, no significant differences were detected at any time point by our model.
Thus, our model seems to confirm the conclusions of \citetlatex{gilbody2015computerised}. 
Figure \ref{fig: depression_results} (right panel) shows the estimated posterior means and associated $95\%$ point wise credible intervals for three individual specific curves.

\subsection{Childhood Asthma Management Program (CAMP) Data} \label{sec: sm camp data}

The CAMP study \citeplatex{childhood2000long} was a randomized clinical trial for children with asthma. 
We analyze a subsampled and anonymized version of the original data, 
publicly available at the National Institute of Health (NIH) \href{https://biolincc.nhlbi.nih.gov/teaching/}{\texttt{website}}.
We use this semi-synthetic data set for illustrative purposes alone, the conclusions must not be extended to the original study.

The trial's goal was to infer the long-term impact of three treatment assignments (Budesonide, Nedocromil, or placebo) on pulmonary function. 
A total of $n=1,041$ children aged 5-12 years were enrolled and balanced across the treatment groups. 
We use one of the endpoints of the trial as the response variable, namely lung function as measured by the Forced Vital Capacity (FVC). 
Other predictors include the children's gender and ethnicity as well as dummy variables indicating if participants shared their house with pets and/or smokers.
The size of the unstructured model is $T \prod_{j=1}^{p} x_{j,\max} = 16 \times 3 \times 2^{3} \times 4 = 1,536$. 

\begin{figure}[!ht]
	\centering
	\includegraphics[width=.99\linewidth]{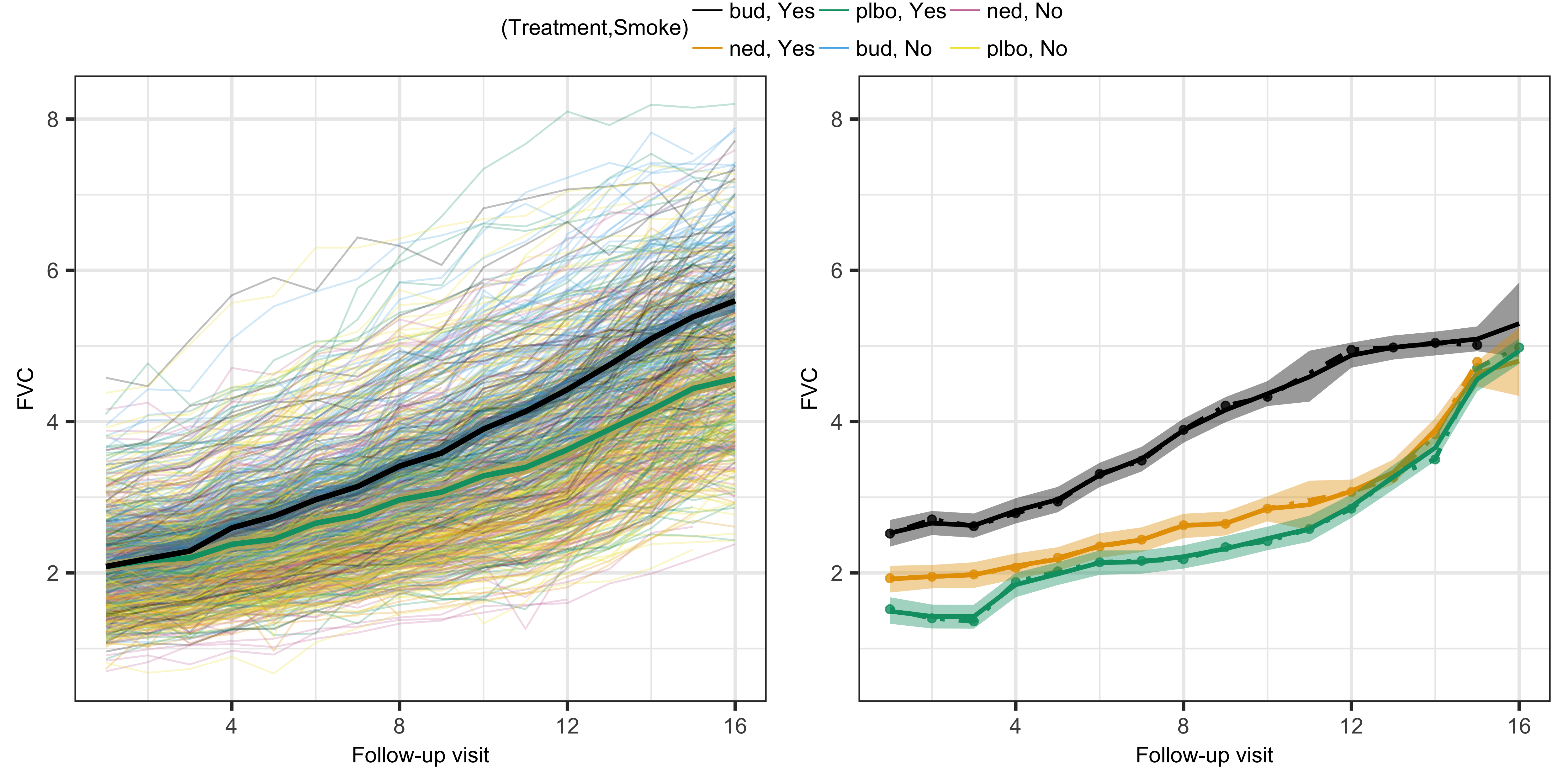}
	\caption{Results for the CAMP data: The left panel shows the estimated posterior means and $95\%$ point wise credible intervals for the fixed effects,
	superimposed on the observed trajectories $y_{i,\ell,t}$. 
	The right panel shows the estimated posterior means and $95\%$ point wise credible intervals for three individual specific curves,
	superimposed on the associated individual responses $y_{i,\ell,t}$.}
	\label{fig: asthma_results}
\end{figure}

Figure \ref{fig: asthma_results} (left panel) shows the estimated posterior means and associated $95\%$ point wise credible intervals 
for the fixed effects curves for each of the six possible paired combinations the predictors, 
namely treatment assignment and presence of smokers in the household.
The only significant predictor in the model is the treatment assignment variable. 
In particular, participants assigned to Budesonide appear to have larger FVC. 
These differences, however, emerge only after the third visit and seem to become more pronounced as time progresses. 
Figure \ref{fig: asthma_results} (right panel) shows the estimated posterior means and associated $95\%$ point wise credible intervals for three individual specific curves, exhibiting a high degree of heterogeneity around the mean profiles.

\subsection{National Longitudinal Survey of Youth Data} \label{sec: sm youth data}

The national longitudinal survey of youth of 1997 \citeplatex[NLSY97,][]{moore2000national} is a longitudinal study that 
follows a nationally representative sample of the American youth born between 1980 and 1984 on various aspects of life.
Participants enter the study between the ages of 12 and 16.
Interviews were conducted annually from 1997 to 2011 and biennially since then.
The NLSY97 collects information on respondents' labor market behavior and educational experiences. 
The survey also includes data on the participants' family backgrounds to help researchers assess the impact of  environmental factors on these labor statistics. 
We use a publicly available version of the data that can be found at the U.S. Bureau of Labor Statistics \href{https://www.bls.gov/nls/getting-started/accessing-data.htm}{\texttt{website}}.

We analyze a subsample of the original data consisting of a random $25\%$ of the participants 
which resulted in $n=2,120$ youths surveyed for a total of $16,188$ questionnaires.
We use yearly income (in $\$10,000$ units) as the response variable and determine the effects of the socio-demographic variables on this outcome. 
The exogenous covariates include the participants' gender and ethnicity. 
The time-varying predictors are region, marital status and a dichotomous variable indicating if the participants live in urban or rural areas.
The size of the unstructured model is $T \prod_{j=1}^{p} x_{j,\max} = 16 \times 2^{2} \times 4^{2} \times 5 = 5,120$.

\begin{figure}[!h]
	\centering
	\begin{center}
		\includegraphics[width=0.85\linewidth]{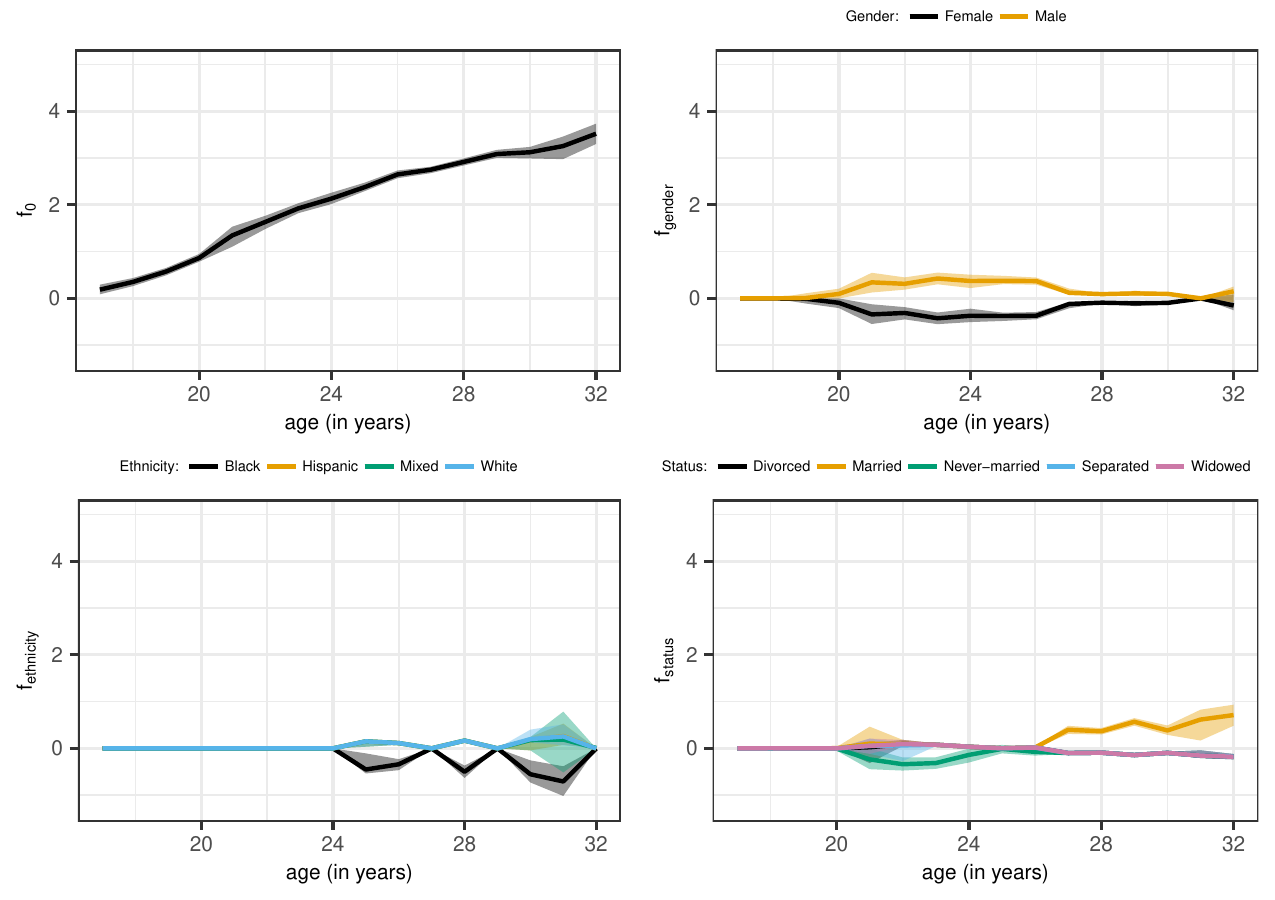}
	\end{center}
	\vskip-15pt
	\caption{Results for the NLSY97 data: the significant predictors were gender, ethnicity and marital Status.
	Showing their estimated posterior means (colored lines) and $95\%$ point wise credible intervals. 
	Clockwise from top left: overall mean; main effects of gender; main effects of ethnicity; and main effects of marital status.
	}
	\label{fig: youth effects}
\end{figure}

Our analysis produced three significant predictors, namely gender, ethnicity and marital status.
Displaying every level combination of these predictors is difficult, 
so we show only their main effects as defined in equation (\ref{eq: main and interaction}) in Section \ref{sec: fixed effects} of the main paper and discussed in Section \ref{sec: sm main and interaction effects} in the supplementary materials. 
Figure \ref{fig: youth effects} shows the estimated posterior means and associated $95\%$ point wise credible intervals for the overall mean (top left) and the predictors' main effects (other panels). 
The top left panel shows that incomes increase as a function of age on average across the entire sample.
The top right panel shows a gender gap that becomes especially important between the ages of 21 and 26, with men earning up to $\$5,000$ more than women.
A racial gap also appears to be significant, with African American participants having lower earnings, as illustrated in the bottom left panel.
Finally, as illustrated in the bottom right panel, married couples seem to have higher incomes compared to single earners, 
perhaps an artifact of their joint filing of taxes.

\baselineskip=14pt
\bibliographystylelatex{natbib}
\bibliographylatex{biblio}

\end{document}